\documentclass[11pt,a4paper]{article}
\pdfoutput=1 
\usepackage[dvipdfmx]{graphicx}
\usepackage[dvipdfmx]{color} % for output of png file

\usepackage{amsmath,amssymb,amsfonts,color,scalefnt}
\usepackage{epsfig,amsthm}
\usepackage{soul}
\usepackage[makeroom]{cancel}
\usepackage{mathrsfs}
\usepackage{epstopdf}
\usepackage[utf8]{inputenc}
\usepackage{hyperref}
\usepackage{slashed}
\usepackage{pict2e}
\usepackage{tikz}
\usepackage{cite}
\usepackage{float}
\usepackage[percent]{overpic}
\usepackage{multirow}
\usepackage{bbold}
\usepackage{physics}
\usepackage{booktabs}
\usepackage{mwe}
\usepackage [autostyle, english = american]{csquotes}
\usepackage{bbm}

\usepackage{mathtools}
\MakeOuterQuote{"}

\newcommand{\beq}{\begin{eqnarray}}
\newcommand{\eeq}{\end{eqnarray}}
\newcommand{\ba}{\begin{eqnarray}}
\newcommand{\ea}{\end{eqnarray}}
\newcommand{\be}{\begin{equation}}
\newcommand{\ee}{\end{equation}}
\newcommand{\bpmatrix}{\begin{pmatrix}}
\newcommand{\epmatrix}{\end{pmatrix}}
\renewcommand{\braket}[1]{\left(#1\right)}

\renewcommand{\Re}{{\rm Re}}
\renewcommand{\Im}{{\rm Im}}

\renewcommand{\Re}{\text{Re}\,}
\renewcommand{\Im}{\text{Im}\,}
\newcommand{\crn}{\nonumber \\}
\newcommand{\fr}{\frac}
\newcommand{\sbeta}{{s_{\beta}}}
\newcommand{\cbeta}{{c_{\beta}}}
\newcommand{\tbeta}{{t_{\beta}}}
\newcommand{\DRb}{{{\overline{\rm DR}}}}
\newcommand{\ZH}{{\bf Z}^{H}}
\newcommand{\ie}{{\it i.e.\;}}
\newcommand{\eg}{{\it e.g.\;}}
\newcommand{\hc}{\text{ h.c.}}

\newcommand{\comment}[1]{\ignorespaces}

\newcommand{\s}{\newline \vspace*{-3.5mm}}

\newcommand{\GeV}{{\text{GeV}}}
\newcommand{\TeV}{{\text{TeV}}}

\newcommand{\mHp}{M_{{H^\pm}}}

\newcommand{\ti}[1]{\tilde{#1}}
\newcommand{\m}[1]{m_{#1}}

\begin{document}

\title{
	\vspace*{-3cm}
\phantom{h} \hfill\mbox{\small IFIRSE-TH-2020-4} 
\\[-1.1cm]
	\phantom{h} \hfill\mbox{\small KA-TP-24-2020}
	\\[1cm]
	%\vspace{13mm}   
	\textbf{One-loop Corrections to the Two-Body Decays of the Charged
          Higgs Bosons \\ in the Real and Complex NMSSM}}

\date{}
\author{
Thi Nhung Dao$^{1\,}$\footnote{E-mail: \texttt{dtnhung@ifirse.icise.vn}},
Margarete M\"{u}hlleitner$^{3\,}$\footnote{E-mail:
	\texttt{margarete.muehlleitner@kit.edu}},
Shruti Patel$^{3,4\,}$\footnote{E-mail: \texttt{shruti.patel@kit.edu}},
Kodai Sakurai$^{3\,}$\footnote{E-mail:  \texttt{kodai.sakurai@kit.edu}}
\\[9mm]
{\small\it
$^1$Institute For Interdisciplinary Research in Science and Education, ICISE,}\\
{\small\it 590000 Quy Nhon, Vietnam.}\\[3mm]
{\small\it
$^3$Institute for Theoretical Physics, Karlsruhe Institute of Technology,} \\
{\small\it Wolfgang-Gaede-Str. 1, 76131 Karlsruhe, Germany.}\\[3mm]
{\small\it$^4$Institute for Nuclear Physics, Karlsruhe Institute of Technology,
76344 Karlsruhe, Germany.}\\[3mm]
}
\maketitle

%
%
%\date{\today}

\begin{abstract}
We evaluate the full next-to-leading order supersymmetric (SUSY) electroweak and
SUSY-QCD corrections to the on-shell two-body decays of the charged
Higgs bosons in the framework of the CP-conserving and CP-violating
Next-to-Minimal Supersymmetric extension of the Standard Model
(NMSSM). Our corrections are implemented in the code {\tt NMSSMCALCEW}
in order to compute the branching ratios of the charged Higgs boson
where we also take into account the 
state-of-the-art QCD corrections already included in the code. 
We investigate the impact of the NLO corrections for each decay mode in a
wide range of the parameter space that is allowed by the theoretical
and experimental constraints. The new version of {\tt NMSSMCALCEW} is made
publicly available.

\end{abstract}

%\maketitle
%\tableofcontents

\thispagestyle{empty}
\vfill
\newpage

%-----------------------------
\section{Introduction}
\label{sec:intro}
%------------------------------------------------
%\input Introduction.tex

% importance of indirect search of Higgs bosons and precise
% predictions
While the Higgs boson with a mass of 125 ${\rm GeV}$ \cite{Aad:2015zhl}
discovered by the LHC experiments ATLAS~\cite{Aad:2012tfa} and
CMS~\cite{Chatrchyan:2012ufa} behaves very Standard Model
(SM)-like~\cite{Aad:2019mbh,Sirunyan:2018koj} the pending open
questions within the SM call for extensions of the Higgs sector. The
lacking direct discovery of any new physics sign so far forces us to
focus more and more on the indirect discovery of physics beyond the
SM which in turn requires precise predictions of the considered
observables. \s
 
Among the most popular and best studied models beyond the SM is
supersymmetry (SUSY) 
\cite{Golfand:1971iw, Volkov:1973ix, Wess:1974tw, Fayet:1974pd,
Fayet:1977yc, Fayet:1976cr, Nilles:1983ge, Haber:1984rc, Sohnius:1985qm,
Gunion:1984yn, Gunion:1986nh}, which requires at least two complex
Higgs doublets. This 
Minimal Supersymmetric SM (MSSM)
\cite{Gunion:1989we,Martin:1997ns,Dawson:1997tz,Djouadi:2005gj} is
extended by an additional
complex singlet superfield in the Next-to-Minimal Supersymmetric SM
(NMSSM) \cite{Fayet:1974pd,Barbieri:1982eh,Dine:1981rt,Nilles:1982dy,Frere:1983ag,Derendinger:1983bz,Ellis:1988er,Drees:1988fc,Ellwanger:1993xa,Ellwanger:1995ru,Ellwanger:1996gw,Elliott:1994ht,King:1995vk,Franke:1995tc,Maniatis:2009re,Ellwanger:2009dp}. With a Higgs sector consisting of seven Higgs bosons, three
neutral CP-even, two neutral CP-odd and two charged Higgs bosons, it
entails a rich phenomenology. In this context, charged Higgs bosons
play an important role.  Their discovery would be a clear
manifestation of extended Higgs sectors. Moreover, their possible decay
channels induce interesting non-SM decay signatures, besides those in
SUSY pairs, in particular modes with a massive gauge boson and a neutral Higgs
boson in the final state. \s

The very SM-like nature of the discovered Higgs boson calls for
sophisticated experimental techniques together with precise
theoretical predictions in order to reveal new physics signs. In the
context of the NMSSM, various radiative corrections to Higgs
observables have been calculated. More specifically in the context of
Higgs boson decay widths, the next-to-leading order (NLO)
SUSY-electroweak (EW) and SUSY-QCD corrections to the
decays of CP-odd Higgs bosons of the CP-conserving NMSSM into stop
pairs have been computed in \cite{Baglio:2015noa}. The full one-loop
renormalization of the CP-conserving NMSSM has been worked out in
\cite{Belanger:2016tqb,Belanger:2017rgu} together with the computation
of the one-loop two-body Higgs decays in the on-shell (OS)
renormalization scheme. In \cite{Goodsell:2017pdq} a generic
calculation of the two-body decays widths at full one-loop level was
provided in the $\overline{\mbox{DR}}$ scheme. The full one-loop
corrections for the neutral Higgs decays into fermions and gauge
bosons was performed in \cite{Domingo:2018uim} in the framework of the CP-violating
NMSSM and combined with the leading QCD corrections. More recently,
the effects of Sudakov logarithms on fermionic decays of heavy Higgs
bosons, which appear though radiative corrections of electroweak gauge
bosons, have been studied in Ref.~\cite{Domingo:2019vit}. In \cite{Baglio:2019nlc},
members of our group completed the evaluation of the NLO SUSY-EW and
SUSY-QCD corrections to the full set of two-body on-shell decays of
the neutral Higgs bosons in the CP-violating NMSSM, including the
state of the art QCD corrections. For the Higgs-to-Higgs decays, the
complete one-loop corrections~\cite{Nhung:2013lpa} and two loop
corrections of order
$\mathcal{O}(\alpha_{s}\alpha_{t})$\cite{Muhlleitner:2015dua} have
been calculated in both the CP-conserving and the CP-violating
NMSSM. As for the charged Higgs bosons, authors of our group have
computed the SUSY-EW corrections to the charged Higgs two-body
decays into a neutral Higgs boson and a $W^+$ boson and studied the
gauge dependence arising from the mixing of different loop orders due
to the inclusion of mass corrections to the external Higgs
boson~\cite{Dao:2019nxi} in order to comply with phenomenological
constraints. The authors of \cite{Domingo:2020wiy} discussed this problem further and
proposed two strategies to preserve or restore gauge invariance. \s

In this work, we complete the one-loop higher-order corrections to the
on-shell charged Higgs boson decays and compute the NLO SUSY-EW and
NLO SUSY-QCD corrections to all charged Higgs boson decays into
two-particle final states in the framework of the CP-conserving and
CP-violating NMSSM, with the exception of Higgs pair final states that
are left for future work. More specifically, we evaluate the corrections to
the decays into SM fermions, electroweakinos, sleptons and squarks. 
The one-loop corrections are based on the
  renormalization schemes that we have introduced and applied 
  in Refs.~\cite{Graf:2012hh,Muhlleitner:2014vsa,Dao:2019qaz,Muhlleitner:2015dua,Baglio:2019nlc}.
%are renormalized in the mixed OS-
%$\overline{\rm DR}$ scheme, which has been introduced and applied
%in~Refs.~\cite{Ender:2011qh,Graf:2012hh,Muhlleitner:2014vsa,Dao:2019qaz}.
Our newly computed higher-order corrections to the charged Higgs
decays as well as those obtained previously in \cite{Dao:2019nxi} are
included in the Fortran code 
{\tt NMSSMCALCEW} \cite{Baglio:2019nlc}. The program is based on {\tt
  NMSSMCALC}~\cite{Baglio:2013iia} that has been derived
by extending the {\tt Fortran} code {\tt
  HDECAY}~\cite{Djouadi:1997yw,Djouadi:2018xqq} to the NMSSM and as
such contains the state-of-the-art QCD corrections and 
relevant off-shell decays as well as higher-order SUSY corrections
through effective couplings. In the first version of {\tt
  NMSSMCALCEW} we already included the SUSY-EW 
corrections to the neutral NMSSM Higgs boson decays of the
CP-violating NMSSM published in
Ref.~\cite{Baglio:2019nlc} as well as their SUSY-QCD corrections to the coloured
final states. The code {\tt NMSSMCALCEW} with now also the higher-order
corrections to the charged Higgs boson decays is available at the url: 
\begin{center}
{\tt http.//www.itp.kit.edu/$\sim$maggie/NMSSMCALCEW/}
\end{center}
With our implemented corrections in {\tt
  NMSMMCALCEW} we compute the decay widths and
branching ratios of the charged Higgs bosons in the CP-conserving and
CP-violating NMSSM and investigate the impact of the SUSY-EW and
SUSY-QCD corrections, taking into account the recent 
theoretical and experimental constraints. \s

The paper is organized as follows. In section~\ref{sec:nmssmtree}, we
introduce the NMSSM at tree level to fix our notation for the
higher-order computations.  In section~\ref{sec:renorm}, we briefly
discuss the renormalization of the 
Higgs sector, the electroweakino and the squark sector. 
In section~\ref{sec:calc} we move on to the calculations of the radiative
corrections to the decay widths of the on-shell two-body decays of the
charged Higgs boson. Our numerical analysis is presented in
section~\ref{sec:numerics}. Conclusions are given in section~\ref{sec:conclusions}.

%------------------------------------------------
\section{The NMSSM at Tree Level \label{sec:nmssmtree}}
%------------------------------------------------
%\input Lagrangian.tex

In this section, we briefly introduce the model and set up the
notation used throughout the paper. Our notation and conventions
follow those already used and explained in detail in our previous
works
\cite{Ender:2011qh,Graf:2012hh,Nhung:2013lpa,Muhlleitner:2014vsa,Dao:2019qaz,Muhlleitner:2015dua,Dao:2019nxi,Baglio:2019nlc}. We work in the framework of
the scale-invariant NMSSM, in which the superpotential is constrained
by a $\mathbb{Z}_3$ symmetry. The NMSSM is obtained from the MSSM by
adding a gauge-singlet chiral superfield ${\hat S}$ 
to the MSSM field content. Since the new additional field is not
charged under the $SU(2)_L \times U(1)_Y $ symmetry, the modification of
the supersymmetric Lagrangian with respect to the MSSM only arises in the
superpotential, 
\begin{align}
\mathcal{W}_{\rm NMSSM}=\mathcal{W}_{\rm MSSM} -\epsilon_{ij} \lambda
  \hat S \hat H_d^i\hat H_u^j+\frac{1}{3}\kappa {\hat S}^3, 
\end{align}
where ${\hat H}_u, {\hat H}_d$ are the Higgs doublet superfields and $\epsilon_{ij}$ ($i,j=1,2$) is  the totally antisymmetric tensor, with
$\epsilon_{12}= \epsilon^{12}=1$, and $i,j$ denote the indices of the
fundamental $SU(2)_L$ representation. Here and in the following we sum
over repeated indices. The dimensionless NMSSM-specific parameters
$\lambda$ and $\kappa$ are complex in the CP-violating NMSSM. The MSSM
superpotential $\mathcal{W}_{\rm MSSM}$ is  written in terms of the quark and lepton chiral superfields ${\hat Q}, {\hat L},{\hat U}, {\hat D}$ and ${\hat E}$ as 
\begin{align}
\mathcal{W}_\text{MSSM} = \epsilon_{ij} [y_e \hat{H}^i_d \hat{L}^j
\hat{E}^c + y_d \hat{H}_d^i \hat{Q}^j \hat{D}^c - y_u \hat{H}^i_u
\hat{Q}^j \hat{U}^c] \,,
\label{eq:mssmsuperpot}
\end{align}
where, for simplicity, we neglect colour and generation
indices. Assuming flavour conservation, the Yukawa couplings $y_u,\
y_d$ and $y_e$ are diagonal 3$\times$3 matrices in flavour space. In
accordance with the applied $\mathbb{Z}_3$ symmetry, the MSSM
parameter $\mu$ has been set to zero. \s

In the NMSSM, the soft supersymmetry (SUSY) breaking Lagrangian reads
\begin{align}\label{eq:breaking_term}\notag
\mathcal{L}_{\rm soft,NMSSM} = & -m_{H_d}^2 H_d^\dagger H_d - m_{H_u}^2
H_u^\dagger H_u -
m_{\tilde{Q}}^2 \tilde{Q}^\dagger \tilde{Q} - m_{\tilde{L}}^2 \tilde{L}^\dagger \tilde{L}
- m_{\tilde{u}_R}^2 \tilde{u}_R^*
\tilde{u}_R - m_{\tilde{d}_R}^2 \tilde{d}_R^* \tilde{d}_R
\nonumber      \\\nonumber
& - m_{\tilde{e}_R}^2 \tilde{e}_R^* \tilde{e}_R - (\epsilon_{ij} [y_e A_e H_d^i
\tilde{L}^j \tilde{e}_R^* + y_d
A_d H_d^i \tilde{Q}^j \tilde{d}_R^* - y_u A_u H_u^i \tilde{Q}^j
\tilde{u}_R^*] + \mathrm{h.c.})      \\
& -\frac{1}{2}(M_1 \tilde{B}\tilde{B} + M_2
\tilde{W}_j\tilde{W}_j + M_3 \tilde{G}\tilde{G} + \mathrm{h.c.}) \\ \nonumber
& - m_S^2 |S|^2 +
(\epsilon_{ij} \lambda
A_\lambda S H_d^i H_u^j - \frac{1}{3} \kappa
A_\kappa S^3 + \mathrm{h.c.}) \;,
\end{align}
where $\tilde{Q}=(\tilde{u}_L,\tilde{d}_L)^T$,
$\tilde{L}=(\tilde{\nu}_L,\tilde{e}_L)^T$, $H_d$ and $H_u$
are the complex scalar components of the chiral superfields $\hat{Q}$,
$\hat{L}$, $\hat H_d$ and $\hat H_u$, respectively. Similarly,
$\tilde{u}_R$, $\tilde{d}_R$ and $\tilde{e}_R$ denote the complex
scalar components of the right-handed quark and lepton chiral
superfields.  
Moreover, $\tilde{B},\ \tilde{W}_i$ $(i=1,2,3)$ and $\tilde{G}$ represent the bino, wino and gluino fields, with masses $M_j\ (j=1,2,3)$, respectively. Finally, the parameters $m_X^2$ are the squared soft SUSY-breaking masses of the fields $X=S, H_d, H_u, \tilde{Q},
\tilde{u}_R, \tilde{d}_R, \tilde{L}, \tilde{e}_R$. 
The terms $A_{\alpha}\ (\alpha=u, d, e, \lambda,\kappa)$ are the soft
SUSY-breaking trilinear couplings. In the CP-violating NMSSM, the
trilinear couplings and the gaugino mass parameters $M_j\ (j=1,2,3)$
are complex.  

\paragraph{The Higgs Sector}
The tree-level Higgs potential derived from the $F$- and $D$-terms in
the supersymmetric Lagrangian, and the soft SUSY-breaking Lagrangian,
Eq.\eqref{eq:breaking_term}, reads 
\begin{align}
\label{eq:potential}
\notag
V_{H}&=|\lambda S|^2\left(H_{u}^\dagger H_u+H_{d}^\dagger H_d\right)
       +\left|-\epsilon_{ij}\lambda\left(H_d^iH_u^j\right)+\kappa S^2\right|^2
        \\ \notag
&+\frac{1}{2}g_2^2\left|H_u^\dagger
  H_d\right|^2+\frac{1}{8}(g_1^2+g_2^2)\left(H_u^\dagger
  H_u-H_d^\dagger H_d\right)^2 \\  
&+m_{H_u}^2H_u^\dagger H_u+m_{H_d}^2H_d^\dagger
  H_d+m_S^2\left|S\right|^2+
  \left[
 -\epsilon_{ij}\lambda A_\lambda\left(H_d^iH_u^j\right)S
  +\frac{1}{3}\kappa A_\kappa S^3+{\rm h.c.}\right], 
%\label{eq:HiggsPotential}
\end{align}
where $g_1$ and $g_2$ are the gauge couplings of the $U(1)_Y$ and
$SU(2)_L$ symmetry, respectively. The complex scalar Higgs doublet 
fields $H_u$ and $H_d$ and the complex singlet field $S$ are expressed
in terms of the  component fields and vacuum expectation values (VEVs)
as 
\begin{align}
H_u=
e^{i\varphi_u}\begin{pmatrix}
h_u^+\\
\frac{1}{\sqrt{2}}(v_u+h_u+ia_u)
\end{pmatrix},\ 
H_d=
\begin{pmatrix}
\frac{1}{\sqrt{2}}(v_d+h_d+ia_d) \\
h_d^-
\end{pmatrix},\ 
S=\frac{1}{\sqrt{2}}e^{i\varphi_s}(v_s+h_s+ia_s) \;, \label{eq:HiggsVEVexpand}
\end{align}
where $v_u, v_d $ and $v_s$ are the VEVs of $H_u$, $H_d$, and $S$, respectively. 
The two CP-violating phases $\varphi_u$ and $\varphi_s$ describe the phase
differences between the VEVs. Following
Refs.\cite{Graf:2012hh,Muhlleitner:2014vsa,Dao:2019qaz,Muhlleitner:2015dua},
we set the phases of the Yukawa couplings to zero and rephase the
left- and right-handed up-quark fields as $u_L \to e^{-i \varphi_u}
u_L$ and $u_R \to e^{i\varphi_u} u_R$, so that the quark and lepton
mass terms yield real masses. Inserting Eq.~(\ref{eq:HiggsVEVexpand}) into
the gauge sector of the Lagrangian, we obtain the masses the charged
gauge bosons $W^\pm$ and the neutral $Z$ boson, respectively, 
\begin{align}
M_W=\frac{1}{2}vg_2 \quad \mbox{and} \quad M_Z=\frac{1}{2}v\sqrt{g_1^2+g_2^2},
\end{align}
with the SM VEV $v$ $\simeq 246 \,{\rm GeV}$ being related to $v_d$ and $v_u$ as 
\begin{align}
v^2=v_u^2+v_d^2 \;,
\end{align}
with 
\beq
\tan\beta = \frac{v_u}{v_d} \;.
\eeq
The weak mixing angle $\theta_W$ is defined as
\be 
\cos \theta_W = \fr{M_W}{M_Z} \,.
\ee
Substituting Eq.~(\ref{eq:HiggsVEVexpand}) into
Eq.~(\ref{eq:potential}), we can express the Higgs potential as 
\beq 
V_H &=& V_H^{\text{const}} + t_{\phi_i} \phi_i
+  \begin{pmatrix} h_d^+ & h_u^+ \end{pmatrix}
\mathbf{M_{H^{\pm}}} \begin{pmatrix} h_d^- \\ h_u^- \end{pmatrix} \crn 
&& + \frac{1}{2} \phi \mathbf{M_{\phi\phi}} \phi^T +
\lambda_{ijk}^{\phi^3}\phi_i\phi_j\phi_k +  V_H^{\phi^4}\,,  
\eeq 
where  $\phi=(h_d,h_u,h_s,a_d,a_u,a_s)$,  $i,j,k=1,...,6$. Explicit
expressions for the tadpoles $t_{\phi_i}$ and the mass matrix squared $\mathbf{M_{\phi\phi}}$ can be found in Refs.\cite{Ender:2011qh,Graf:2012hh}. 
The trilinear couplings $\lambda_{ijk}^{\phi^3}$ have been derived
in Refs.\cite{Nhung:2013lpa,Muhlleitner:2015dua}. The constant and
quartic terms are summarized in $V_H^{\text{const}}$ and $
V_H^{\phi^4}$, respectively. \s

The charged Higgs mass matrix in the 't\,Hooft-Feynman gauge, see
\cite{Baglio:2019nlc}, is given by 
\begin{align} 
\mathbf{M_{H^{\pm}}}= \frac 12 \bpmatrix \tbeta & 1 \\
                                1 & 1/\tbeta \epmatrix  \bigg[& M_W^2 s_{2\beta} +
\fr{\abs{\lambda}v_s}{\cos( \varphi_\lambda+\varphi_u+ \varphi_s) }
\braket{\sqrt{2} \,\Re A_\lambda+\abs{\kappa}v_s \cos(\varphi_\kappa+3\varphi_s) } \crn  
& -\fr{2 |\lambda|^2 M_W^2 s_{\theta_W}^2}{e^2}s_{2\beta} \bigg]  + M_W^2
\bpmatrix \cbeta^2 & -\cbeta\sbeta \\
                                -\cbeta\sbeta & \sbeta^2 \epmatrix\;, 
\end{align}
where $\varphi_\lambda$ and $\varphi_\kappa$ are the angular arguments
of $\lambda$ and $\kappa$, respectively. Here and in the following, 
we use the shorthand notation $t_x\equiv\tan x, s_x \equiv \sin x, c_x
\equiv \cos x$. Rotating the interaction states $h_d^\pm,h_u^\pm$  by
a rotation with the angle $\beta_c=\beta$, we obtain the charged Higgs
boson $H^\pm$ with mass 
\be 
M_{H^\pm}^2 = M_W^2 +\fr{\abs{\lambda}v_s}{s_{2\beta}\cos(
  \varphi_\lambda+\varphi_u+ \varphi_s) } \braket{\sqrt{2} \Re
  A_\lambda+\abs{\kappa}v_s \cos( \varphi_\kappa+ 3\varphi_s)
}-\fr{2 |\lambda|^2 M_W^2 s_{\theta_W}^2}{e^2} 
\ee  
and the charged Goldstone boson $G^\pm$ mass given by the charged $W^+$
mass in the 't\,Hooft-Feynman gauge that we apply. \s

The neutral Higgs boson mass eigenstates are obtained via the diagonalization  of the mass matrix squared $\mathbf{M_{\phi\phi}}$ by an orthogonal matrix $\mathcal R$,
\beq
{\text{diag}}(m_{h_1}^2, m_{h_2}^2,m_{h_3}^2,m_{h_4}^2,
m_{h_5}^2,M_Z^2)&=& \mathcal R \mathbf{M_{\phi\phi}} \mathcal R^T,\\ 
(h_1,h_2,h_3,h_4,h_5,G^0)^T& = & \mathcal R (h_d, h_u,
h_s, a_d, a_u, a_s)^T, \label{eq:rotgaugemasstree}
\eeq
where $G^0$ is the neutral Goldstone boson with its mass given by the
neutral $Z$ boson mass in the applied 't\,Hooft-Feynman gauge, and the tree-level neutral
Higgs boson masses are ordered as $m_{h_1} \leq ... \leq m_{h_5}$. \s

The tree-level Higgs sector of the CP-violating NMSSM is described by
eighteen independent input parameters which we choose as
\begin{align}
& m_{H_d}^2, m_{H_u}^2, m_S^2,M_W^2, M_Z^2, e, 
  \tan\beta, v_s, \varphi_s, \varphi_u, 
 \abs{\lambda}, \varphi_\lambda, \abs{\kappa},
\varphi_\kappa, \Re A_{\lambda},\Im A_\lambda, \Re A_{\kappa},
\Im A_\kappa~, \label{eq:orgparset}
\end{align}
where the three Lagrangian parameters $g_1$, $g_2$ and $v$ have been
traded for the three physical observables $M_W$, $M_Z$ and the
electric coupling $e$. In the following, the three
soft SUSY-breaking mass parameters $m_{H_d}^2, m_{H_u}^2, m_S^2$ as well as $\Im
  A_\lambda$ and $\Im A_\kappa$ will be replaced by the tadpole parameters $t_{\phi_i}$
  ($\phi_i=h_d, h_u, h_s, a_d, a_s$)\footnote{The tadpole parameter
    $t_{a_u}$ is not an independent quantity since it is related to
    $t_{a_d}$ as $t_{a_u}=\frac{t_{a_d}}{t_\beta}$.}. At tree level
  these tadpoles vanish at the minimum of the Higgs potential. They
  are, however, non zero at one-loop level, and hence kept for the renormalization
procedure. There is also the option to replace the
  parameter $\Re A_{\lambda}$ by the charged Higgs boson mass.

%------------------------------------------------
\paragraph{The Electroweakino Sector \label{sec:GSS}}
%------------------------------------------------ 
The fermionic partners of the neutral Higgs bosons, the neutral
higgsinos $\tilde{H}_u$, $\tilde{H}_d$ and the singlino
$\tilde{S}$, mix with the neutral gauginos $\tilde{B}$ and
$\tilde{W}^3$, resulting in five neutralinos. The mass term of the
neutralinos is given in the basis of the Weyl spinor field
$\psi^0=(\tilde{B},\ \tilde{W}^3,\ \tilde{H}_d,\ \tilde{H}_u,
\tilde{S})^T$ as  
\begin{align}
\mathcal{L}^{\psi^0}_{\rm Mass}=\frac{1}{2}(\psi^0)^TM_N\psi^0+{\rm h.c.}\,,
\end{align}
with the $5\times 5$ symmetric mass matrix $M_N$, 
\begin{align}
M_N=
\begin{pmatrix}
M_1 & 0&-c_\beta M_Z s_{\theta_W}& s_\beta M_Z s_{\theta_W} e^{-i \varphi_u}& 0 \\
0 & M_2&M_W c_{\beta}&- s_\beta M_W e^{-i \varphi_u}  & 0 \\
-c_\beta M_Z s_{\theta_W}& M_W c_\beta  &0&\fr{-\lambda v_S e^{i
    \varphi_s}}{\sqrt{2}}& \fr{-\lambda s_\beta ve^{i
    \varphi_u}}{\sqrt{2}} \\ 
s_\beta M_Z s_{\theta_W} e^{-i \varphi_u}&- s_{\beta}M_W e^{-i
  \varphi_u} &\fr{-\lambda v_S
e^{i \varphi_s}}{\sqrt{2}}& 0 &\fr{-\lambda c_\beta v}{\sqrt{2}}\\
0&0 &\fr{-\lambda s_\beta v e^{i \varphi_u}}{\sqrt{2}}& \fr{-\lambda
  c_\beta v}{\sqrt{2}}& \sqrt{2}\kappa v_s e^{i \varphi_s} 
\end{pmatrix}.
\end{align}
By introducing the  $5\times 5$ orthogonal matrix $N$, the mass matrix
is diagonalized by performing the
  rotation\footnote{We choose the mass eigenvalues to be positive,
  possible complex phases are absorbed into the rotation
  matrix.}   
\be
{\rm diag}(m_{\tilde{\chi}^0_1}, m_{\tilde{\chi}^0_2}, m_{\tilde{\chi}^0_3}, m_{\tilde{\chi}^0_4}, m_{\tilde{\chi}^0_5})=N^*M_NN^\dagger,
\ee
where the matrix $N$ transforms the fields $\psi^0_i \equiv
(\psi^0)_i$ into the mass eigenstates $\chi^0_i \equiv
(\chi^0)_i$ $(i=1,..., 5)$, i.e., 
\begin{align}
 \chi^0=N\psi^0. 
 \end{align}
The neutralino mass eigenstates are given by the Majorana fields
$\tilde{\chi}^0_i=(\chi^0_i,\overline{\chi^0_i})^T$ $(i=1,...,5)$. 
 By convention, the mass ordering of the  $\tilde{\chi}^0_i$ is chosen as 
$m_{\tilde{\chi}^0_1}\leq...\leq m_{\tilde{\chi}^0_5}$.
We choose $M_1$ and $M_2$ as input parameters, so that the masses
$m_{\tilde{\chi}^0_i}$ are derived quantities.  \s
%Whereas in our work we have performed the diagonalization numerically,
%the analytical expressions in the specified parameter choice can be
%found in Refs.~\cite{Pandita:1994vw, Pandita:1994ms,
%  Choi:2004zx}. \textcolor{red}{{\bf Question:} What do you mean to
%  say by 'in the specified parameter choice'?}\KS{ in these references, approximate expressions are written in specific parameter regions.}  \s
 
Similarly, the charged higgsinos $\tilde{H}^\pm_d$ and
$\tilde{H}^\pm_u$ mix with the charged gauginos $\tilde{W}^\pm$
resulting in the charginos.  
In the basis of the spinors $\psi^-_R=(\tilde{W}^-, \tilde{H}^-_d)^T,\
\psi^+_L=(\tilde{W}^+, \tilde{H}^+_u)^T$, built from
  the Weyl spinors $\tilde{H}_u^\pm$, $\tilde{H}_d^\pm$,
  $\tilde{W}^\pm$, the chargino mass terms  are expressed as 
 \begin{align}
 \mathcal{L}_{\rm Mass}^{\psi_{L/R}}=(\psi_R^-)^TM_C\psi^+_L+{\rm h.c.},
 \end{align}
 with 
 \begin{align}
 M_C=
 \begin{pmatrix}
 M_2 & \sqrt{2}s_\beta M_We^{-i\varphi_{u}}\\
 \sqrt{2}c_\beta M_W &\lambda v_s e^{i\varphi_{S}} /\sqrt{2}\\
 \end{pmatrix}. 
 \end{align}
 The spinors $\psi^+_L,\ \psi^-_R$ are rotated into the mass
 eigenstates $\ti{\chi}^+_L=(\ti{\chi}^+_{L_1},\ti{\chi}^+_{L_2})^T,\
 \ti{\chi}^-_R=(\ti{\chi}^-_{R_1},\ti{\chi}^-_{R_2})^T$ by 
 \begin{align}
 \ti{\chi}_L^+=V\psi_L^+,\ \ \ \ti{\chi}_R^-=U\psi_R^- \;, 
 \end{align}
where $U$ and $V$ are $2\times 2$ unitary matrices.   
Thereby, the mass matrix for the charginos can be diagonalized as 
\beq
{\rm diag}(m_{\tilde{\chi}_1^{\pm}}, m_{\tilde{\chi}_2^{\pm}})=U^\ast
M_C V^\dagger \;,
\eeq 
with the charginos described by the Dirac spinors ($i=1,2$)
\beq
\tilde{\chi}_i^+ = \left( \begin{array}{c} \ti{\chi}_{L_i}^+ \\[0.1cm]
                             \overline{\ti \chi_{R_i}^-} \end{array}\right) \;.
\eeq
The convention used for the mass ordering is the same as that for the
neutralinos, namely, $m_{\tilde{\chi}_1^{\pm}}\leq
m_{\tilde{\chi}_2^{\pm}}$.  
%Mass eigenstates for charginos are written as Dirac spinors
%$\tilde{\chi}_i^+=(\chi_{L_i}^+, \overline{\chi_{R_i}^-})^T,\
%(i=1,2)$.  
In total, the electroweakino sector requires four
additional input parameters which are the absolute values of the
gaugino masses and their corresponding two complex phases, hence
\beq
\abs{M_1} ,\abs{M_2}, \varphi_{M_1},\varphi_{M_2} \;.
\eeq
 
%------------------------------------------------
\paragraph{The Sfermion Sector}
%------------------------------------------------
In the sfermion sector, we only consider the third generation
which is the relevant one for our calculation of the charged Higgs
boson decays. The mass terms for the left-handed and right-handed
sfermions are derived from the soft-breaking Lagrangian
Eq.~\eqref{eq:breaking_term} as well as from the $F$- and 
$D$-terms of the supersymmetric Lagrangian. 
This yields the mass matrices for the stops and sbottoms, 
\begin{align}
\label{eq:squarkmassmat}
M_{\ti{t}}&=
\begin{pmatrix}
m_{\ti{Q}_3}^2+m_t^2+c_{2\beta}M_Z^2(I_{t}-Q_{t}s_{\theta_W}^2) & m_t(A_t^* e^{- i \varphi_u}-
\mu_{\rm eff}/t_\beta)\\
m_t(A_t e^{i \varphi_u}-
\mu_{\rm eff}^{*}/t_\beta) &m_t^2+ m^2_{\ti{t}_R}+Q_{t}c_{2\beta}M_Z^2s_{\theta_W}^2
\end{pmatrix}, \nonumber\\
M_{\ti{b}}&=
\begin{pmatrix}
m_{\ti{Q}_3}^2+m_b^2+c_{2\beta}M_Z^2(I_{b}-Q_{b}s_{\theta_W}^2) & m_b(A_b^*-
e^{i \varphi_u}\mu_{\rm eff}t_\beta)\\
m_b(A_b-
e^{-i \varphi_u}\mu_{\rm eff}^*t_\beta) &m_b^2+ m^2_{\ti{b}_R}+Q_{b}c_{2\beta}M_Z^2s_{\theta_W}^2
\end{pmatrix},
\end{align}
with the isospins $I_{t}$, $I_{b}$ and the electric charges $Q_{t}$,
$Q_{b}$ given by
$I_{t}=\frac{1}{2}$, $I_{b}=-\frac{1}{2}$, $Q_{t}=\frac{2}{3}$, $Q_{b}=-\frac{1}{3}$,
 and 
\begin{align}\label{eq:mueff}
\mu_{\rm eff}=\frac{\lambda v_s e^{i\varphi_s} }{\sqrt{2}} =
  \frac{|\lambda| v_s }{\sqrt{2}} e^{i(\varphi_\lambda + \varphi_s)} \;. 
\end{align}
The mass matrix which mixes left- and right-handed staus reads
\be M_{\ti{\tau}}=
\begin{pmatrix}
m_{\ti{L}_3}^2+m_\tau^2+c_{2\beta}M_Z^2(I_{\tau}-Q_{\tau}s_{\theta_W}^2) & m_\tau(A_\tau^*-
e^{i \varphi_u} \mu_{\rm eff}t_\beta)\\
m_\tau(A_\tau-
e^{-i \varphi_u} \mu_{\rm eff}^*t_\beta) &m_\tau^2+
m^2_{\ti{\tau}_R}+Q_{\tau}c_{2\beta}M_Z^2s_{\theta_W}^2 
\end{pmatrix}\;,\ee
with $Q_{\tau}=-1, I_{\tau}=-1/2$. The sneutrino masses are given by
\be 
m^2_{\ti \nu_i} =\fr{1}2 M_Z^2 c_{2\beta} + m_{\ti{L}_i}^2 \;, 
\ee
with the generation indices $i=1,2,3$. In our setting, only left-handed
neutrinos exist and hence the neutrinos are massless. \s
 
Using a $2\times 2$ rotation matrix for the sfermions, $U^{\ti{f}}$,
which relates $\ti{f}_L$, $\ti{f}_R$ to the mass eigenstates
$\ti{f}_1$ and $\ti{f}_2$, we get 
\beq
\begin{pmatrix}
\ti{f}_1\\
\ti{f}_2
\end{pmatrix}=
U^{\ti{f}}
\begin{pmatrix}
\ti{f}_L\\
\ti{f}_R
\end{pmatrix}.
\eeq
The sfermion mass matrices are diagonalized as $(\ti{f}=\ti{t}, \ti{b},\ti{\tau})$
\beq
{\rm diag}(\m{\ti{f}_1}^2,
\m{\ti{f}_2}^2)=U^{\ti{f}}M_{\ti{f}}U^{\ti{f}\dagger} \;. 
\eeq
In the calculation of the mass matrices $M_{\ti{t}},M_{\ti{b}}$ and
$M_{\ti{\tau}}$, the soft SUSY-breaking masses $\m{\ti{Q}_3}^2$,
$\m{\ti{t}_R}^2$, $\m{\ti{b}_R}^2$ and the trilinear soft SUSY-breaking
couplings $A_t,\ A_b, A_\tau$ are chosen as input parameters, so that
the mass eigenvalues are outputs, for which  the convention
$\m{\ti{f}_1}\leq \m{\ti{f}_2}$ is used as before. \s

In addition to the input parameters of the Higgs and electroweakino
sectors, we have seven more input parameters for the third generation
of the squarks, 
\be 
m_t,m_b,m_{\ti Q_3}^2, m_{\ti t_R}^2,m_{\ti b_R}^2,A_t,A_b \;, 
\ee
and four parameters for the third-generation of the sleptons
\be 
m_\tau,m_{\ti L_3}^2, m_{\ti \tau_R}^2,A_\tau \;. 
\ee 

%------------------------------------------------
\section{Renormalization of the NMSSM \label{sec:renorm}}
%------------------------------------------------
%\input Renormalization_N.tex

In order to obtain UV-finite results at one-loop level, 
the renormalization of the parameters and external fields is mandatory.
In particular, the bare parameters $p_0$ of the Lagrangian are
replaced by the corresponding renormalized parameters, $p$, and the
counterterms, $\delta p$, as
\be 
p_0= p +\delta p
\ee
and the bare fields $\phi_0$ are expressed via the renormalized fields $\phi$ and the wave-function renormalization constants (WFRCs) $Z_\phi$ as
\be 
\phi_0 = \sqrt{Z_\phi}\phi = \braket{1 + \fr{\delta Z_\phi}{2}}\phi. 
\ee
In our previous studies~\cite{Graf:2012hh,Muhlleitner:2014vsa,Dao:2019qaz,Muhlleitner:2015dua,Baglio:2019nlc},
we have established  the renormalization schemes in the complex
NMSSM. We will adopt these procedures for the computation of the NLO
corrections to the two-body decays of the charged Higgs bosons and
summarize here the main points. 
\begin{itemize}
\item The complex phases
  $\varphi_u,\varphi_s,\varphi_\lambda,\varphi_\kappa, \varphi_{M_1},
  \varphi_{M_2}$ do not
need to be renormalized at one-loop level, see \cite{Graf:2012hh,Baglio:2019nlc}. 
\item The tadpole counterterms are chosen in such way that the minimum
  of the Higgs potential does not change at one-loop level, leading to
\be 
\delta t_{\phi_i} = t_{\phi_i}^{(1)}, \quad \phi_i=h_d, h_u, h_s, a_d,
a_s,
\ee
 where $t_{\phi_i}^{(1)}$ are the tadpole contributions at one-loop level.
\item The SM electroweak parameters $e, M_W,M_Z$, inspired by their
  experimental measurements, are renormalized in the on-shell (OS)
  scheme. The $W$ and $Z$ mass counterterms are given by
\beq
\delta M_Z^2 = \mbox{Re} \Sigma_{ZZ}^T (M_Z^2), \quad
\delta M_W^2 = \mbox{Re} \Sigma_{WW}^T (M_W^2) \;,
\eeq
and the electric charge counterterm reads
\beq
\delta Z_e &=& \frac{1}{2} \left.\frac{\partial \Sigma^T_{\gamma\gamma}
   (k^2)}{\partial k^2}\right|_{k^2=0} +
\frac{s_W}{c_W} \frac{\Sigma^T_{\gamma Z} (0)}{M_Z^2}  - \fr{1}{2}\Delta\alpha(M_Z^2),
\\
\Delta\alpha(M_Z^2)&=&\fr{\partial \Sigma^{\text{light},T}_{\gamma\gamma}}{\partial 
  k^2}\bigg\vert_{k^2=0}-\fr{ \mbox{Re} 
  \Sigma^{\text{light},T}_{\gamma\gamma}(M_Z^2)}{M_Z^2} \;,  
 \;
\label{eq:delze}
\eeq  
where $ \Sigma_{VV^\prime}^T$, $V,V^\prime=\gamma,W,Z$, denote the
transverse part of the respective self-energies. The notation
'$\mbox{Re}$' means that we take only the real part of the loop
integrals. The above definition of the electric charge counterterm 
has been chosen to avoid the dependence of the results on large logarithms
$\ln m_f$ from the light fermion $f\ne t$ contributions \cite{Denner:1991kt}. It
corresponds to the input of the fine structure constant at the $Z$ 
boson mass. Therefore, in Eq.~\eqref{eq:delze} the photon
self-energy $\Sigma^{\text{light},T}_{\gamma\gamma}$
includes only the light fermion contributions. 
\item The charged Higgs mass in the calculation of the NLO decay
  widths is always renormalized in the OS scheme and its counterterm 
is defined via the charged Higgs self-energy,
 \beq
\delta M_{H^\pm}^2 = \mbox{Re} \Sigma_{H^\pm H^\mp} (M_{H^\pm}^2) \;.
\eeq  
Note that in the computation of the loop-corrected Higgs
masses in {\tt NMSSMCALCEW} if the charged Higgs mass is an input
parameter it is renormalized in the OS scheme. In case
  the $\DRb$ parameter $\mbox{Re} A_\lambda$ is an input parameter,
  however, the charged Higgs mass is a derived quantity. The derived
value will then be used as input in the decay part as an OS parameter.  
\item  The parameters
  $\abs{\lambda},\abs{\kappa},v_s,\tan\beta$, and
  Re$A_\kappa$ are renormalized in the $\DRb$ scheme. The counterterms  
for  $\abs{\lambda}$ ,$\abs{\kappa}$ and $v_s$ can be defined from the
electroweakino sector \cite{Ender:2011qh,Graf:2012hh} or from the
neutral Higgs boson sector. Both derivations lead to the same results
which has been confirmed explicitly by our previous calculations. 
\item The gaugino mass parameters $\abs{M_1}, \abs{M_2}$ can be
  renormalized in the OS or the $\DRb$ scheme. In Ref.~\cite{Baglio:2019nlc}
we worked with three possible renormalization schemes which we called
OS1, OS2 and $\DRb$. In OS1, the OS conditions are imposed on the
wino-like chargino 
and the bino-like neutralino, while in OS2 they are applied to the wino-
and the bino-like neutralinos. All three schemes lead to small loop
corrections to the masses of the remaining neutralinos and charginos,
{\it cf.}~Ref.~\cite{Baglio:2019nlc}.  
\item The renormalization of the squark sector, {\it i.e.}~of the
  parameters $ m_t,m_b,m_{\ti
    Q_3}^2, m_{\ti t_R}^2, m_{\ti b_R}^2,A_t,$ and $A_b$  can be performed in the
  OS or the $\DRb$ scheme. In Ref~\cite{Baglio:2019nlc} we provided   
the counterterms for both schemes and we also presented our evaluation
of the one-loop SUSY-QCD and SUSY-EW
corrections to the masses of the 
squarks. We observed large loop corrections to the squark masses and the decay widths 
of the heavy neutral Higgs bosons into squarks \cite{Baglio:2015noa,Baglio:2019nlc}. 
The counterterms of the slepton sector can 
be derived similarly to the squark sector and are
  implemented both for the OS and the $\DRb$ scheme.
%We restrict, however, the
%renormalization of the slepton sector to the OS scheme since we did not
%observed large loop-correction in this sector. 
%
\item The neutral Higgs fields are renormalized in the  $\DRb$
  scheme. The OS property of an external neutral Higgs boson is
  ensured by using the wave-function renormalization factors (WFRFs)
  $\ZH$. Thereby potentially large corrections arising from external
  lines with neutral Higgs bosons can be resummed into the decay
  widths via WFRFs. The expressions for $\ZH$ can be found in \eg
  Ref.~\cite{Dao:2019nxi} for the real NMSSM and
  Ref.~\cite{Baglio:2019nlc} for the complex case. 
\item The charged Higgs fields are also renormalized in the $\DRb$
  scheme. 
Their WFRCs are related to the WFRCs of the neutral Higgs
  fields as they are in the same doublet
    \cite{Muhlleitner:2014vsa,Dao:2019qaz}. Unlike the neutral 
  Higgs field, we cannot use the resummed 
  WFRFs for external charged Higgs bosons since they contain 
an infrared (IR) divergence due to the contribution from the massless
photon. We therefore expand the charged Higgs WFRFs to take only into 
account the pure one-loop term which is equivalent to the use of an OS
charged Higgs wave function renormalization constant $\delta Z_{H^\pm}$, 
\begin{align}\label{eq:dZhpm_OS}
\delta Z_{H^\pm}&=-\left. \frac{\partial \Sigma_{H^+H^- }}{\partial p^2}\right|_{p^2=\mHp^2}.
\end{align}
\item The SM fermionic fields and the neutralino and chargino fields
  are renormalized in the OS scheme defined at the respective 
tree-level masses of the fields. The counterterm expressions for the
neutralinos and charginos can be found in
Ref~\cite{Baglio:2019nlc}. The OS property of the external fermionic
line is justified if the corresponding mass of the  
fermion is also renormalized in the OS scheme. This is the case for the
SM fermionic fields. However, in case of the neutralinos and 
 charginos, some or all of them\footnote{In case the
     neutralinos are renormalized OS none of the charginos can be
     renormalized OS.} cannot be renormalized
 OS. As mentioned in the sixth point in this list,  the loop corrections to
 the masses of
 the neutralinos and charginos in the studied renormalization schemes,
 OS1, OS2 and $\DRb$, are small. This allows us to use the OS WFRCs
 defined at the tree-level masses in the decay processes. In the cases
 where loop-corrections to particles on the external line are large, 
\ie squarks, we have to use the modified squark WFRCs
which are defined at the loop-corrected masses, \cite{Baglio:2019nlc},  
\be 
\delta Z^{\ti q}_{ii}(M_{\ti q_i}^2) = - \fr{\partial
  \Sigma^{\ti q\, \text{div}}_{ii}(p^2)}{\partial p^2}\bigg{|}_{p^2 =
  m_{\ti q_i}^2}- \fr{\partial \hat\Sigma^{\ti q}_{ii}(p^2)}{\partial
  p^2}\bigg{|}_{p^2 = M_{\ti q_i}^2}, 
\quad i=1,2 \;, \label{eq:diagWFRsquark}
\ee
for the diagonal WFRCs and 
\be  
\delta Z^{\ti q}_{ij}(M_{\ti q_k}^2) = \fr{\Sigma^{\ti q
    \,\text{div}}_{ij}(m_{\ti q_{i}}^2)- \Sigma^{\ti q\,
    \text{div}}_{ij}(m_{\ti q_{j}}^2)}{m_{\ti q_{j}}^2 -m_{\ti 
    q_{i}}^2} -2 \fr{\hat\Sigma^{\ti q}_{ij}(M_{\ti
    q_k}^2)}{M_{\ti q_k}^2- m_{\ti q_i}^2},\quad i,j,k=1,2, \quad i\ne j,
\label{eq:offdiagWFRsquark}
\ee
for the off-diagonal WFRCs. Here $\Sigma^{\ti q}_{ij}$ denotes the
unrenormalized self-energy for the $\ti q^*_i\rightarrow 
\ti q^*_j $ transition and the $\hat\Sigma^{\ti q} $ stands for the
renormalized one. The capital letter $M_{\ti q}$ is used for 
the loop-corrected mass while the small letter $m_{\ti q}$ for the
tree-level mass. The superscript '$\text{div}$' means that we take only
the UV divergent part. $M_{\ti q_k}$ stands for the
  loop-corrected mass of the external squark where
    $\tilde{q}_k$ can be either $\tilde{q}_i$ or $\tilde{q}_j$.
The renormalization of the slepton fields is performed
in the OS scheme in accordance with their mass renormalization. 

\end{itemize}
Given our flexibility in choosing different renormalization schemes,
many particles may get loop-corrections to their masses, \eg Higgs
bosons, neutralinos, charginos, squarks, and sleptons. Loop corrections to the Higgs boson masses have been computed at  the two-loop  order
${\cal O}(\alpha_t \alpha_s + \alpha_t^2)$  \cite{Muhlleitner:2014vsa,Dao:2019qaz} while for the neutralinos, charginos, squarks and sleptons
loop corrections are calculated at one-loop level
\cite{Baglio:2019nlc}. These corrections, depending on the chosen
renormalization scheme, have been
implemented in the code {\tt NMSSMCALCEW}
\cite{Baglio:2019nlc}. 
We use the loop-corrected masses in the initial and in the final
states of the decay processes. However, we use the tree-level masses
and tree-level couplings, if not specified otherwise, in the internal
lines of the loop diagrams to ensure the cancellation of the UV
divergences.  

%------------------------------------------------
\section{Higher-Order Corrections to the Two-Body Decays of the Charged Higgs Bosons \label{sec:calc}}
%------------------------------------------------
%\input Calculations_N.tex

In this section we describe the computation of the
NLO EW, QCD as well as SUSY-EW, and SUSY-QCD
corrections to various two-body decays of the positively
charged Higgs boson $H^+$. The NLO corrections for the decay widths of the
  negatively charged Higgs boson are derived in a similar way. In the
  real NMSSM the decay widths of the $H^+$ and $H^-$ are the same,
  however, in the presence CP-violating phases they may not be
  equal. 
  In the following, we will discuss the decay modes $H^{+}\to q
  q^{\prime}$, $H^{+}\to \ell \nu$, $H^{+}\to \ti{\chi}^{+}
  \ti{\chi}^{0}$, $H^{+}\to \ti{q} \ti{q}^{\prime}$, and $H^{+}\to
  \ti{\ell} \ti{\nu}^{\prime}$. 
The  NLO SUSY-EW corrections to Higgs boson decays into neutral
Higgs bosons and a $W^+$ boson as well as their gauge dependences have
been discussed in Ref.~\cite{Dao:2019nxi}. Higher-order corrections to
the two-body decays of the neutral Higgs 
bosons in the CP-conserving and CP-violating NMSSM have
been presented in Refs.~\cite{Baglio:2019nlc,Nhung:2013lpa,Muhlleitner:2015dua}. \s
 
We used the public codes {\tt FeynArts}~\cite{Kublbeck:1990xc,Hahn:2000kx}, {\tt FormCalc}~\cite{Hahn:1998yk}  and \texttt{FeynCalc}~\cite{FeynCalc,Shtabovenko:2016sxi} in order to generate the Feynman diagrams and calculate the squared amplitudes. 
The NMSSM model file for {\tt FeynArts} has been generated with the
help of {\tt SARAH}~\cite{Staub:2009bi,Staub:2010jh,Staub:2012pb,Staub:2013tta}.
We have performed two independent calculations for all processes
studied here and the results are in full agreement. \s

%As mentioned above, some of the UV divergences encountered in the
%loop corrections are renormalized in the $\overline{{\rm DR}}$
%scheme~\cite{delAguila:1998nd,Capper:1979ns} which at one-loop order
%is equivalent to the constrained differential renormalization scheme
%(CDR)~\cite{Siegel:1979wq} as was shown in Ref.~\cite{Hahn:1998yk}. 
The UV divergent diagrams are regularized by the dimensional reduction
scheme~\cite{Siegel:1979wq} which is equivalent to the constrained
differential renormalization scheme~\cite{delAguila:1998nd,Capper:1979ns} 
 at one loop level as was shown in Ref.~\cite{Hahn:1998yk} and
 preserves SUSY at this level. 

%------------------------------------------------
%------------------------------------------------
\subsection{Charged Higgs Boson Decays into Fermions}
%------------------------------------------------
%------------------------------------------------
In our previous implementation in {\tt NMSSMCALC}, the charged Higgs
boson decay width includes higher-order
QCD corrections for the 
quark final states\footnote{The QCD corrections are the same as in the
  MSSM and can be taken over from the corresponding MSSM results
  \cite{Mendez:1990jr,Li:1990ag,Djouadi:1994gf}.} 
while the decays into lepton  
final states were evaluated at tree level. We have taken
into account, however, the universal large corrections proportional to
$\tan\beta$ that arise from the SUSY-QCD (for the top-bottom final
state) and the SUSY-EW (for the top-bottom and the $\tau$-neutrino final
states) corrections by absorbing them into effective Yukawa couplings
as described in our publication \cite{Baglio:2013iia}. 
In this section we improve the calculation of the decay widths by
including the missing one-loop SUSY-QCD and SUSY-EW
corrections, {\it i.e.}~their finite remainders. 
We present here the decay into the third generation fermions. Other
decays into the first and second generation are kept as implemented in {\tt
  NMSSMCALC}, {\it i.e.}~apart from the absorption
  into effective couplings where applicable no further one-loop SUSY-QCD and
  SUSY-EW corrections have been included, since they are suppressed
by the smallness of couplings due to the light quark masses and/or
small CKM matrix elements. We note that in the setting of {\tt
  NMSSMCALC} the CKM matrix is set to unity  in the
calculation of the loop corrections. This setting is valid for the
third generation since the mixing with the first and second
generations is small.  We, however, keep the corresponding CKM matrix element in
front of the charged Higgs decay widths as a factor and do not
renormalize it. \s

The Yukawa interaction terms of the charged Higgs boson with
third generation quarks and leptons read 
\beq 
V_{tb} \bar t\braket{ g^{L}_{H^+  b\bar t} P_L +  g^{R}_{H^+  b\bar t} P_R } b H^+ + 
\bar \nu_\tau \braket{ g^{L} _{H^+  \tau\bar\nu_\tau} P_L +  g^{R}_{H^+
    \tau\bar\nu_\tau} P_R } \tau H^+ +\hc , 
\eeq
where the projection operators $P_{L/R}$ are given by $P_{L/R} = (1\mp
\gamma_5)/2$ and $V_{tb}$ is the CKM matrix element. The
explicit expressions of the tree-level couplings are given by
\begin{align}
\label{eq:g_R_LO}
g^{R,{\rm tree}}_{H^+ b\bar
  t}&=e^{i\fr{\varphi_u}{2}}\frac{\sqrt{2}m_b }{v}\tan\beta \;,&\quad
g^{L,{\rm tree}}_{H^+ b\bar t}&=e^{i\fr{\varphi_u}{2}}\frac{\sqrt{2}
                                m_t}{v}\cot\beta \;,\\
g^{R,{\rm tree}}_{H^+ \tau\bar\nu_\tau}&=\frac{\sqrt{2}
                                         m_\tau}{v}\tan\beta \;,&\quad
g^{L,{\rm tree}}_{H^+ \tau\bar\nu_\tau}&=0 \;.
\end{align}
%\textcolor{red}{\st{The decay width of $H^+\to t\bar b$ at NLO SM QCD
%    was implemented {\tt NMSSMCALC} which is based on {\tt HDECAY}
%   .} \cite{Djouadi:1997yw, Djouadi:2018xqq,Spira:2016ztx}
In {\tt NMSSMCALC} we included higher-order QCD corrections to the
decay width $H^+\to t\bar b$ based on {\tt HDECAY}
    \cite{Djouadi:1997yw, Djouadi:2018xqq,Spira:2016ztx}.
In {\tt HDECAY}, there is an interpolation between the decay 
width that is valid near the threshold
$m_t+m_b-\delta<M_{H^\pm}<m_t+m_b+\delta$ ($\delta=2 \,\GeV$) and the
one that is valid far above the threshold $M_{H^\pm}\ge
m_t+m_b+\delta$. We include the newly computed one-loop SUSY-QCD and
SUSY-EW corrections in the latter case. Our implemented
loop-corrected decay width is hence given by 
\begin{eqnarray}
\Gamma [H^+\rightarrow t\bar{b}\,] &=& \frac{3 G_F
M^3_{H^\pm}}{4\sqrt{2}\pi}
|V_{tb}|^2 \, \beta^{1/2}(\mu_t,\mu_b) \, \left\{ (1-\mu_t -\mu_b) \left[
\frac{\mu_t}{ \tan^2
\beta } \left( 1+ \frac{4}{3} \frac{\alpha_s}{\pi} \delta_{tb}^+ \right)
\right. \right. \nonumber\\
&& \left. \left. +\mu_b^2 \tan^2 \beta R^2  \left( 1+ \frac{4}{3}
\frac{\alpha_s}
{\pi} \delta_{bt}^+ \right) \right]
-4\mu_t\mu_b R \left( 1+ \frac{4}{3}
\frac{\alpha_s}{\pi} \delta_{tb}^- \right)
   \right\} \label{eq:htotb} \\\crn
&& + |V_{tb}|^2 \Gamma _{H^+\rightarrow t\bar{b}}^{\text{SUSYQCD}}+
|V_{tb}|^2 \Gamma _{H^+\rightarrow t\bar{b}}^{\text{SUSYEW}}\,] \;,\nonumber
\end{eqnarray}
where $\mu_i = m_i^2/M_{H^\pm}^2$ ($i=t,b$), with $m_{t/b}$ being pole masses and 
\be 
\beta^{1/2}(\mu_t,\mu_b) =(1-\mu_t-\mu_b)^2-4\mu_t\mu_b 
\ee
denotes the two-body phase space factor. Explicit expressions for
$\delta_{tb}^+,\delta_{bt}^+$ and $\delta_{tb}^-$ can be found in 
 Ref.~\cite{Spira:2016ztx}. The universal SUSY-QCD and SUSY-EW
 corrections that become large in the large $\tan\beta$ regime
 are resummed into effective bottom Yukawa couplings. These
$\Delta_b$ corrections are incorporated into the decay width through the $R$ factor,
\beq
R = \frac{1}{1+\Delta_b} \left[ 1 -
  \frac{\Delta_b}{\tan^2\beta} \right],
\eeq
where the explicit expression for $\Delta_b$ is given by
\cite{Hall:1993gn,Hempfling:1993kv,Carena:1994bv,Pierce:1996zz,Carena:1998gk,Carena:1999py,Carena:2002bb,Guasch:2003cv,Spira:2016ztx}
%Guasch:2001wv,DAmbrosio:2002vsn,Buras:2002vd,Barger:2009me,Christensen:2012ei,Spira:2016ztx}}
\beq 
\Delta_b &=& \frac{\Delta_b^{\text{QCD}}+\Delta_b^{\text{elw}}}{1+\Delta_1} \;,
\label{eq:deltabcorr} 
\eeq
with the one-loop corrections for the complex NMSSM \cite{Baglio:2013iia} 
\beq
\Delta_b^{\text{QCD}} &=& 
\frac{C_F}{2} \, \frac{\alpha_s (\mu_R)}{\pi} \, M_3^* \,
\mu_{\mbox{\scriptsize eff}}^* \, \tan\beta \, I (m_{\tilde{b}_1}^2,
m_{\tilde{b}_2}^2 , m_{\tilde{g}}^2 ) \;,  \label{eq:deltabcorr_cp1} \\ 
\Delta_b^{\text{elw}} &=& \frac{\alpha_t (\mu_R)}{4\pi} \, A_t^* \,
\mu_{\mbox{\scriptsize eff}}^* \, \tan\beta \, I (m_{\tilde{t}_1}^2, 
m_{\tilde{t}_2}^2 , |\mu_{\mbox{\scriptsize eff}}|^2) \;, \label{eq:deltabcorr_cp2}\\
\Delta_1&=& - \frac{C_F}{2} \, \frac{\alpha_s (\mu_R)}{\pi}\, M_3^*\,
A_b \, I (m_{\tilde{b}_1}^2,
m_{\tilde{b}_2}^2 , m_{\tilde{g}}^2 )\;,
\label{eq:deltabcorr_cp3}
\eeq
where $\alpha_t = y_t^2/(4\pi)$ with $y_t = \sqrt{2} m_t /(v
\sin\beta)$ is the top-Yukawa coupling and $C_F=4/3$. The generic
function $I$ is defined as 
\beq
I(a,b,c) &=& \frac{ab \, \log \frac{a}{b} + bc \, \log \frac{b}{c} +
  ca \, \log \frac{c}{a}}{(a-b)(b-c)(a-c)} \;.
\eeq
The remaining one-loop SUSY-QCD and SUSY-EW corrections are denoted by
$\Gamma _{H^+\rightarrow t\bar{b}}^{\text{SUSYQCD}}$ 
and 
$\Gamma _{H^+\rightarrow t\bar{b}}^{\text{SUSYEW}}$ in
Eq.\eqref{eq:htotb}, respectively. The SUSY-EW correction can be
expressed as 
\begin{align}
\label{eq:Gam_hpmtb_NLO_imp}
\Gamma_{H^+\to t\bar b}^{\text{SUSYEW}} &=
\frac{3\beta^{1/2}\left(\mu_t,\mu_b\right) M_{H^\pm}}{8\pi}\Bigg[ 
\left(1-\mu_t-\mu_b\right) 
{\rm Re}\left[ g_{ H^+ b\bar{t}}^{L, {\rm tree}}\left(\Delta_{ H^+ b\bar{t}}^{L, {\rm SUSYEW}}\right)^*+
g_{ H^+ b\bar{t}}^{R, {\rm tree}}\left(\Delta_{ H^+ b\bar{t}}^{R, {\rm SUSYEW}}\right)^*
\right] \nonumber \\
&-2\sqrt{\mu_t\mu_b}\ {\rm Re}\left[{g}_{ H^+ b\bar{t}}^{L,\rm tree}\left(\Delta_{ H^+ b\bar{t}}^{R, \rm SUSYEW}\right)^*
+{g}_{ H^+ b\bar{t}}^{R,\rm tree}\left(\Delta_{ H^+b\bar{t}}^{L,\rm
  SUSYEW}\right)^*\right]\Bigg]+ \Gamma_{H^+\to t\bar b \gamma} \;,
\end{align}
where $\Delta_{ H^+ b\bar{t}}^{L/R, {\rm SUSYEW}}$ denote the left-
and right-handed form factors of the renormalized one-loop amplitude
and include the following contributions
\begin{align}
\Delta_{ H^+ b\bar{t}}^{L/R, {\rm SUSYEW}} = \Delta_{ \rm
  SUSYEW}^{L/R, {\rm vert}}+\Delta_{\rm SUSYEW}^{L/R, {\rm
  CT}}+\Delta_{ \rm SUSYEW}^{L/R, {\rm GWmix}}+\Delta_{\rm
  SUSYEW}^{L/R, {\rm sub}}\,.\label{eq:refhtbCL} 
\end{align}
The  $\Delta_{\rm SUSYEW }^{L/R, {\rm vert}}$ contributions are
computed from the corresponding one-loop triangle
diagrams.\footnote{We do not include any explicit formulae for our
  results as they are quite lengthy. They can be extracted, however, from the code {\tt
      NMSSMCALCEW}.}
%------------------------------------------------
%\begin{figure}[htbp]
%   \centering
%%   \includegraphics[scale=1.3]{diagHpmud.eps} % requires the graphicx package
%   \includegraphics[scale=1.3]{diagHpmud-eps-converted-to.pdf} % requires the graphicx package
%   \caption{Feynman diagrams for the $H^{-}u \bar{d}$ vertex at one-loop order.   
%   }
%   \label{fig:diagHpmud}
%\end{figure}
%------------------------------------------------
%--------------------------------------------------------------
%   \begin{figure}[htbp]
%        \centering
%%        \includegraphics[scale=1.4]{diagHptbgam.eps}
%        \includegraphics[scale=1.4]{diagHptbgam-eps-converted-to.pdf}
%        \vspace{1.5cm}
%      \caption{Feynman diagrams for $H^\pm \to \bar{t}b\gamma$.}
%\label{FIG:Hptbgam}
%\end{figure}
%--------------------------------------------------------------
 The counterterm contributions read
\begin{align}
\label{eq:quarkCTR}
 \Delta^{R, {\rm CT}}_{\rm SUSYEW}&= g^{R, {\rm tree}}_{H^+b\bar{t}}
\left(\frac{\delta m_b}{m_b}-\frac{\delta v}{v}-\frac{\delta c_\beta}{c_\beta}+\frac{1}{2}\delta \bar{Z}_R^t+\frac{1}{2}\delta Z_L^b+\frac{1}{2}\delta Z_{H^\pm}+ \fr{ c_\beta^3}{ s_\beta} \delta t_\beta \right) ,\\
\label{eq:quarkCTL}
 \Delta^{L, {\rm CT}}_{\rm SUSYEW}&=g^{L, {\rm tree}}_{H^+b\bar{t}}
\left(\frac{\delta m_t}{m_t}-\frac{\delta v}{v}-\frac{\delta s_\beta}{s_\beta}+\frac{1}{2}\delta \bar{Z}_L^t+\frac{1}{2}\delta Z_R^b+\frac{1}{2}\delta Z_{H^\pm} - c_{\beta}s_{\beta} \delta t_\beta\right),
\end{align}
where $\delta v$ is given in terms of the counterterms
  for $e$, $M_W$ and $M_Z$ as,
\beq
\frac{\delta v}{v} = \frac{\delta e}{e} + \frac{c_W^2}{2s_W^2} \left(
  \frac{\delta M_Z^2}{M_Z^2} - \frac{\delta M_W^2}{M_W^2} \right) +
\frac{\delta M_W^2}{2 M_W^2} \;.
\eeq
The contributions from the transition of the charged Higgs boson on the
external line to the charged $W^+$ boson and the charged Goldstone
boson, respectively, do not vanish. They are denoted by  $\Delta_{\rm
  SUSYEW}^{L/R, {\rm GWmix}}$ and given by
\begin{align}
\label{eq:quarkWGmixL}
 \Delta^{L, {\rm GWmix}}_{\rm SUSYEW}&=\fr{e m_t  e^{i\varphi_u/2}}{\sqrt{2}s_W M_W^2}\hat\Sigma_{H^+W^+}(M_{H^\pm}^2)\;,\\
\label{eq:quarkWGmixR}
 \Delta^{R, {\rm GWmix}}_{\rm SUSYEW}&= -\fr{e m_b  e^{i\varphi_u/2}}{\sqrt{2}s_W M_W^2}\hat\Sigma_{H^+ W^+}(M_{H^\pm}^2) \;,
\end{align}
where $\hat\Sigma_{H^+W^+}(M_{H^\pm}^2)$ is the renormalized
self-energy of the transition $H^+\to W^+$ computed at $p^2 =
M_{H^\pm}^2$. The last term in Eq.\eqref{eq:refhtbCL}  
represents the subtraction terms which are needed to avoid double
counting arising from the $\Delta_b$ corrections. They read
\begin{align}
\label{eq:quarsub}
 \Delta^{L, {\rm sub}}_{\rm SUSYEW}&=0,\\
%\label{eq:quarkCTL}
 \Delta^{R, {\rm sub}}_{\rm SUSYEW}&= g^{R, {\rm tree}}_{H^+b\bar{t}}\fr{\Delta_b^{\rm elw}}{s_\beta^2}
                                     \;,
\end{align}
with $\Delta_b^{\rm elw}$ given in Eq.~(\ref{eq:deltabcorr_cp2}).
The contribution from real photon emission,
$\Gamma_{H^+\to t\bar b \gamma}$ in Eq.~\eqref{eq:Gam_hpmtb_NLO_imp},
is needed to cancel the infrared divergences (IR) 
from the triangle contributions. 
The decay width for the real emission of a photon reads
\begin{align}\label{eq:formula_htbam}\notag
\Gamma_{H^+\to t \bar{b}\gamma}&=\fr{\alpha N_C  M_{H^\pm}}{36\pi^2 v^2}\Bigg [
 \left( \mu_b t^2_\beta+ \frac{\mu_t}{t^2_\beta}\right) \bigg(-8 I_{21}^{00}+6 
I_{01}^{22}-14 I_1^{0}+8 I_1^{2}-17 I_2^{0}-3 I_0^{1}\crn
&-7 
I_2^{1}+3 I_ 0^{2}-26 I \bigg) +4M_{H^\pm}^2 
\braket{(\mu_b+\mu_t-1)\left(\mu_b t^2_\beta+ \frac{\mu_t}{t^2_\beta}\right)+ 4 \mu_t\mu_b}\crn
& \times\bigg(2 
\braket{M_{H^\pm}^2 -m_t^2 -m_b^2}  I_{21}+6 \braket{M_{H^\pm}^2 +m_t^2 -m_b^2} 
I_{01} +3 \braket{M_{H^\pm}^2 -m_t^2 +m_b^2}  I_{02}\crn
&+9 M_{H^\pm}^2 I_{00}+4  m_t^2 
I_{11}+m_b^2 I_{22}+9 I_0+4I_1+I_2\bigg)
\Bigg],
\end{align}
where we have dropped the
 arguments $(M_{H^\pm}^2,m_{t}^2,m_{b}^2)$ of the $I$ functions for
 better readability. 
In the above formulae, the $I$ functions are the Bremsstrahlung
integrals defined  as\cite{Denner:1991kt} 
\begin{align}\label{eq:Ifunc}
I_{i_1...i_n}^{j_1...j_m}=\frac{1}{\pi^2}\int\frac{d^3k_1}{2k_1^0}\frac{d^3k_2}{2k_2^0}\frac{d^3k_\gamma}{2k_g^0}\delta^{(4)}(k_0-k_1-k_2-k_\gamma)
\frac{(\pm2k_\gamma\cdot k_{j_1})\ .\  .\  . \ (\pm2k_\gamma\cdot k_{j_m})}{(\pm2k_\gamma\cdot k_{i_1})\ .\  .\  . \ (\pm2k_\gamma\cdot k_{i_n})},
\end{align}
where $k_{0}$, $k_{1}$ and $k_{2}$ stand for the incoming momentum
  of the charged Higgs boson and the outgoing momentum of top and bottom quark,
  respectively, while $k_{\gamma}$ is the outgoing momentum of the photon. 
 For the sign of the momentum products $k_\gamma\cdot k_{i_n}$ and
 $k_\gamma\cdot k_{j_m}$, the minus sign is chosen only when the
 indices $i_n, \ j_m$ are 0.  
%Here as IR divergent quantities  $I_{11}$, $I_{22}$, and $I_{12}$ are appeared.
%Above result is consistent with that of Ref.\cite{Bartl:1995tx}
In the $I$ functions defined above, the integrals  $I_{00},\ I_{01},\
I_{02},\ I_{11},\ I_{22},$ and $I_{12}$ are IR divergent.  
We checked that the analytical expression
Eq.~\eqref{eq:formula_htbam} is consistent with the generic result
for the real corrections to $S\to F\bar{F}$ ($S$ denotes a scalar, $F$
a fermion) calculated in Ref.~\cite{Goodsell:2017pdq}. \s

The SUSY-QCD contribution, $\Gamma _{H^+\rightarrow
  t\bar{b}}^{\text{SUSYQCD}}$, can be computed in a similar way as the
SUSY-EW corrections,  
\begin{align}\label{eq:Gam_hpmtb_NLOQCD}
\Gamma_{H^+\to t\bar b}^{\text{SUSYQCD}} &=
\frac{3\beta^{1/2}\left(\mu_t,\mu_b\right)M_{H^\pm}}{8\pi}\Bigg[ 
\left(1-\mu_t-\mu_b\right) 
{\rm Re}\left[g_{ H^+ b\bar{t}}^{L, {\rm tree}}\left(\Delta_{ H^+
                                           b\bar{t}}^{L, {\rm
                                           SUSYQCD}}\right)^*
                                   \right.        \nonumber \\
&\left. +
g_{ H^+ b\bar{t}}^{R, {\rm tree}}\left(\Delta_{ H^+ b\bar{t}}^{R, {\rm SUSYQCD}}\right)^*
\right] \notag \\
&-2\sqrt{\mu_t\mu_b}\ {\rm Re}\left[\tilde{g}_{ H^+ b\bar{t}}^{L,\rm tree}\left(\Delta_{ H^+ b\bar{t}}^{R, \rm SUSYQCD}\right)^*+\tilde{g}_{ H^+ b\bar{t}}^{R,\rm tree}\left(\Delta_{ H^+b\bar{t}}^{L,\rm SUSYQCD}\right)^*\right]\Bigg],
\end{align}   
where the left- and right-handed form factors of the renormalized 
vertex $g_{ H^+b\bar{t}}^{L/R,\rm SUSYQCD}$ are given by
\beq
\Delta_{ H^+ b\bar{t}}^{L/R, {\rm SUSYQCD}} = \Delta_{\rm SUSYQCD}^{L/R, {\rm vert}}+\Delta_{\rm SUSYQCD}^{L/R, {\rm CT}}
+\Delta_{\rm SUSYQCD}^{L/R, {\rm sub}}\,.\label{eq:refhtbQCD}
\eeq   
The contribution $\Delta_{\rm SUSYQCD}^{L/R, {\rm vert}}$ comes from
one-loop diagrams containing a virtual gluino. The counterterm
contributions $\Delta_{\rm SUSYQCD}^{L/R, {\rm CT}}$ are given by
Eqs.~\eqref{eq:quarkCTR} and \eqref{eq:quarkCTL} by setting the
counterterms $\delta v, \delta t_\beta, \delta s_\beta, \delta c_\beta, \delta
Z_{H^\pm}$ to zero  and including in the computation of $\delta
m_{b,t}$ and $\delta Z_{L,R}^{t,b}$ instead of the SUSY-EW the
SUSY-QCD corrections. The subtraction terms of the SUSY-QCD
contributions read 
\begin{align}
\label{eq:quarsubQCD}
 \Delta^{L, {\rm sub}}_{\rm SUSYQCD}&=0,\\
\label{eq:quarkCTLQCD}
 \Delta^{R, {\rm sub}}_{\rm SUSYQCD}&= g^{R, {\rm tree}}_{H^+b\bar{t}}\fr{\Delta_b^{\rm QCD}+\Delta_1}{s_\beta^2},
\end{align}
with $\Delta_b^{\rm QCD}$ and $\Delta_1$ given in
Eqs.~\eqref{eq:deltabcorr_cp1} and \eqref{eq:deltabcorr_cp3}, respectively.\s

The higher-order corrections for the charged Higgs decays into leptons
and neutrinos can be expressed in the same way as the decays into
quarks, $H^+\to t\bar b$.  
Since the left-handed form factor of the  vertex $H^+l\bar \nu$
vanishes the formula for the NLO width is given by  
\begin{align}
\label{eq:Gam_lnu_NLO_imp}
\Gamma_{H^+\to  \nu\bar{\tau}}^{\rm NLO}&=
\frac{M_{H^\pm}}{16\pi} 
\left(1-\frac{m_\tau^2}{M_{H^\pm}^2}\right)^2 
\left(\abs{{ {\ti g}}_{ H^+ \tau\bar{\nu}_\tau}^{R, {\rm tree}}}^2+2{\rm Re}\left[{g}_{ H^+
\tau\bar{\nu}_\tau}^{R, {\rm tree}}\big(\Delta_{R}^{\rm vert}+\Delta_{R}^{\rm CT}
+\Delta_{R}^{\rm GWmix}+\Delta_{R}^{\rm sub}\big)^*\right)\right] \\
& +\Gamma (H^+\to  \nu\bar{\tau} \gamma) \;,  \nonumber
\end{align}
where the effective tree-level coupling resumming the large
$\Delta_\tau$ corrections reads
\beq 
{\tilde{g}}_{ H^+ \tau\bar{\nu}_\tau}^{R, {\rm tree}} ={g}_{ H^+
  \tau\bar{\nu}_\tau}^{R, {\rm tree}} \frac{1-\Delta_\tau/t_\beta^2}{1
  +\Delta_\tau}\,,  
\eeq
with
\be \
\label{eq:deltatau}
\Delta_\tau= \fr{\alpha_1}{4\pi} M_1^* \mu_{\rm eff}^* t_\beta
I(m_{\tilde{\tau}_1}^2,m_{\tilde{\tau}_2}^2, \abs{M_1}^2)+
\fr{\alpha_2}{4\pi} M_2^* \mu_{\rm eff}^* t_\beta
I(m_{\tilde{\nu}_\tau}^2,m_{\tilde{\tau}_2}^2, \abs{M_2}^2)
\ee
and
\be
\alpha_{1,2}= \fr{g_{1,2}^2}{4\pi} \;.
\ee 
The vertex correction $\Delta_{R}^{\rm vert}$ is given by the
corresponding genuine one-loop triangle diagrams. The
counterterm contribution, which is needed to cancel the UV divergences
arising from the genuine one-loop triangle contribution, is denoted
by $\Delta_{R}^{\rm CT}$ and reads 
\beq
\Delta_{R}^{\rm CT}= {g}_{ H^+ \tau\bar{\nu}_\tau}^{R, {\rm
    tree}}\left( \fr{\delta m_\tau}{m_\tau} - \fr{\delta v}{v}-
  \fr{\delta c_\beta}{c_\beta} + \fr 12 \delta Z_{H^\pm } +
  \fr{c_\beta^3}{s_\beta}\delta t_\beta +\fr 12 \delta
  Z_\tau+\fr 12 \delta \bar{Z}_{\nu_\tau}  \right) \;.
\eeq
For the one-loop $H^\pm$-$W/G$ mixing contributions $\Delta^{\rm
  GWmix}_{R}$ to the effective coupling we get
 \beq
 \Delta_{R}^{\rm GWmix} =-\frac{
   g_{2}m_\tau}{\sqrt{2}}\frac{\hat{\Sigma}_{H^+W^+}}{M_W^2} \,, 
 \eeq
and the subtraction term to avoid double counting reads
\beq 
\Delta_{R}^{\rm sub} = {g}_{ H^+ \tau\bar{\nu}_\tau}^{R, {\rm tree}}
\Delta_\tau \left( 1 + \fr{1}{t_\beta^2}\right).  
\eeq
%
%\KS{This is given in \eqref{eq:deltatau}.}
The real photon emission $\Gamma_{H^+\to  \nu\bar{\tau} \gamma}$ can
be cast into the form
\begin{align}
\label{eq:formula_hnuellgam}\notag
\Gamma_{H^+\to  \nu\bar{\tau} \gamma}&=\fr{\alpha t^2_\beta m_\tau^2}{4 \pi^2 v^2 M_{H^{\pm}} }
\bigg(-2 (M_{H^{\pm}}^2 -m_\tau^2 ) \big(M_{H^{\pm}}^2 \left(I_{00}+I_{20}\right)\crn
&+m_\tau^2  \left(I_{20}+I_{22}\right)+I_0+I_2\big)-I_2^0-I \bigg),
\end{align}
where the arguments $(M_{H^{\pm}}^2,0,m_{\tau}^2)$ of the bremsstrahlung functions have been dropped.
%%%%%%%%%%%%%%%%%%%%%%%%%%%%%%%%%%%%%%%%%%%%%%%%%%%%%%%%%%%%%
%----------------------------------------
\subsection{Charged Higgs Boson Decays into Electroweakinos}
%----------------------------------------
In this section, we present the results for charged Higgs bosons
decaying into a pair of electroweakinos. The Lagrangian describing the
interaction between a charged particle $H_k^+$ ($k=1$ for the
charged Goldstone boson $G^+$ and $k=2$ for the charged Higgs boson $H^+$)
 and a neutralino $\ti \chi_i^0$ ($i=1,...,5$) and a chargino 
$\ti \chi_j^-$ ($j=1,2$) reads
\beq
\bar \psi_{\ti\chi_i^0} \left(g^{R}_{H^+_k  \ti{\chi}_i^0\ti{\chi}^-_j} P_R+g^{L}_{H^+_k  \ti{\chi}_i^0 \ti{\chi}^-_j }P_L\right)\psi_{\ti\chi_j^-} H^+_k +\hc. 
\eeq
The tree-level couplings are given by 
\begin{align}
g_{H^+_k  \ti{\chi}_i^0\ti{\chi}^-_j}^{R,{\rm tree}}&=
-\frac{1}{\sqrt{2}}{V}_{j2}\Big((g_1N_{i1}+g_2N_{i2})Z^{H^\pm}_{k2}e^{i \varphi_u}+\sqrt{2}\lambda^\ast N_{i5}Z^{H^\pm}_{k1}\Big) 
-g_2 e^{i \varphi_u} {V}_{j1}N_{i4} Z^{H^\pm}_{k2} \,,  
\\ 
g_{H^+_k  \ti{\chi}_i^0\ti{\chi}^-_j}^{L,{\rm tree}}&=
-\frac{1}{\sqrt{2}}{U}_{j2}^\ast \Big({-}g_1N_{i1}^\ast Z^{H^\pm}_{k1} {-}g_2N_{i2}^\ast Z^{H^\pm}_{k1} +\sqrt{2} e^{i \varphi_u}\lambda N_{i5}^\ast Z^{H^\pm}_{k2}\Big) 
-g_2{U}_{j1}^\ast N_{i3}^\ast Z^{H^\pm}_{k1}, \label{eq:lhcoupling}
\end{align}
with the matrix elements 
\beq
Z^{H^\pm}_{11}=-Z^{H^\pm}_{22}=-c_\beta \;, \quad 
Z^{H^\pm}_{12}=Z^{H^\pm}_{21}=s_\beta \;.
\eeq

In the previous version of {\tt NMSSMCALC}, we included only the
tree-level decay width which reads 
\beq
\Gamma^{\rm tree}_{H^+ \to \ti{\chi}_i^0\ti{\chi}^+_j } &= &
\frac{\beta^{1/2}\left(\mu_i,\mu_j\right)M_{H^\pm}}{16\pi}
\Bigg[ 
\left(1-\mu_i-\mu_j \right) \left(\abs{g_{ H^+ \tilde{\chi}_i^0 \tilde{\chi}^{-}_j }^{L, {\rm tree}}}^2+\abs{{g}_{ H^+ \tilde{\chi}_i^0 \tilde{\chi}^{-}_j }^{R, {\rm tree}}}^2\right) \\
&&-2 \sqrt{\mu_i\mu_j}\left({g}_{ H^+ \tilde{\chi}_i^0 \tilde{\chi}^{-}_j }^{L,{\rm tree}}(g_{ H^+ \tilde{\chi}_i^0 \tilde{\chi}^{-}_j }^{R,{\rm tree}})^* 
+{g}_{ H^+ \tilde{\chi}_i^0 \tilde{\chi}^{-}_j }^{R,{\rm tree}}(g_{ H^+ \tilde{\chi}_i^0 \tilde{\chi}^{-}_j }^{L,{\rm tree}})^*
\right)\Bigg]\,,\notag
\eeq
where $\mu_i = M_{\tilde{\chi}_i^0}^2/M_{H^+}^2$ and $\mu_j =
M_{\tilde{\chi}_j^+}^2/M_{H^+}^2$.\footnote{Note that by capital letters we denote the loop-corrected final state particle masses while the masses of the
  particles running in the loops are tree-level masses.} We now
include the NLO SUSY-EW corrections to the decay width as 
\be  
\Gamma^{\rm NLO}_{H^+ \to \ti{\chi}_i^0\ti{\chi}^+_j }= \Gamma^{\rm
  tree}_{H^+ \to \ti{\chi}_i^0\ti{\chi}^-_j } +\Gamma^{\rm
  SUSYEW}_{H^+ \to \ti{\chi}_i^0\ti{\chi}^-_j }\,, \label{eq:ewino1}
\ee 
with
\beq
\Gamma^{\rm SUSYEW}_{H^+ \to \ti{\chi}_i^0\ti{\chi}^+_j }  
 &= &
\frac{\beta^{1/2}\left(\mu_i,\mu_j\right)M_{H^\pm}}{8\pi}
\Bigg[ 
\left(1-\mu_i-\mu_j \right) {\rm Re}\left[g_{ H^+ \tilde{\chi}_i^0 \tilde{\chi}^{-}_j }^{L, {\rm tree}}\Delta_{L}^*+{g}_{ H^+ \tilde{\chi}_i^0 \tilde{\chi}^{-}_j }^{R, {\rm tree}}\Delta_{R}^*\right] \label{eq:ewino2}\\
&&-2 \sqrt{\mu_i\mu_j}\ {\rm Re}\left[{g}_{ H^+ \tilde{\chi}_i^0 \tilde{\chi}^{-}_j }^{L,{\rm tree}}\Delta_{R}^* 
+{g}_{ H^+ \tilde{\chi}_i^0 \tilde{\chi}^{-}_j }^{R,{\rm tree}}\Delta_{L}^*
\right]\Bigg]
+ \Gamma_{H^+\to \tilde{\chi}_i^0\tilde{\chi}^{+}_j\gamma}\,.\notag
\eeq
The left- and right-handed form factors $\Delta_{L/R}$ consist of  the
genuine triangle one-loop diagram contributions $\Delta_{L/R}^{\rm
  vert}$, the counterterms $\Delta_{L/R}^{\rm CT}$ and the
contributions $\Delta_{L/R}^{\rm GWmix}$ from the transitions of
$H^+\to G^+/W^+$ on the external line,
\beq 
\Delta_{L/R}&=& \Delta_{L/R}^{\rm vert} + \Delta_{L/R}^{\rm CT} +
\Delta_{L/R}^{\rm GWmix} \;. 
\eeq 
The counterterm contributions read
\beq
\Delta_{R}^{\rm CT}&=&
-\frac{1}{\sqrt{2}}{V}_{j2}\Big((\delta g_1N_{i1}+ \delta g_2N_{i2})c_\beta e^{i \varphi_u}+\sqrt{2}\delta \lambda^\ast N_{i5}s_\beta \Big) 
- \delta g_2 e^{i \varphi_u} {V}_{j1}N_{i4} c_\beta\\
&& + \fr 1 2 \sum_{m=1}^2 g_{H^+_m  \ti{\chi}_i^0\ti{\chi}^-_j}^{R,{\rm tree}}\delta Z^{H^\pm}_{m k} +
  \fr 1 2 \sum_{m=1}^5 g_{H^+_k  \ti{\chi}_m^0\ti{\chi}^-_j}^{R,{\rm tree}}\delta \bar Z^{\ti{\chi}^0}_{L,im} + \fr 1 2 \sum_{m=1}^2 g_{H^+_k  \ti{\chi}_i^0\ti{\chi}^-_m}^{R,{\rm tree}}\delta Z^{\ti{\chi}^+}_{R,mj}\crn 
\Delta_{L}^{\rm CT}&=& -\frac{1}{\sqrt{2}}{U}_{j2}^\ast \Big({-}\delta g_1N_{i1}^\ast Z^{H^\pm}_{k1} {-}\delta g_2N_{i2}^\ast Z^{H^\pm}_{k1} +\sqrt{2} e^{i \varphi_u}\delta \lambda N_{i5}^\ast Z^{H^\pm}_{k2}\Big) 
-\delta g_2{U}_{j1}^\ast N_{i3}^\ast Z^{H^\pm}_{k1}\\
&& + \fr 1 2 \sum_{m=1}^2 g_{H^+_m  \ti{\chi}_i^0\ti{\chi}^-_j}^{L,{\rm tree}}\delta Z^{H^\pm}_{m k} +
 \fr 1 2 \sum_{m=1}^5 g_{H^+_k  \ti{\chi}_m^0\ti{\chi}^-_j}^{L,{\rm
     tree}}\delta \bar Z^{\ti{\chi}^0}_{R,im} + \fr 1 2 \sum_{m=1}^2
 g_{H^+_k  \ti{\chi}_i^0\ti{\chi}^-_m}^{L,{\rm tree}}\delta
 Z^{\ti{\chi}^+}_{L,mj} \;.\notag
\eeq
The $\Delta_{L/R}^{\rm GWmix}$ can be expressed in terms of the 
renormalized self-energy  $\hat\Sigma_{H^+W^+}(M_{H^\pm}^2)$ as 
\beq
\Delta_{L}^{\rm GWmix} &=& \left( M_{\ti\chi^+_j} g_{W^+\ti\chi_i^0\ti\chi_j^-}^L- 
 M_{\ti\chi^0_i} g_{W^+\ti\chi_i^0\ti\chi_j^-}^R\right) \fr{\hat\Sigma_{H^+W^+}(M_{H^\pm}^2)}{M_W^2}\,,\\
\Delta_{R}^{\rm GWmix} &=& \left(  M_{\ti\chi^+_j} g_{W^+\ti\chi_i^0\ti\chi_j^-}^R- 
 M_{\ti\chi^0_i} g_{W^+\ti\chi_i^0\ti\chi_j^-}^L\right) \fr{\hat\Sigma_{H^+W^+}(M_{H^\pm}^2)}{M_W^2}\,,
\eeq
where 
\beq 
g_{W^+\ti\chi_i^0\ti\chi_j^-}^R&=&g_2\left({V}_{j1} N^*_{i2} - {V}_{j2} N^*_{i4}/\sqrt{2}\right)\,, \\
g_{W^+\ti\chi_i^0\ti\chi_j^-}^L&=&g_2 \left({U}_{j1}^* N_{i2} + {U}_{j2}^* N_{i3}/\sqrt{2}\right) \,.\eeq

The IR divergences which appear in the $\Delta_{L/R}^{\rm vert}$ vertex
corrections are removed by  the contribution from the real photon
emission. It is given by
\begin{align}\label{eq:formula_hpchipchi0gam}\notag
 \Gamma_{H^+\to \tilde{\chi}_i^0\tilde{\chi}^{+}_j\gamma}&=\fr{2\alpha \Gamma^{\rm tree}_{H^+ \to \ti{\chi}_i^0\ti{\chi}^-_j }}{\pi \beta^{1/2}\left(\mu_i,\mu_j\right) } \bigg[ -2 \left(M_{H^\pm}^2 I_{00}+I_{01} (M_{H^\pm}^2+M_{\ti{\chi}_{i}^{-}}^2-M_{\ti{\chi}_{i}^{0}}^2 )+M_{\ti{\chi}_{i}^{-}}^2  I_{11}
+I_0 +I_1 \right) \bigg]\crn
&+\fr{\alpha }{8\pi^2 M_{H^\pm} }\left(g_{H^+ \tilde{\chi}_i^0 \tilde{\chi}^{-}_j}^{R, {\rm tree}}
(g_{H^+ \tilde{\chi}_i^0 \tilde{\chi}^{-}_j }^{L, {\rm tree}})^* + g_{ H^+ \tilde{\chi}_i^0 \tilde{\chi}^{-}_j }^{L, {\rm tree}}
(g_{ H^+ \tilde{\chi}_i^0 \tilde{\chi}^{-}_j }^{R, {\rm tree}})^* \right) \left(I_1^{0}+I\right),
\end{align}
where  the arguments
$(\mHp^2,M_{\ti{\chi}_{j}^{+}}^2,M_{\ti{\chi}_{i}^{0}}^2)$ of the $I$
functions have been dropped. 

 %----------------------------------------

%%%%%%%%%%%%%%%%%%%%%%%%%%%%%%%%%%%%%%%%%%%%%%%%%%%%%%%%%%%%
 %----------------------------------------
\subsection{Charged Higgs Decays into Squarks and Sleptons}
%----------------------------------------
In this section, we give the expressions for the
loop-corrected decays of the charged Higgs boson into squarks and
sleptons. The Lagrangian describing the interactions between the
charged particle $H^+_{k}$ (with
$H_1 = G^+$ and $H_2 = H^+$) and the squarks and sleptons reads 
\beq 
i g_{H^+_k   \ti t^*_i\ti b_j} H^+_k  \ti t^*_i\ti b_j + i g_{H^+_k
  \ti \nu_\tau \ti \tau_i^*} H^+_k  \ti \nu_\tau
\ti \tau_i +\hc \;, 
\eeq 
with $i,j=1,2$ and the tree-level couplings 
\beq
g_{H^+_k   \ti t^*_i\ti b_j }&=&\sum_{a=1}^2\sum_{b=1}^2
U^{\ti{b}*}_{ja}(C_{H^+_k\ti{b}\ti{t}^\ast })_{ab}(U^{\ti{t}})_{ib}\,, \\
g_{H^+_k  \ti \nu_\tau \ti \tau_i^* }&=&-\frac{v
  U^{\ti \tau *}_{i1} \left(c_{\beta} \left(g_2^2-2  
y_{\tau}^2\right) Z^{H^\pm}_{k1}+g_2^2 s_{\beta} \
Z^{H^\pm}_{k2}\right)}{2 \sqrt{2}}\crn
&&+U^{\ti \tau*}_{i2} y_{\tau}
\left( A_{\tau}^* Z^{H^\pm}_{k1}+e^{i \varphi_u } \
\mu_{\rm eff}   Z^{H^\pm}_{k2}\right) \,,\label{eq:ghptaunu}
\eeq
with
\begin{align}
\label{eq.ct1}
(C_{H^+_k\ti{b}\ti{t}^\ast })_{11}&=-\frac{c_{\beta} v \left(g_2^2-2 y_b^2\right) Z^{H^\pm}_{k1}}{2 
\sqrt{2}}-\frac{s_{\beta} v \left(g_2^2-2 y_t^2\right) 
Z^{H^\pm}_{k2}}{2 \sqrt{2}} \;,
\\\label{eq.ct2}
(C_{H^+_k\ti{b}\ti{t}^\ast})_{12}&=
e^{i\varphi_u} y_t A_t Z^{H^\pm}_{k2}+\mu_{\rm eff}^* y_t 
Z^{H^\pm}_{k1} \;,
\\\label{eq.ct3}
(C_{H^+_k\ti{b}\ti{t}^\ast})_{21}&=
y_b A_b^* Z^{H^\pm}_{k1}+e^{i \varphi_u } \mu_{\rm eff}  y_b
                                   Z^{H^\pm}_{k2} \;,
\\\label{eq.ct4}
(C_{H^+_k\ti{b}\ti{t}^\ast })_{22}&=\frac{c_{\beta} e^{i\varphi_u} v y_b y_t 
Z^{H^\pm}_{k2}}{\sqrt{2}}+\frac{e^{i\varphi_u} s_{\beta} v y_b y_t 
Z^{H^\pm}_{k1}}{\sqrt{2}}\,. 
\end{align}
The fermion Yukawa couplings are defined as 
\beq
y_t= \fr{\sqrt 2 m_t}{v s_\beta } \quad \mbox{ and } \quad
y_{b/\tau}= \fr{\sqrt 2 m_{b/\tau}}{v c_\beta } \;.
\eeq

The NLO decay width for the process $H^+\to \ti t_i \ti b_j^* $ consists
of the tree-level decay width and
the NLO SUSY-QCD\footnote{This correction arises from Feynman
    diagrams having a gluon or a gluino in the loop
    corrections and the gluon radiations.} and SUSY-EW corrections,
\beq 
\Gamma^{\rm NLO}_{H^+\to \ti t_i \ti b_j^*}= \Gamma^{\rm tree}_{H^+\to \ti t_i \ti b_j^*}+ \Gamma^{\rm SUSYQCD}_{H^+\to \ti t_i\ti b_j^* }+
\Gamma^{\rm SUSYEW}_{H^+\to \ti t_i \ti b_j^*}\,, \label{eq:hosbotstop1}
\eeq
where the tree-level decay width was implemented in the {\tt NMSSMCALC} as
\beq  
\Gamma^{\rm tree}_{H^+\to \ti t_i \ti b_j^*}&=&3R_2\abs{g_{H^+_k   \ti
    t^*_i\ti b_j }} ^2 \;,
\eeq 
with the two-body phase space factor
$R_2=\beta^{\frac{1}{2}}\left(\mu_i,\mu_j\right)/(16\pi M_{H^\pm})$, $\mu_i=
M_{\ti t_i}^2/M_{H^\pm}^2 $, $\mu_j= M_{\ti b_j}^2/M_{H^\pm}^2
$. The newly computed NLO SUSY-QCD and SUSY-EW corrections are given by
\beq 
 \Gamma^{\rm SUSYQCD}_{H^+\to \ti t_i \ti
   b_j^*}&=&6R_2g_{H^+_k   \ti 
   t^*_i\ti b_j } \Delta^{\rm SUSYQCD}_{H^+_k   \ti t_i^*\ti
   b_j }  +\Gamma_{H^+\to \ti t_i \ti b_j^*g}\,, 
\label{eq:hosbotstop2} \\
 \Gamma^{\rm SUSYEW}_{H^+\to \ti t_i \ti b_j^*}&=&6R_2g_{H^+_k   \ti
   t_i^*\ti b_j }\Delta^{\rm SUSYEW}_{H^+_k   \ti t^*_i\ti
   b_j }  +\Gamma_{H^+\to \ti t_i \ti b_j^*\gamma}\,. \label{eq:hosbotstop3}
\eeq
 The $\Delta^{\rm SUSYQCD/SUSYEW}_{H^+_k   \ti t^*_i\ti b_j }$ can be decomposed into 
\beq 
\Delta^{\rm SUSYQCD}_{H^+_k   \ti t^*_i\ti b_j } &=& 
\Delta_{\rm SUSYQCD}^{\rm vert} +
 \Delta_{\rm SUSYQCD}^{\rm CT} \,,\\
\Delta^{\rm SUSYEW}_{H^+_k   \ti t^*_i\ti b_j } &=& 
\Delta_{\rm SUSYEW}^{\rm vert} +
 \Delta_{\rm SUSYEW}^{\rm CT} +  \Delta_{\rm SUSYEW}^{\rm GWmix}\,.
\eeq
The contributions to $ \Delta_{\rm SUSYQCD}^{\rm vert} $ ($
\Delta_{\rm SUSYEW}^{\rm vert} $) arise 
from one-loop triangle diagrams containing at least a gluon or a
gluino (all possible particles except for gluons and
gluinos). The SUSY-EW counterterm contribution can be expressed as
 \beq
\Delta_{\rm SUSYEW}^{\rm CT}&=& \sum_{a=1}^2\sum_{b=1}^2 U^{\ti{b}*}_{ja}(\delta C_{H^+_2\ti{b}\ti{t}^\ast })_{ab}(U^{\ti{t}})_{ib} + \fr 12 \sum_{m=1}^2 g_{H^+_m   \ti t^*_i\ti b_j }^{\rm tree}\delta Z^{H^\pm}_{2m}\crn
&&+ \fr 12 \sum_{m=1}^2 g_{H^+_2   \ti t^*_m\ti b_j }^{\rm tree}\delta \bar Z^{\ti t}_{im}
+ \fr 12 \sum_{m=1}^2 g_{H^+_2   \ti t^*_i\ti b_m }^{\rm tree}\delta  Z^{\ti b}_{jm}\,,
\eeq
where the $(\delta C_{H^+_2\ti{b}\ti{t}^\ast })_{ab}$ are obtained
from Eqs.~\eqref{eq.ct1}--\eqref{eq.ct4}
%,\eqref{eq.ct2},\eqref{eq.ct3},\eqref{eq.ct4} 
by differentiating these expressions with respect to their parameters
except for the rotation matrix $Z^{H^\pm}$, resulting in
\begin{align}
\label{eq.dct1}
(\delta C_{H^+_2\ti{b}\ti{t}^\ast })_{11}&=\frac{c_{\beta}^2 \delta t_{\beta} v \left(s_{\beta}^2 \left(g_2^2-2 
y_b^2\right)-c_{\beta}^2 \left(g_2^2-2 y_t^2\right)\right)}{2 
\sqrt{2}}\crn
&+\frac{c_{\beta} s_{\beta} \left(2 v (\delta y_b y_b+\delta 
y_t y_t-\delta g_2 g_2)+\delta v \left(y_b^2+y_t^2-g_2^2\right)\right)}{\sqrt{2}} \;,
\\\label{eq.dct2}
(\delta  C_{H^+_2\ti{b}\ti{t}^\ast})_{12}&=c_{\beta} \delta A_t
e^{i\varphi_u} y_t+\delta \mu_{\rm eff}^* s_{\beta} y_t+\delta y_t \left(A_t c_{\beta} 
e^{i\varphi_u}+\mu_{\rm eff}^* s_{\beta}\right) \;,
\\\label{eq.dct3}
(\delta  C_{H^+_2\ti{b}\ti{t}^\ast})_{21}&=
c_{\beta} \delta \mu_{\rm eff} e^{i \varphi_u } y_b+\delta 
A_b^* s_{\beta} y_b+\delta y_b \left( A_b^* s_{\beta}+c_{\beta} 
e^{i \varphi_u } \mu_{\rm eff} \right) \;,
\\\label{eq.dct4}
(\delta  C_{H^+_2\ti{b}\ti{t}^\ast })_{22}&=\frac{\delta v e^{i\varphi_u} y_b y_t }{\sqrt{2}}+\frac{\delta y_b e^{i\varphi_u} v y_t 
}{\sqrt{2}}+\frac{\delta y_t e^{i
\varphi_u} v y_b }{\sqrt{2}}\,, 
\end{align}
with 
\begin{align}
\frac{\delta \mu_{\rm eff}}{\mu_{\rm eff}}=\frac{\delta \lambda}{\lambda}+\frac{\delta v_s}{v_s}\,. 
\end{align}
The SUSY-QCD counterterm $\Delta_{\rm
  SUSYQCD}^{\rm CT}$ is similar to the counterterm
$\Delta_{\rm SUSYEW}^{\rm CT}$, with the modification
  that the counterterms $\delta g_2$, $\delta v$, $\delta t_\beta$, $\delta
  \mu_{\rm eff}$ do not receive QCD contributions and are hence set to
  zero then, and the remaining counterterms are
    obtained from SUSY-QCD loop corrections instead of SUSY-EW corrections.
The mixing contributions with a $W^+$ boson and a charged Goldstone
boson are given by 
\begin{align} \label{eq:mixstsb}
    \Delta^{\rm GWmix}_{\rm SUSYEW}
    =\frac{g_2\hat{\Sigma}_{H^{+}W^{+}}(M_{H{\pm}}^{2}) }{\sqrt{2} M_{W}^{2}}
\big(M_{ \ti{t}_{i}}^{2}- M_{ \ti{b}_{j}}^{2}\big) U^{\ti t}_{i1} U^{{\ti b}*}_{j1}\,.
   \end{align}  

Finally the real photon and  gluon emission
contributions, necessary to cancel the IR divergences
  arising from the vertex corrections, are given by
\begin{align}
\notag
\label{eq:chpsusdgam}
\Gamma_{H^+\to \ti t_i \ti b_j^*\gamma}&=
\frac{3\alpha}{4\pi^2 M_{H^{\pm}} }\abs{g_{H^+_k   \ti t^*_i\ti b_j }^{\rm tree}}^2 
\bigg[-\frac{2}{3}  (M_{H^{\pm}}^2 +M_{\ti{t}_i}^2-M_{\ti{b}_j}^2)I_{10}+\frac{1}{3}  (-M_{H^{\pm}}^2+M_{\ti{t}_i}^2-M_{\ti{b}_j}^2)I_{20}\crn
&
-\frac{2}{9}  (M_{H^{\pm}}^2-M_{\ti{t}_i}^2-M_{\ti{b}_j}^2)I_{21}-
M_{H^{\pm}}^2 I_{00}-\frac{4}{9} M_{\ti{t}_i}^2 I_{11}-\frac{1}{9} 
M_{\ti{b}_j}^2 I_{22}-I_0-\frac{4}{9} I_1-\frac{1}{9} I_2\bigg] \;,
\\ \notag
\label{eq:chpsusdg}
\Gamma_{H^+\to \ti t_i\ti b_j^*g}&=
\frac{3C_F\alpha_{\rm S}}{4\pi^2M_{H^{\pm}}}\abs{g_{H^+_k   \ti t^*_i\ti b_j }^{\rm tree}}^2 
\Big((M_{H^{\pm}}^2-M_{\ti{t}_i}^2-M_{\ti{b}_j}^2)I_{12}-M_{\ti{b}_j}^2I_{22}-M_{\ti{t}_i}^2I_{11}
 \\ 
&-I_1-I_2\Big),
\end{align}
where  the arguments $(M_{H^{\pm}}^2,M_{\ti{t}_i}^2,M_{\ti{b}_j}^2)$ of the $I$ functions have been dropped. 
The full expressions for the real photon and gluon emission
contributions Eqs.~\eqref{eq:chpsusdgam} and \eqref{eq:chpsusdg} are
in agreement with those of Ref.~\cite{Goodsell:2017pdq}. \s

The NLO decay width for the decay $H^{+}\to \ti{\tau}^\ast_i \ti{\nu}$
($i=1,2$) is composed of the tree-level contribution and the SUSY-EW corrections
and given by
\beq 
\Gamma^{\rm NLO}_{H^+\to \ti \tau_i^\ast \ti \nu_\tau }= \Gamma^{\rm
  tree}_{H^+\to \ti \tau_i^\ast \ti \nu_\tau }+ \Gamma^{\rm
  SUSYEW}_{H^+\to \ti \tau_i^\ast \ti \nu_\tau}\,, 
\label{eq:hostausneut}
\eeq
with the tree-level decay width 
\beq  
\Gamma^{\rm tree}_{H^+\to \ti \tau_i^*\ti \nu_\tau
}&=&R_2\abs{g_{H^+_{2} \ti \nu_\tau  \ti  \tau_i^* }} ^2,
\eeq 
in terms of the coupling $g_{H^+_{2} \ti \nu_\tau  \ti  \tau_i^* }$
given in Eq.~(\ref{eq:ghptaunu})
and where in the phase space factor $R_{2}$ the
loop-corrected masses of the staus 
and the sneutrino are used,  hence, $\mu_{i}=M^{2}_{\ti
  \tau_{i}}/M_{H^+}^{2},\ \mu_{j}=M^{2}_{\ti \nu_\tau}/M_{H^+}^{2}$.
The NLO SUSY-EW corrections are given by
\beq 
 \Gamma^{\rm SUSYEW}_{H^+\to \ti \tau_i^*\ti \nu_\tau }&=&2R_2g_{H^+_{2}  \ti \nu_{\tau} \ti
   \tau _i^* }^{\rm tree} \Delta^{\rm SUSYEW}_{H^+  \ti
   \nu_\tau \ti \tau_i^* }  +\Gamma_{H^+\to \ti \tau_i^*\ti \nu_\tau\gamma}\,, \label{eq:hostausnugam}
\eeq
 where the $\Delta^{\rm SUSYEW}_{H^+  \ti \nu_\tau \ti \tau_i^* }$ are given by
\beq 
\Delta^{\rm SUSYEW}_{H^+  \ti \nu_\tau \ti \tau_i^* } = \Delta_{\rm SUSYEW}^{\rm vert} +
 \Delta_{\rm SUSYEW}^{\rm CT} +  \Delta_{\rm SUSYEW}^{\rm GWmix} \;.
\eeq
 While $\Delta_{\rm SUSYEW}^{\rm vert}$ denotes the contributions
 from the one-loop triangle diagrams of the loop-corrected decay
 $H^{+}\to \ti{\tau}^{\ast}_i \ti{\nu}_{\tau} $, the counterterm
 contribution $\Delta_{\rm SUSYEW}^{\rm CT}$ reads 
 \beq
\Delta_{\rm SUSYEW}^{\rm CT}&=& \sum_{a=1}^2 U^{\ti{\tau *}}_{ia}(\delta
C_{H^+\ti{\nu}\ti{\tau}^* })_{a} + \fr 12 \sum_{m=1}^2 g_{H^+_m   \ti
  \nu_{\tau} \ti \tau_{i}^*  }\delta Z^{H^\pm}_{2m}\crn 
&&+ \fr 12  g_{H^+_2  \ti \nu_{\tau} \ti \tau_{i}^* }\delta
\bar Z^{\ti{\nu}_{\tau}} 
+ \fr 12 \sum_{m=1}^2 g_{H^+_2   \ti \nu_{\tau} \ti \tau_{m}^* }\delta
Z^{\ti \tau}_{im} \;,  
\eeq
where
\begin{align}
%\label{eq.dct1}
(\delta C_{H^+\ti{\tau}^*\ti{\nu}_{\tau} })_{1}&=\frac{c_{\beta}^2 \delta t_{\beta} v \left(s_{\beta}^2 \left(g_2^2-2 
y_\tau^2\right)-c_{\beta}^2 g_2^2\right)}{2 
\sqrt{2}}\crn
&+\frac{c_{\beta} s_{\beta} \left(2 v (\delta y_\tau y_\tau-\delta g_2 g_2)+\delta v \left(y_\tau^2-g_2^2\right)\right)}{\sqrt{2}} ,
\\
%\label{eq.dct3}
(\delta  C_{H^+\ti{\tau}^*\ti{\nu}_{\tau}})_{2}&={c_{\beta} \delta \mu_{\rm eff} e^{i \varphi_u } y_\tau}+\delta 
A_\tau^* s_{\beta} y_\tau+\delta y_\tau \left( A_\tau^* s_{\beta}+{c_{\beta} 
e^{i \varphi_u } \mu_{\rm eff} }\right)\;.
\end{align} 
The contribution of the $G/W$ mixing reads
\begin{align}
 \Delta_{\rm SUSYEW}^{\rm GWmix}&=\fr{g_2U^{\ti\tau *}_{i1} \hat{\Sigma}_{H^{+}W^{+}}(M_{H^+}^{2})}{M_W^2\sqrt{2}}(M^{2}_{\ti\nu_{\tau}}-M^{2}_{\ti\tau_{i}}) \,,
\end{align}
 and the real photon emission
is given by
\begin{align}
\notag
\label{eq:chpstausneugam}
\Gamma_{H^+\to \ti \tau^*_i \ti \nu_\tau\gamma}&= \fr{\alpha}{4\pi^2 M_{H^{\pm}}}\bigg(
 (-M_{H^{\pm}}^2-M^{2}_{\ti\tau_{i}}+M^{2}_{\ti\nu_{\tau}})I_{01}
-M_{H^{\pm}}^2 I_{00}-M^{2}_{\ti\tau_{i}}  I_{11}-I_0-I_1  \bigg),
\end{align}
where again the arguments
$(M_{H^{\pm}}^2,M^{2}_{\ti\tau_{i}},M^{2}_{\ti\nu_{\tau}})$ of the $I$
functions have been dropped. 
%%%%%%%%%%%%%%%%%%%%%%%%%%%%%%%%%%%%%%%%%%%%%%%%%%%%%%%%%%%%%%
\section{Numerical Results \label{sec:numerics}}
In the following we will discuss the impact of the computed
higher-order corrections on the charged Higgs boson decays and
branching ratios. In order to get an overall picture we performed a
scan in the NMSSM parameter space and only retained those data sets
whose phenomenology is in accordance
with the most recent experimental results. For this purpose, the
parameter points were checked against compatibility with the experimental
constraints from the Higgs data by using the programs {\tt
   HiggsBounds}~\cite{Bechtle:2008jh,Bechtle:2011sb,Bechtle:2013wla}
and {\tt HiggsSignals}~\cite{Bechtle:2013xfa}. 
The effective couplings of the Higgs bosons normalized to the corresponding SM
values, that are required as input for these programs, have been
obtained with the Fortran code {\tt NMSSMCALCEW} \cite{Baglio:2019nlc}. 
One of the neutral
CP-even Higgs bosons is identified with the SM-like Higgs boson and
will be called $h$ from now on. Its mass is required to lie in the
range 
\beq
123 \mbox{ GeV } \le m_h \le 127 \mbox{ GeV} \;.
\eeq
The SM input parameters have been chosen
as~\cite{PhysRevD.98.030001,Dennerlhcnote}   
\begin{equation}
\begin{tabular}{lcllcl}
\quad $\alpha(M_Z)$ &=& 1/127.955, &\quad $\alpha^{\overline{\mbox{MS}}}_s(M_Z)$ &=&
0.1181\,, \\
\quad $M_Z$ &=& 91.1876~GeV\,, &\quad $M_W$ &=& 80.379~GeV \,, \\
\quad $m_t$ &=& 172.74~GeV\,, &\quad $m^{\overline{\mbox{MS}}}_b(m_b^{\overline{\mbox{MS}}})$ &=& 4.18~GeV\,, \\
\quad $m_c$ &=& 1.274~GeV\,, &\quad $m_s$ &=& 95.0~MeV\,,\\
\quad $m_u$ &=& 2.2~MeV\,, &\quad $m_d$ &=& 4.7~MeV\,, \\
\quad $m_\tau$ &=& 1.77682~GeV\,, &\quad $m_\mu$ &=& 105.6584~MeV\,,  \\
\quad $m_e$ &=& 510.9989~KeV\,, &\quad $G_F$ &=& $1.16637 \cdot 10^{-5}$~GeV$^{-2}$\,.
%\label{eq:param1} 
\end{tabular}
\end{equation} 
For the NMSSM sector we follow the SUSY Les Houches Accord
(SLHA) format~\cite{Skands:2003cj} in which the soft SUSY breaking
masses and trilinear couplings are understood as $\DRb$ parameters at
the scale
\be 
\mu_R = M_s= \sqrt{m_{\ti Q_3} m_{\ti t_R}} \;.  \label{eq:renscale}
\ee
This is also the renormalization scale that we use in the computation
of the higher-order corrections. 
The code {\tt NMSSMCALCEW} provides the option to choose either
$A_\lambda$ or $M_{H^\pm}$ as input parameter. We adopted the latter
choice and used the charged Higgs boson mass as an OS input
parameter. The computation of the ${\cal 
  O}(\alpha_t\alpha_s +\alpha_t^2)$ corrections to the Higgs boson
masses is done in the $\DRb$ renormalization scheme of the top/stop
sector \cite{Muhlleitner:2014vsa,Dao:2019qaz}. 
In Table~\ref{tab:nmssmscan} we summarize the ranges applied in the 
parameter scans. In order to ensure perturbativity we apply the rough
constraint 
\beq
\lambda^2 + \kappa^2 < 0.7^2 \;.
\eeq
The bottom trilinear coupling has been fixed to 
\beq
 \quad A_b= 2 \mbox{ TeV}\;.
\eeq
The mass parameters of the first and second generation sfermions are
chosen as
\beq   
m_{\tilde{u}_R,\tilde{c}_R} = 
m_{\tilde{d}_R,\tilde{s}_R} =
m_{\tilde{Q}_{1,2}}= m_{\tilde L_{1,2}} =m_{\tilde e_R,\tilde{\mu}_R}
= 3\;\mbox{TeV} \;.
\eeq
\begin{table}
\begin{center}
\begin{tabular}{l|ccc|ccccccccccc} \toprule
& $t_\beta$ & $\lambda$ & $\kappa$ & $M_1,M_2$ & $M_3$ & $A_t,A_\tau$ &
 $m_{\tilde{Q}_3}$ & $m_{\tilde{t}_R},m_{\tilde{b}_R}$ & $m_{\tilde{\tau}_R},m_{\tilde{L}_3}$ & $M_{H^\pm}$
& $A_\kappa$ & $\abs{\mu_{\text{eff}}}$ \\
& & & & \multicolumn{9}{c}{in TeV} \\ \midrule
min &  1 &   0 & -0.7& 0.5 & 1.8 & -6 &  1 & 1 & 0.4 & 0.5 &-2 & 0.2 \\
max & 20 & 0.7 & 0.7 & 1   & 2.5 & 6  & 2.5   &  2.5 & 3   & 3   & 2 & 1 \\ \bottomrule
\end{tabular}
\caption{Input parameters for the NMSSM scan. All parameters have been 
varied independently between the given minimum and maximum
values. \label{tab:nmssmscan}}
\end{center}
\end{table}
From the scan we retain those points that
have a $\chi^2$ computed by {\tt HiggsSignals}-2.2.3 that is consistent with
an SM $\chi^2$ within $2\sigma$.\footnote{In {\tt HiggsSignals}-2.2.3,
  the SM $\chi^2$ obtained with the latest data set is 84.44. We
  allowed the NMSSM $\chi^2$ to be in the range $[78.26,90.62]$.}  
For the scan and the results
presented in the first part of the numerical analysis we keep the
CP-violating phases equal to zero. In the second part we turn on
various CP-violating phases in order to study their impact
individually. Note that in all our CP-conserving scenarios it is
  the lightest CP-even Higgs boson $H_1$ that is SM-like and has a
  mass around 125~GeV.
\s

All the branching ratios shown in the following have been calculated
by implementing the higher-order corrections to the
various charged Higgs boson decay widths in {\tt NMSSMCALCEW}. In this
way the newly computed corrections are combined with the
state-of-the-art higher-order QCD 
corrections already included in {\tt NMSSMCALCEW}. Note, however, 
that the (SUSY-)EW and SUSY-QCD corrections are only taken into
account if the respective decay is kinematically allowed. Otherwise,
the corresponding decay width without the higher-order corrections
discussed in this paper, which only apply for on-shell decays, is used in the
computation of the total decay width and branching ratios. \s

For the computation of radiative corrections we used the following
renormalization schemes unless stated differently (see also
Section~\ref{sec:renorm}): In the 
electroweakino sector we used the OS1 
renormalization scheme. For the SUSY-EW corrections to the decays into
stop-sbottom and stau-sneutrino pairs the $\overline{\mbox{DR}}$
scheme was applied for the stop, sbottom sector and the OS scheme for
the stau sector. For details
on the definition of the schemes, we refer to Ref.~\cite{Baglio:2019nlc}.

%%%%%%%%%%%%%%%%%%%%%%%%%%%%%%%%%%%%%%%%%%%%%%%%%%%%%%%%%%%%%%
\subsection{The Branching Ratios}
\begin{figure}[htbp] 
   \centering
   \includegraphics[scale=0.7, trim=0mm 8mm 3mm 14mm, clip]{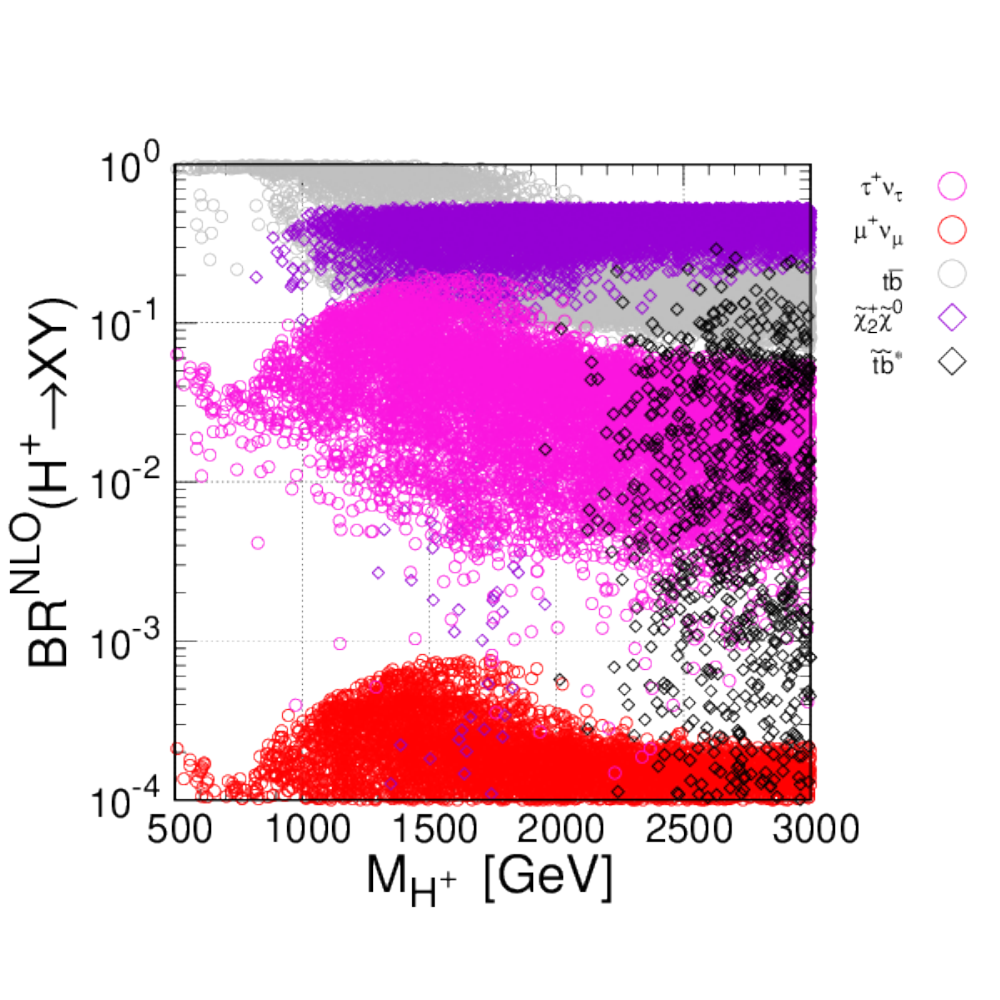} 
   \includegraphics[scale=0.7, trim=0mm 8mm 3mm 14mm, clip]{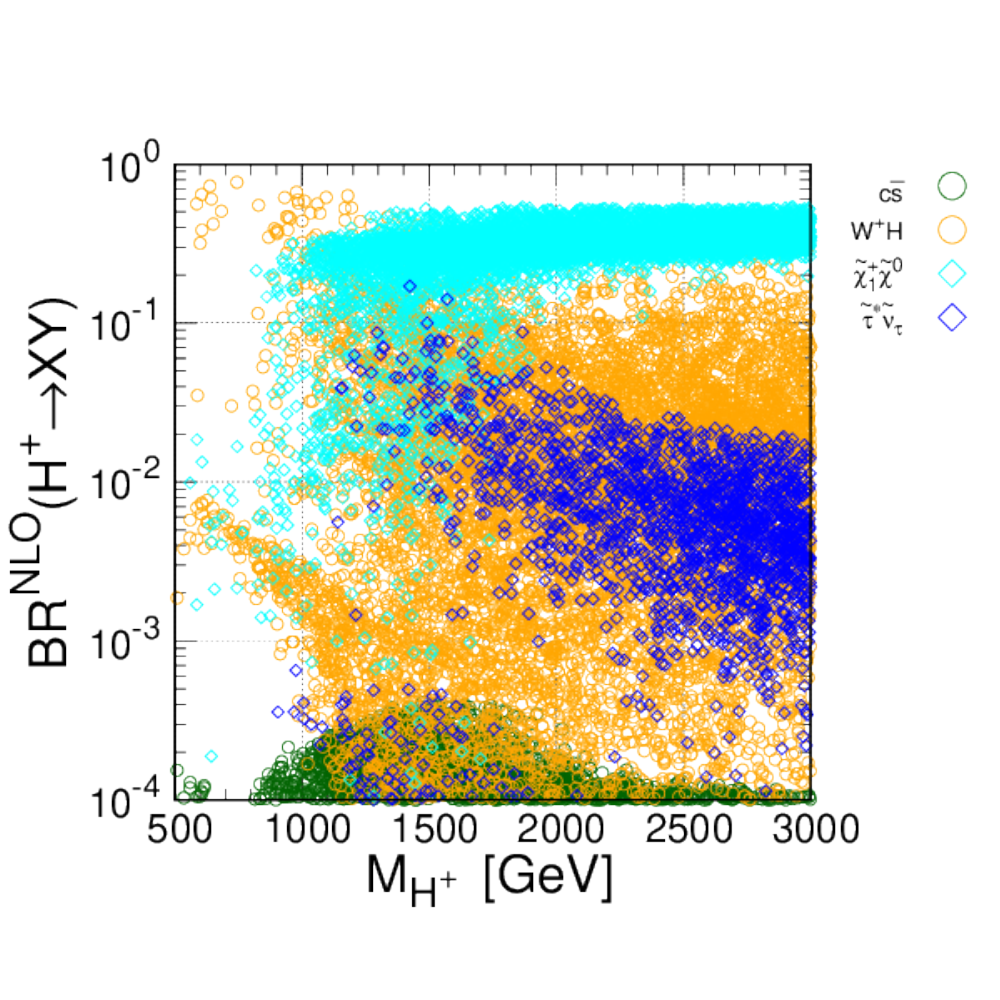} 
%\vspace*{-8cm}
   \caption{Branching ratios of the charged Higgs boson $H^+$ into
     various final states including the NLO corrections as described
     in the text as a function of $M_{H^\pm}$ for all scan parameter points
     passing the constraints. Left: $t\bar{b}$ (gray diamond), $\tilde{\chi}_2^+
     \tilde{\chi}^0$ (violet diamond), $\tau^+ \nu_\tau$ (pink circle),
     $\tilde{t}\tilde{b}^*$ (black diamond), $\mu^+
     \nu_\mu$ (red circle); right: $W^+H$ (orange circle) $\tilde{\chi}_1^+
     \tilde{\chi}^0$ (cyan diamond), $\tilde{\tau}^* \tilde{\nu}_\tau$ (blue
     diamond), $c\bar s$ (green circle).}
   \label{fig:loandnlobrs}
\end{figure}

In Fig.~\ref{fig:loandnlobrs} we show as function of the charged Higgs
boson mass the NLO branching ratios of the
charged Higgs boson decays into the various possible final states,
namely into the SM-like final states $t\bar{b}$, $\tau^+ \nu_\tau$, $\mu^+
\nu_\mu$ and $c\bar{s}$, and the new physics final states $W^+H$,
$\tilde{\chi}^+_1 \tilde{\chi}^0$, $\tilde{\chi}^+_2 \tilde{\chi}^0$,
$\tilde{t}\tilde{b}^\ast$ and $\tilde{\tau}^\ast \tilde{\nu}_\tau$. In the
decays into a charged $W^+$ boson plus Higgs final state we have summed
up all $W^+H_i$ final states, so that the branching ratio $\mbox{BR}(H^+ \to W^+ H)$ is
given by\footnote{Note that
  the here investigated parameter sets do not include CP violation so that the five
  neutral Higgs states are CP eigenstates dividing up in three neutral CP-even states
  $H_i$ and two neutral CP-odd bosons $A_j$.}  
%------------------------------------------------
%------------------------------------------------
\beq
\mbox{BR}(H^+ \to W^+ H) \equiv \sum_{i=1}^{3} \mbox{BR}(H^+ \to W^+
H_i) + \sum_{j=1}^{2} \mbox{BR}(H^+ \to W^+ A_j) \;.
\eeq
Analogously, in the branching ratios into the electroweakino
final states we have summed over the neutralinos, hence 
\beq
\mbox{BR}(H^+ \to \tilde{\chi}^+_{1,2} \tilde{\chi}^0) \equiv \sum_{i=1}^5
\mbox{BR}(H^+ \to \tilde{\chi}^+_{1,2} \tilde{\chi}^0_i) \;.
\eeq
Also the decays into the sfermion final states are summed over so that 
\begin{align}
\mbox{BR}(H^+ \to \tilde{t}\tilde{b}^*) &\equiv \sum_{i=1}^2 \sum_{j=1}^2
\mbox{BR}(H^+ \to \tilde{t}_i \tilde{b}_j^*) \,,\\
\mbox{BR}(H^+ \to \tilde{\tau}^*\tilde{\nu}_\tau) &\equiv \sum_{i=1}^2
\mbox{BR}(H^+ \to \tilde{\tau}_i^* \tilde{\nu}_{\tau})\;.
\end{align} 

The NLO branching ratios include the higher-order corrections to the
Higgs decays widths as presented in the draft, namely the SUSY-EW
corrections as well as the QCD and SUSY-QCD corrections for the 
coloured final states. More specifically, the formulae for the
loop-corrected decay widths are given in Eq.~(\ref{eq:htotb}) for the
decay in the top-bottom final state, in Eq.~(\ref{eq:Gam_lnu_NLO_imp}) for the decay
into $\tau^+ \nu_\tau$, in Eqs.~(\ref{eq:ewino1}) for the decays into 
electroweakino pairs, in Eqs.~(\ref{eq:hosbotstop1}) for the decays into 
stop-sbottom pairs, and in Eq.~(\ref{eq:hostausneut}) for those into
stau-sneutrino pairs. The implemented higher-order corrections to the
decays into charged $W^+$ boson plus Higgs final states have been
described in Ref.~\cite{Dao:2019nxi} for the CP-even Higgs bosons in
the CP-conserving NMSSM. We have extended this to the
CP-violating case. The decays into the SM-like final states include the
higher-order QCD and resummed SUSY corrections as specified in the
manual for {\tt HDECAY} \cite{Djouadi:1997yw,Djouadi:2018xqq} and extended to the
NMSSM case in \cite{Baglio:2013iia}. 
As mentioned above the SUSY-EW and SUSY-QCD
corrections are only included in on-shell decays. Otherwise, (where
applicable) only QCD corrections and resummed corrections through
effective couplings are included. We furthermore include in the decays
with on-shell neutral Higgs bosons in the external states the resummed $\ZH$
factors, {\it cf.}~Sec.~\ref{sec:renorm}. \s

As can be inferred from the plots,   the
largest branching ratios are given by the decays into top-bottom final
states (gray circles) with values of up to almost 100\% for charged
Higgs mass values below $2\,\TeV$. For larger values of the charged
Higgs mass the decays into electroweakinos become dominant. For $\mHp<
1.2\,\TeV$, the decays into a charged $W$ plus a neutral Higgs boson
provide the second largest branching ratio for some parameter
points. As stated above 
we show here the sum over all possible neutral Higgs bosons. The
resulting branching ratio, indicated by the orange circles, can reach
up to 98\%. The decays into the electroweakinos
(cyan diamonds for $\tilde{\chi}_1^+ \tilde{\chi}^0$, violet diamonds
for $\tilde{\chi}_2^+ \tilde{\chi}^0$) can reach up to
54\% (55\%) for $\tilde{\chi}_1^+ \tilde{\chi}^0$ ($\tilde{\chi}_2^+
\tilde{\chi}^0$) when summed 
up. The summed-up branching ratios into stau-sneutrino pairs (blue diamonds) can
go up to 17\% for charged Higgs masses below 1.4 TeV and specific parameter
configurations, whereas the decays into stop-sbottom pairs (black
diamonds) become important for large charged Higgs masses and can have
branching ratios of up to 29\% in their sum. The branching
ratios into $\tau^+ \nu_\tau$ (pink circles) reach 20\%. The branching
ratios into $\mu^+ \nu_\mu$ (red circles) and 
$c\bar{s}$ (green circles) attain at most $7\cdot 10^{-2}$\% and
$4 \cdot 10^{-2}$\%, respectively. 
The comparison of these scatter plots with the corresponding ones for the
leading order (LO) branching ratios shows that 
the overall pattern of the distribution of the branching ratios does
not change when NLO corrections are included. For individual parameter
points the changes can be substantial, however. In the following, we
will discuss the impact of the higher-order corrections for the
various final states separately.

%%%%%%%%%%%%%%%%%%%%%%%%%%%%%%%%%%%%%%%%%%%%%%%%%%%%%%%%%%%%%%
\subsection{Impact of Higher-Order Corrections}
For the discussion of the impact of the NLO corrections on the decay
width of the decay $H^+ \to X Y$ we introduce the
  relative correction of the partial width as
\begin{align}
\label{eq:delGamdef}
\delta_{\Gamma}(H^{+}X Y)&=\frac{\Gamma(H^{+}\to X Y)^{\rm
                                NLO} }{\Gamma(H^{+}\to X Y)^{{\rm
                                LO}} }-1   \;.
\end{align}
We furthermore define the relative change in the branching ratio for
the decay $H^+ \to X Y$ as
\begin{align}
\Delta_{\rm BR}^{\rm }(H^{+}X Y)=\frac{{\rm BR}^{\rm NLO} (H^{+}\to X Y) 
-{\rm BR}^{\rm LO}(H^{+}\to X Y)}
{ {\rm max}( {\rm BR}^{\rm NLO} (H^{+} \to X Y),{\rm BR}^{\rm
  LO}(H^{+} \to X Y) ) } \;.
\label{eq:relchange}
\end{align}
This quantity allows us to directly identify large corrections in the
branching ratios that are not 'artificially' enhanced because of tiny
LO branching ratios. \s
%and
%\begin{align}
%\label{eq:delBRdef}
%\delta_{\rm BR}(H^{\pm}X_iX_j)&=\frac{{\rm BR}(H^{\pm}\to
%                                X_iX_j)^{{\rm NLO }} }{{\rm
%                                BR}(H^{\pm}\to X_iX_j)^{{\rm LO}} }-1  \;,
%\end{align}
%respectively. \s

We have to specify what we mean by LO widths and branching
ratios. They are the LO quantities calculated with 'Higgs effective
tree-level couplings', which means that the Higgs tree-level rotation matrix
elements have been replaced by the loop-corrected ones. For decays with neutral
Higgs bosons in the final state, additionally the
improved resummed $\ZH$ factor as described in
Ref.~\cite{Baglio:2019nlc} is included in the LO decay widths and
branching ratios. Note, that these 'LO'
quantities also include the QCD corrections
and resummed SUSY-EW and SUSY-QCD corrections in effective quark
couplings as already implemented in the first release of {\tt
  NMSSMCALC} \cite{Baglio:2013iia} and described there. In fact the
use of the word 'LO' in essence means that we thereby refer to the old
implementation in {\tt NMSSMCALC} without the genuine SUSY-EW and
SUSY-QCD vertex corrections. This means that the definitions
Eqs.~(\ref{eq:delGamdef}) and (\ref{eq:relchange}) give us information
on the effects of the newly computed corrections, namely the SUSY-EW
and SUSY-QCD vertex corrections, respectively their finite remainders,
compared to the previous 
implementation in {\tt NMSSMCALC} which only uses the improved
LO decay widths as defined here. 

\subsection{Decays into Fermion Pairs}
\begin{figure}[htbp]
   \centering
   \includegraphics[scale=0.7]{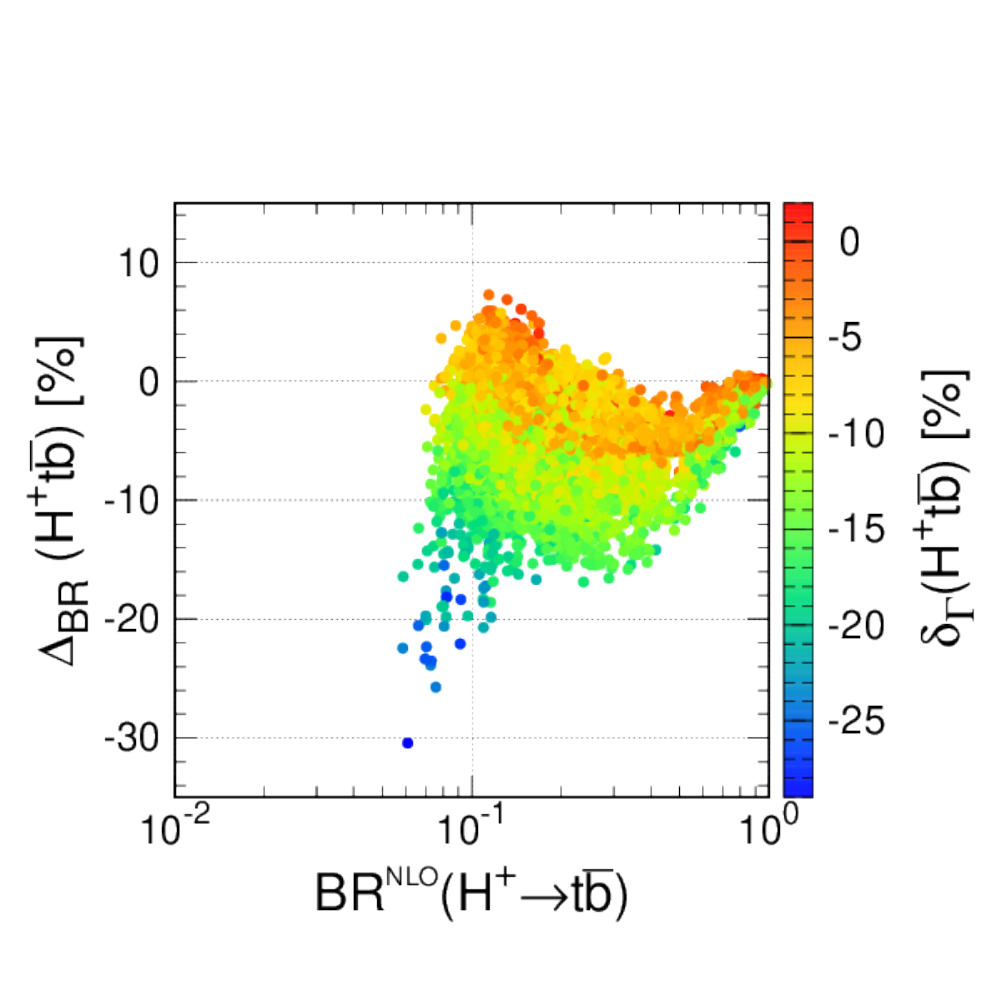} 
   \includegraphics[scale=0.7]{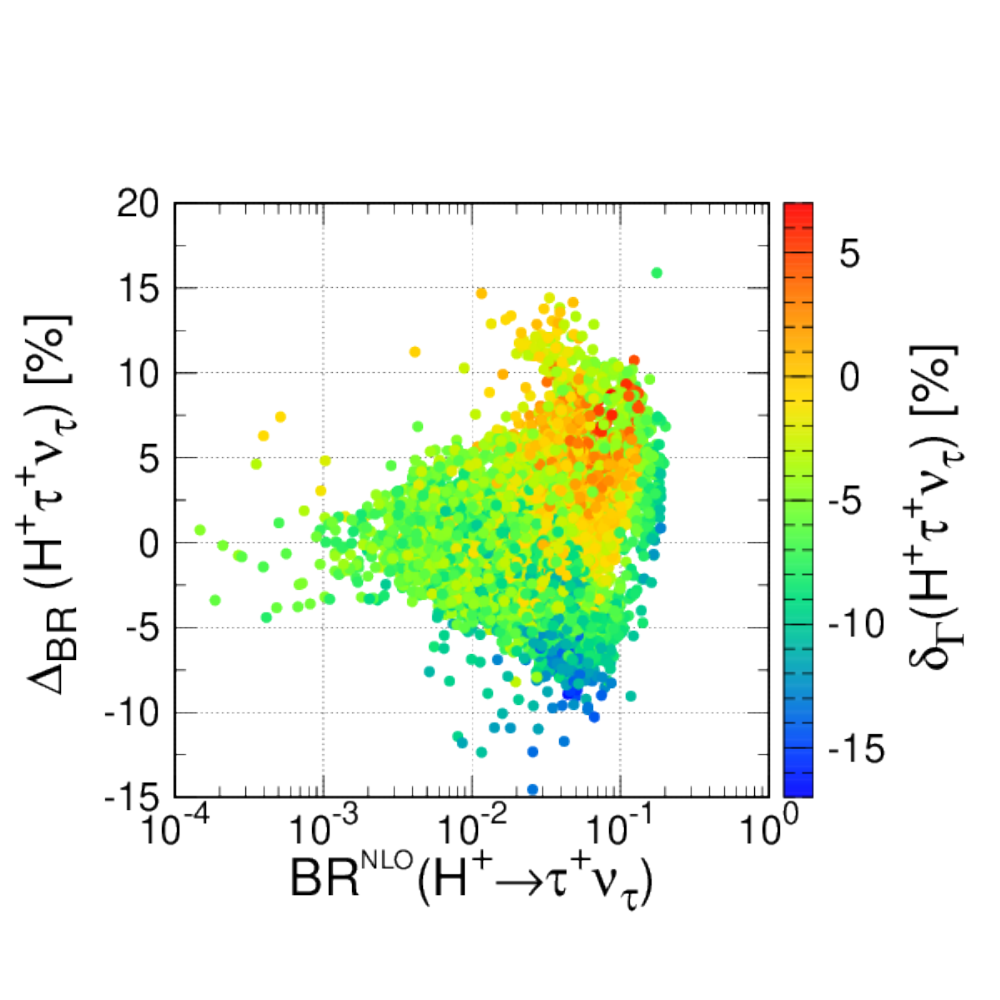} 
   \caption{Relative change in the branching ratio as defined in 
     Eq.~(\ref{eq:relchange}) for the $H^+$
     decays into $t\bar{b}$ (left) and and $\tau^+ \nu_\tau$ (right) as a
     function of the respective NLO-corrected (SUSY-QCD and SUSY-EW for
     the former and SUSY-EW for the latter decay) branching ratio for all
     scan parameter points passing the constraints. The color code
     indicates the relative correction in the partial width as defined in
     Eq.~(\ref{eq:delGamdef}).}  
   \label{fig:hofermions}
\end{figure}

Figure~\ref{fig:hofermions} displays the relative, $\Delta_{\rm
  BR}$, due to the impact of the SUSY-QCD and SUSY-EW corrections on
the branching ratio 
${\rm BR} (H^+ \to t\bar{b})$ (left) and the impact of the SUSY-EW
corrections on ${\rm BR} (H^+ \to \tau^+ \nu_\tau)$ (right) as a
function of their respective NLO branching ratios. The color code
indicates the respective relative corrections to the partial width in
per cent for the newly computed 
SUSY-QCD and SUSY-EW corrections. It shows that the impact of the
corrections on the partial width for the decay into $t\bar{b}$ is of
moderate size, ranging between -20\% to +2\% with a
few outliers going down to -29\%. The relative change in the
branching ratio is of similar size with values between
-20\% and +8\% and a few outliers going down to about -30\%.
%The reduction in the
%branching ratio is due to the fact that the decay into $t\bar{t}$ is
%dominating the total width entering the branching ratio. 
Splitting up the contributions, we find that apart from a few outliers
the relative SUSY-EW corrections to the partial width (branching ratio) lie between
-16\%  and -2\% (-12\% and +6\%), whereas the relative SUSY-QCD corrections
range between -11\%  and +7\% (-10\%  and +4\%). The SUSY-EW
corrections on the width are 
negative and of comparable size as the SUSY-QCD ones which underlies
the importance of including both types of corrections. Large relative
corrections to the decay widths of up 
  to about -30\% arise from the sum of SUSY-QCD and SUSY-EW
  corrections with same sign. The dominant contributions to both the
  SUSY-EW corrections and the SUSY-QCD corrections stem from
  one-particle irreducible triangle diagrams. The SUSY-EW corrections to the
decay width into $\tau^+ \nu_\tau$ mostly lie between -17\%  and +7\% 
and between -10\% and +15\% for $\Delta_{\rm BR}$ for the bulk of the
points.  

%------------------------------------------------
%------------------------------------------------

%%%%%%%%%%%%%%%%%%%%%%%%%%%%%%%%%%%%%%%%%%%%%%%%%%%%%%%%%%%%%%
\subsection{Decays into Gauge plus Higgs Boson Pairs}
\begin{figure}[htbp]
   \centering
   \includegraphics[scale=0.6, trim=0mm 28mm 3mm 35mm, clip]{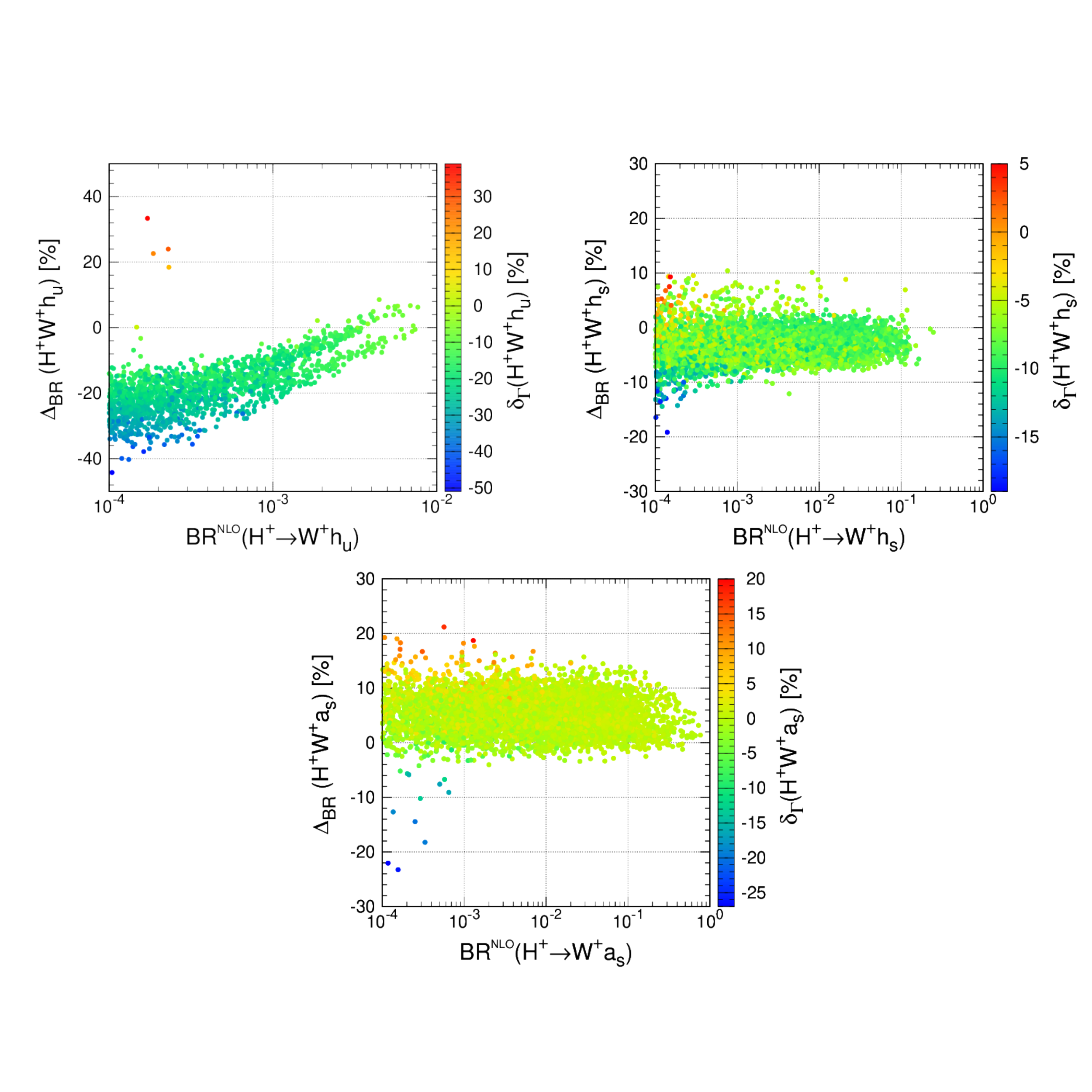} 
\vspace*{0cm}
   \caption{Relative change in the branching ratio as defined in 
     Eq.~(\ref{eq:relchange}) for the $H^+$
     decays into $W^+h_u,$ $W^+h_s$, and $W^+a_s$ (going clockwise
     from upper left to lower middle) as a
     function of the respective SUSY-EW corrected branching ratio for all
     scan parameter points passing the constraints. The color code
     indicates the relative correction in the partial width as defined in
     Eq.~(\ref{eq:delGamdef}).}  
   \label{fig:whiggsdecay}
\end{figure}

In Fig.~\ref{fig:whiggsdecay} we show the relative change in the
branching ratios due to our newly computed genuine SUSY-EW 
corrections and as colour code the relative correction for the partial decay widths of
the $H^+$ decays into charged $W^+$ boson plus Higgs final
states as a function of the corresponding NLO branching ratio. We do
not classify the Higgs final states by the mass eigenstates but by the gauge eigenstates,
{\it i.e.}~we show final states with mostly $h_u$, $h_s$ and $a_s$
final states (decays into $W^+ h_d$ and $W^+ a_d$ are kinematically 
closed). Mostly $j$-like ($j=h_u,h_s,a_s$) means that the mixing
matrix element squared $|R_{i j}|^2$ of the Higgs eigenstate $H_i$
exceeds 0.5. Note, that the $h_u$-like state 
corresponds to the SM-like Higgs boson as compatibility with the Higgs
data requires a maximum coupling to the top-quark. \s
%------------------------------------------------
%------------------------------------------------

As can be inferred from the upper left plot the bulk of the corrections to the
branching ratio (decay width) for the 
decay into an $h_u$-like, {\it i.e.}~SM-like, Higgs boson together
with the charged $W^+$ boson 
lies between -42\% and +10\% (-35\% and +3\%). There
are a few outliers with somewhat larger corrections. We note that the
branching ratio into $W^+ h_u$ always remains below
$0.75 \%$ and is hence rather unimportant for charged Higgs
decays. \s

The branching ratios into singlet-like CP-even
Higgs final states, $W^+ h_s$, with up to 25\% reach larger values than those into 
$W^+h_u$. The relative corrections to the decay widths are moderate and
range between -18\% and +4\% with the change in the branching ratio
being between -14\% and +10\%. There are two outliers with
$\Delta_{\rm BR}$ reaching up to -20\%. Similarly, the corrections to the decay width in
the CP-odd singlet-like Higgs state are moderate with
corrections to the decay width between -10\% and
  +15\% and to the branching ratio between mostly -2\% and +20\%. A few
outliers can reach corrections to the branching ratio of up to
-23\%. \s

Note that for all scattering plots here we do not consider branching
ratios that are smaller than $10^{-4}$, since they are not
phenomenologically interesting. We can have large relative corrections in these
cases because of the suppression of the tree-level
couplings. 

%%%%%%%%%%%%%%%%%%%%%%%%%%%%%%%%%%%%%%%%%%%%%%%%%%%%%%%%%%%%%%
\subsection{Decays into Electroweakinos}
\begin{figure}[h!]\centering
\includegraphics[width=0.48\linewidth, trim=0mm 5mm 1mm 20mm, clip]{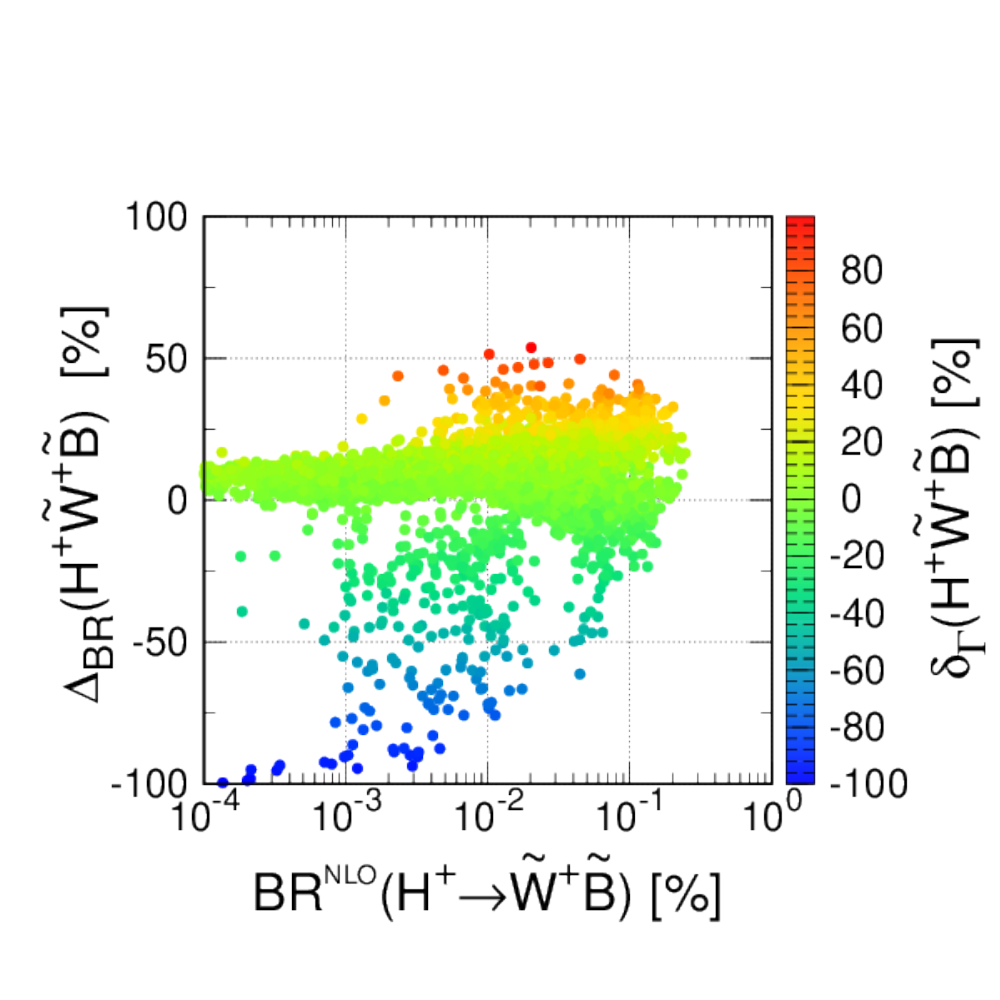}\hspace{5mm}
\includegraphics[width=0.48\linewidth, trim=0mm 5mm 1mm 20mm, clip]{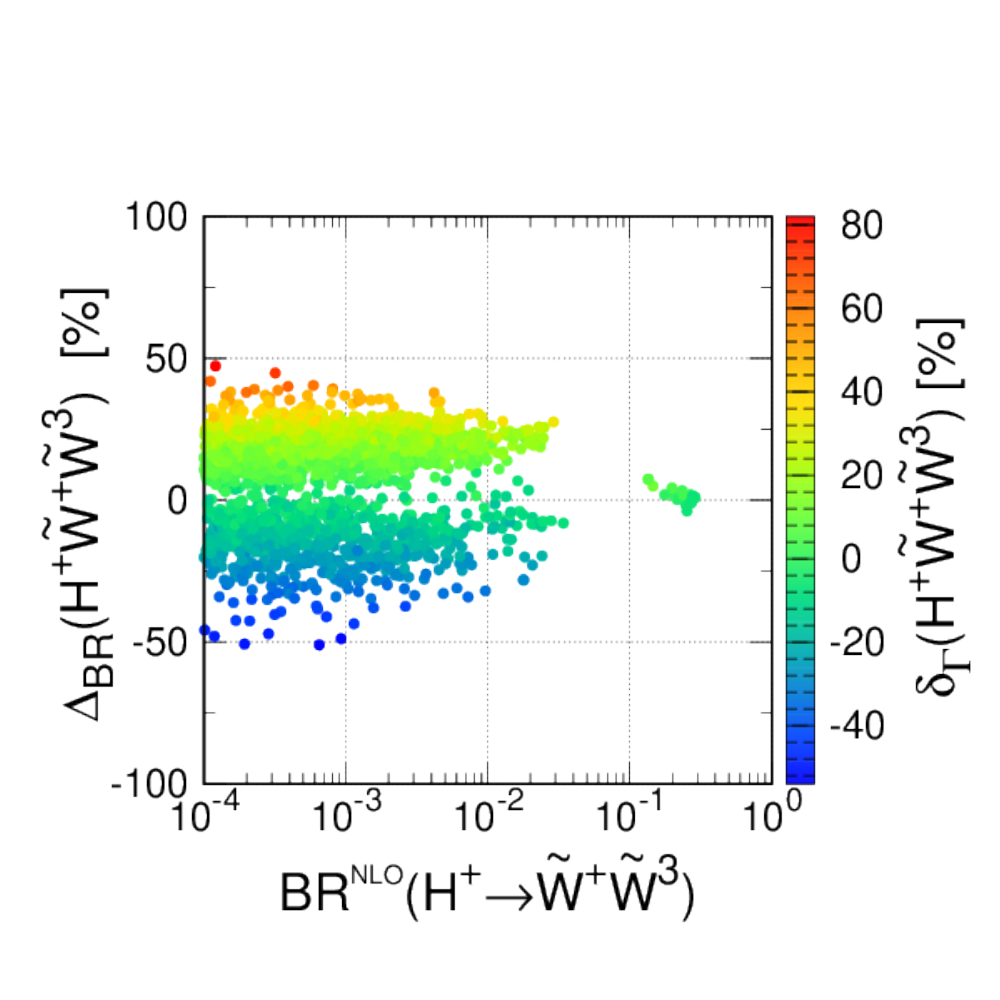} \\
\includegraphics[width=0.48\linewidth, trim=0mm 5mm 1mm 20mm, clip]{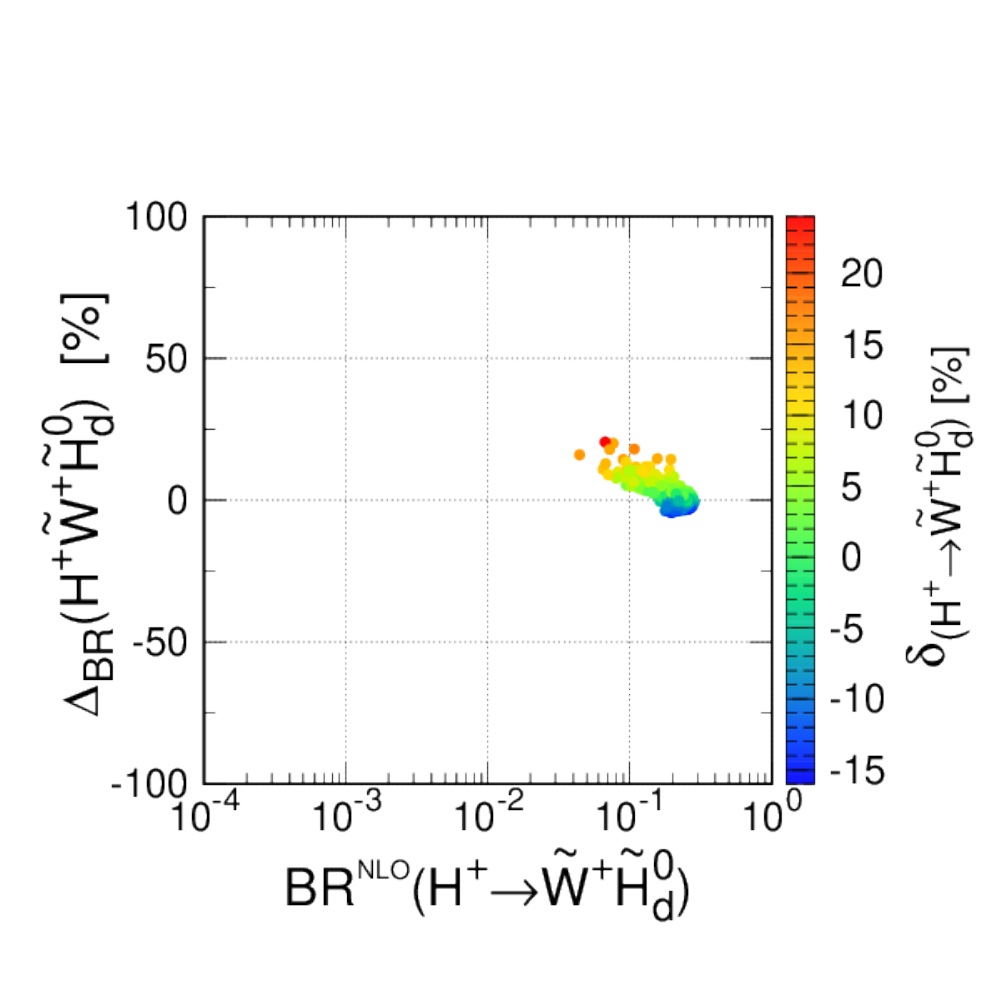}\hspace{5mm}
\includegraphics[width=0.48\linewidth, trim=0mm 5mm 1mm 20mm, clip]{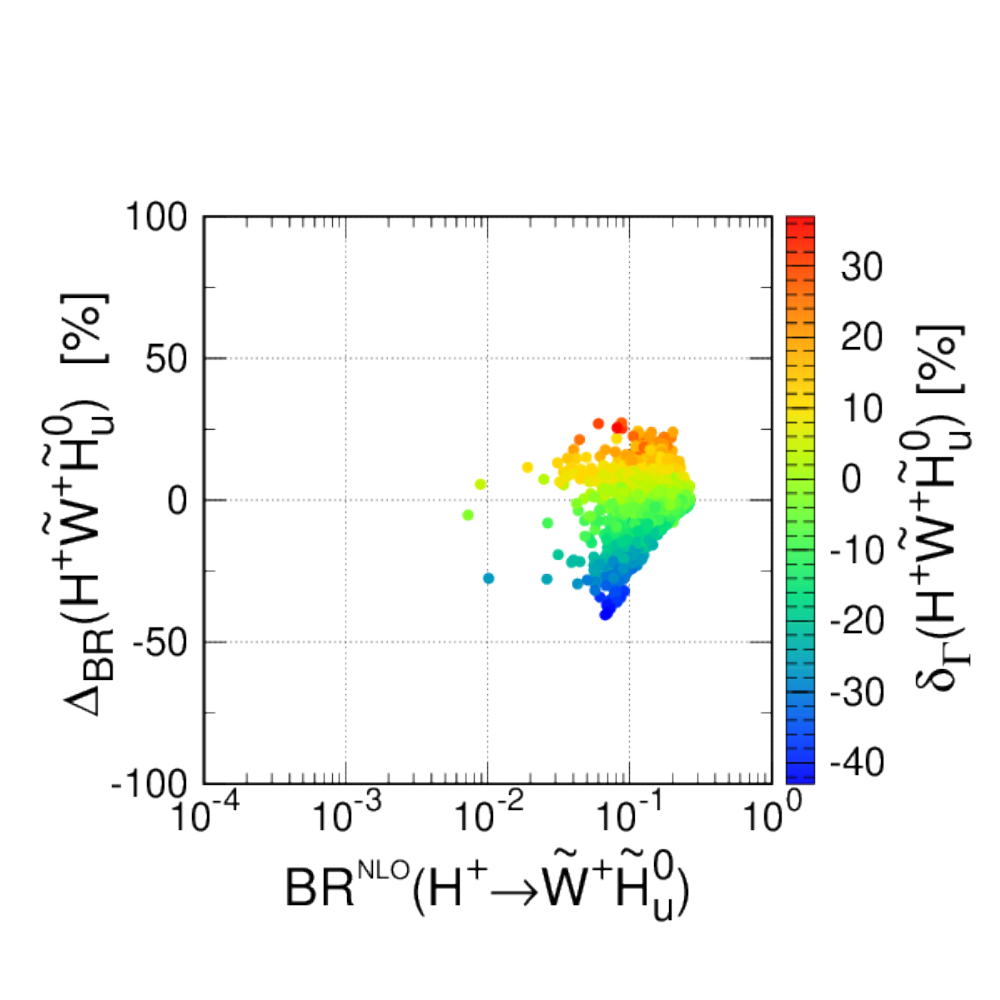} %\\
\includegraphics[width=0.48\linewidth, trim=0mm 5mm 1mm 20mm, clip]{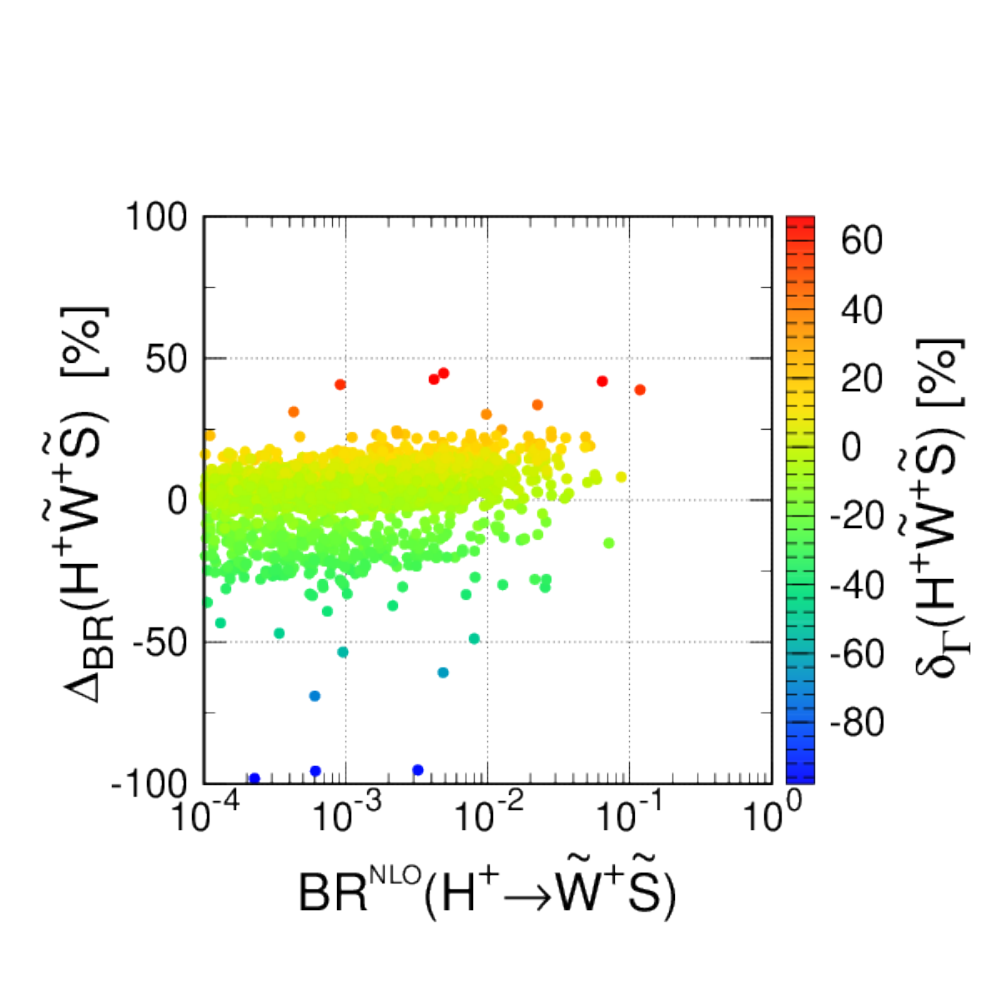}
\caption{Relative change in the branching ratio as defined in 
     Eq.~(\ref{eq:relchange}) for the $H^+$
     decay into  into charged wino plus neutral electroweakino final states
     all in the gauge basis as a
     function of the respective SUSY-EW corrected branching ratio for all
     scan parameter points passing the constraints. The color code
     indicates the relative correction in the partial width as defined in
     Eq.~(\ref{eq:delGamdef}).}
\label{fig:DelBRimp_chi0chia}
\end{figure}

\begin{figure}[ht!]\centering
\includegraphics[width=0.48\linewidth, trim=0mm 5mm 1mm 20mm, clip]{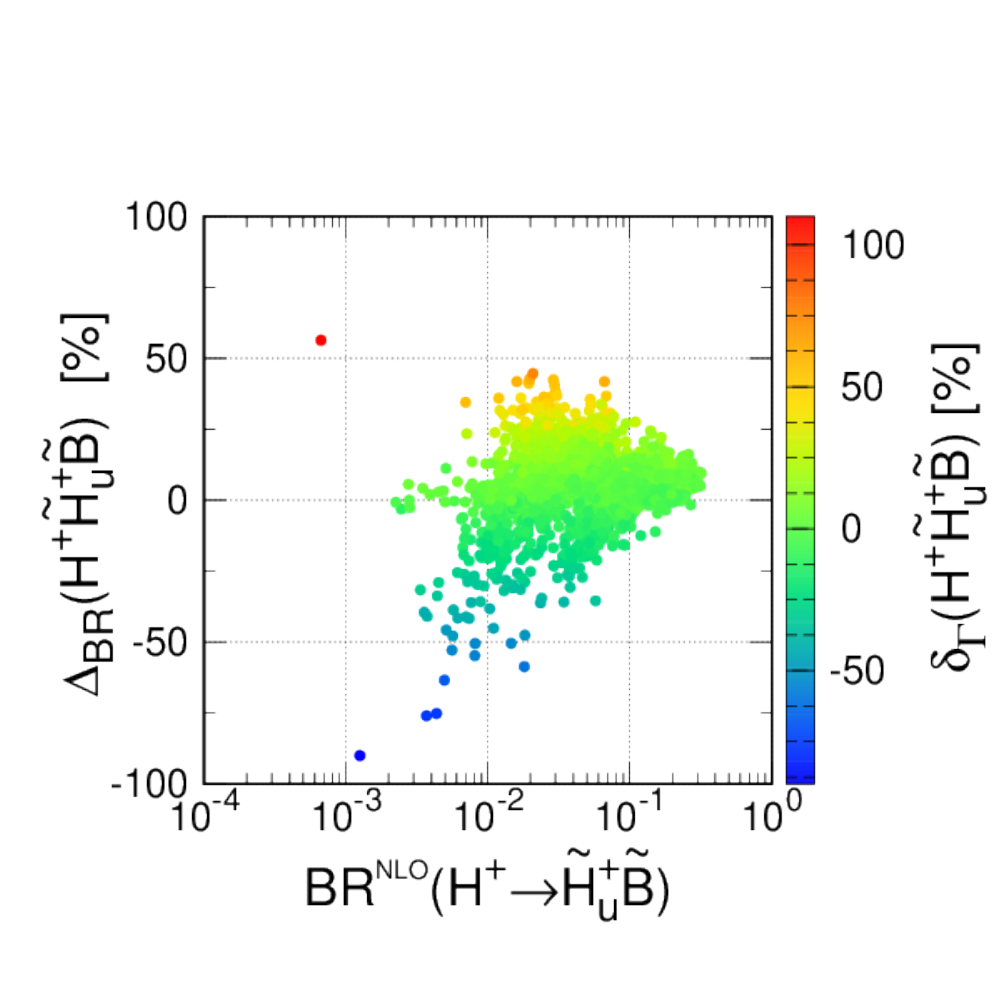}\hspace{5mm}
\includegraphics[width=0.48\linewidth, trim=0mm 5mm 1mm 20mm, clip]{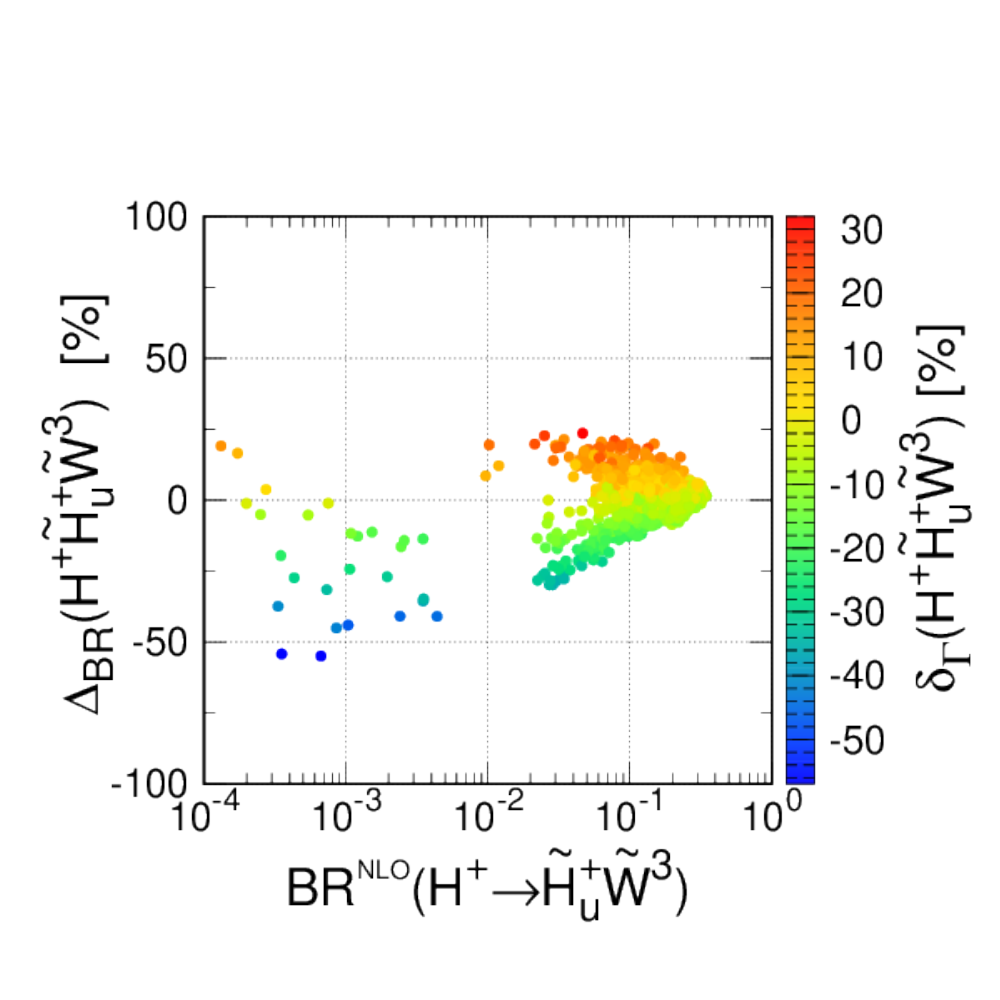} \\
\includegraphics[width=0.48\linewidth, trim=0mm 5mm 1mm 20mm, clip]{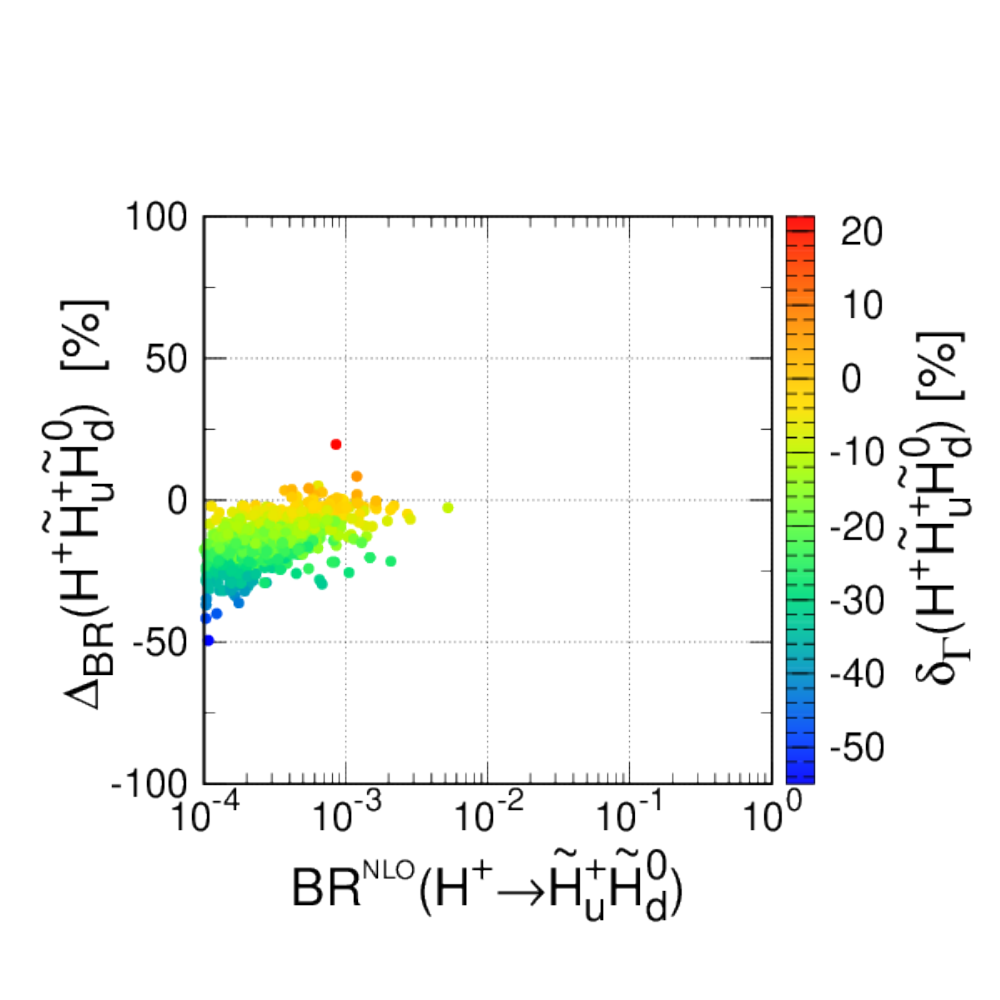}\hspace{5mm}
\includegraphics[width=0.48\linewidth, trim=0mm 5mm 1mm 20mm, clip]{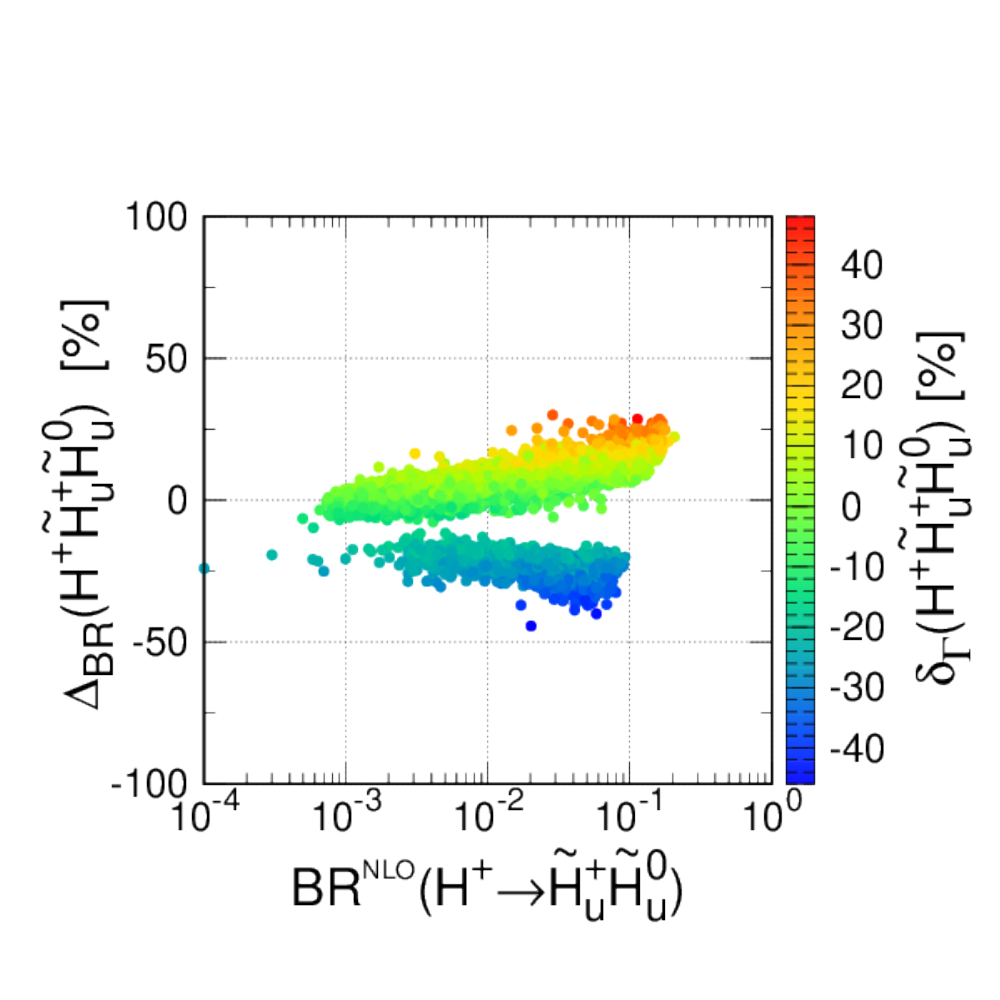} %\\
\includegraphics[width=0.48\linewidth, trim=0mm 5mm 1mm 20mm, clip]{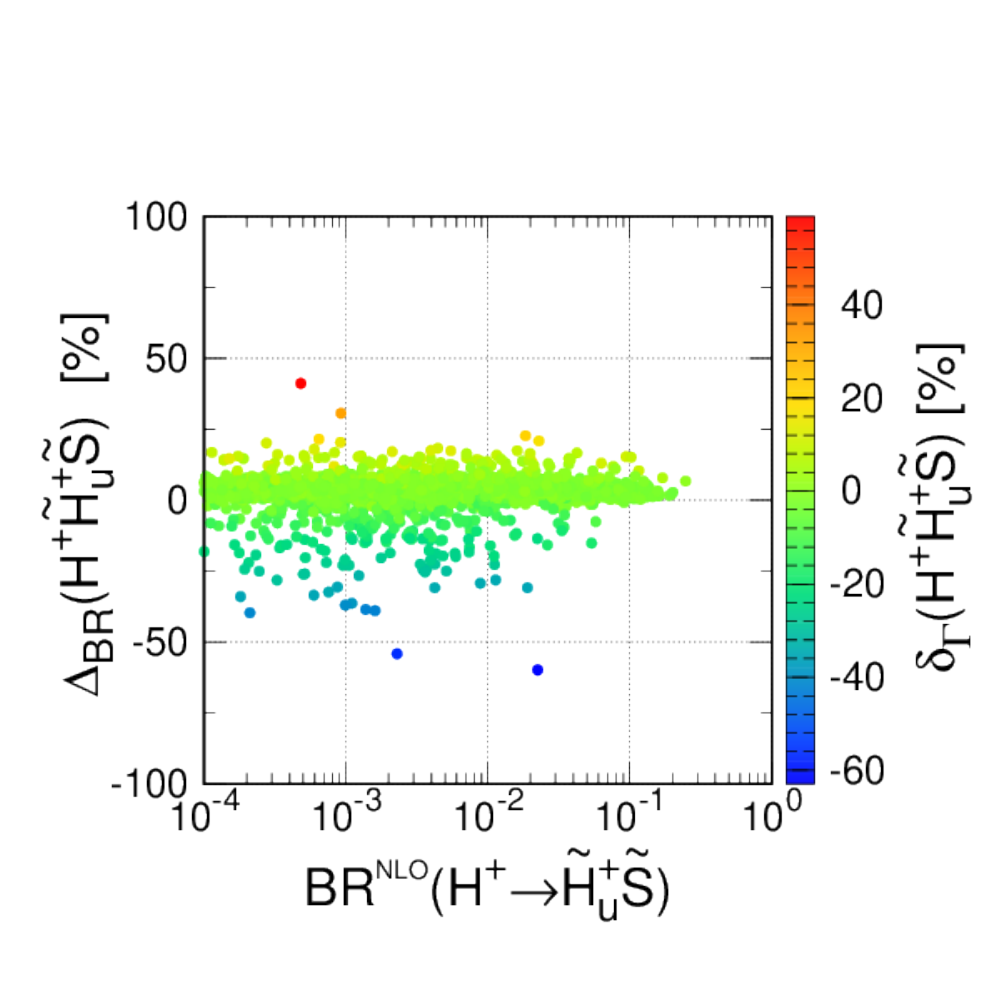}
\caption{Same as Fig.~\ref{fig:DelBRimp_chi0chia} but for the charged
  Higgsino plus neutral electroweakino final states.}
\label{fig:DelBRimp_chi0chib}
\end{figure}

We now turn to the impact of the SUSY-EW corrections on the decays
into electroweakino pairs. The relative changes in the branching
ratios and the relative corrections of the decay widths are shown for
the final states in the 
gauge basis, in Fig.~\ref{fig:DelBRimp_chi0chia} for the charged wino
and in Fig.~\ref{fig:DelBRimp_chi0chib} for the charged higgsino final
state, respectively, together with a neutral electroweakino as
specified in the figure labels. In the plots we show results for the
parameter points passing our constraints after applying the following
cuts: We cut the ratio 
\beq
r = \frac{g_2(U_{i2}^* N_{j2}^*-U_{i1}^* N_{j3}^*)}{g_1 U_{i2}^* N_{j1}^*}
\eeq
to lie in the range
\beq
0.5 \le r \le 1.5 \;.
\eeq
This ensures that there are no large hierarchies among the left-handed
couplings\footnote{The left-handed couplings give the dominant
  contribution.} of the charged Higgs to a neutralino-chargino pair,
{\it cf.}~Eq.~(\ref{eq:lhcoupling}), which would otherwise blow up the NLO
corrections compared to the LO width. If $r\in [0.5 ... 1.5] $, 
there exist cancellations in the tree-level couplings of the decay in question,
that lead to a suppression of the tree-level decay width. Contributions coming
from a  neutralino close
  in mass with an enhanced tree-level coupling will dominate and
can be very large. In {\tt NMSSMCALCEW} we print out a warning if this
case occurs. Furthermore, for the charged 
Higgs decay $H^+ \to \tilde{\chi}_i^+ \tilde{\chi}_j^0$ we impose the
following cuts on the mass differences 
\beq
\begin{array}{ll}
m_{\tilde{\chi}_2^+} - m_{\tilde{\chi}_1^+} > 10 \mbox{ GeV}
  \hspace*{0.3cm} \mbox{ and }\\[0.3cm]
m_{\tilde{\chi}_{j+1}^0} - m_{\tilde{\chi}_{j}^0} > 10 \mbox{ GeV }
\; \wedge \;\; 
m_{\tilde{\chi}_j^0} - m_{\tilde{\chi}_{j-1}^0} > 10 \mbox{ GeV } & \mbox{ if } j=2,3,4 
\\
m_{\tilde{\chi}_{j+1}^0} - m_{\tilde{\chi}_{j}^0} > 10 \mbox{ GeV } & \mbox{ if } j=1
\\
m_{\tilde{\chi}_j^0} - m_{\tilde{\chi}_{j-1}^0} > 10 \mbox{ GeV } & \mbox{ if } j=5
\end{array}
\eeq
to avoid large mixing effects between the close-in-mass electroweakino
masses that also induce huge NLO corrections. Since we use the
OS scheme for the WFR constants of the electroweakinos, if two
neutralinos (charginos) are degenerate then the WFR constant
contributions are dominant and huge.\footnote{Note, that
  huge corrections blowing up for certain renormalization schemes
  in specific corners of the SUSY parameter space are a known feature, see {\it
    e.g.}~Refs.~\cite{Belanger:2016tqb,Belanger:2017rgu}, 
  which requires dedicated treatments tailored to specific parameter
  configurations.} In the code we print out a
warning if degenerate cases are involved. In our study, we fix the renormalization
scheme for all points to be OS thus encountering about one hundred
points\footnote{These points are in the decays of $H^+$ into $\ti W^+
  \ti B$,  $\ti W^+ \ti S$, $\ti  H_u^+ \ti B$ and $\ti  H_u^+ \ti S$. with relative corrections beyond
100\% among 10,000 allowed points. Without the applied cuts the
scattering plots will involve large scales so that
  it becomes difficult to read off the corrections of most of the
points that have mild corrections. After applying these cuts} the
corrections have the typical size of EW corrections that we comment on
in the following. In case of unnaturally large loop
corrections we recommend to change the renormalization scheme. \s
%------------------------------------------------

The maximum relative corrections do not differ much for the charged wino and charged
Higgsino final states when comparing the final states with the same neutral
electroweakino. The smallest corrections are found for neutral
down-type Higgsino production together with a charged wino or
Higgsino, {\it i.e.}~$\tilde{W}^+ \tilde{H}_d^0$ and $\tilde{H}_u^+
\tilde{H}_d^0$. The relative corrections of the partial width lie
between about -16\% and +24\% for the former and -37\% and +2\% for
the latter, for the bulk of the points. A few outliers involve also
corrections of up to -55\% to the $\tilde{H}_u^+
\tilde{H}_d^0$ decay. The relative corrections of the branching ratios 
lie between about -4\% and
+20\% for $\tilde{W}^+ \tilde{H}_d^0$ production and -30\% and 0\% for $\tilde{H}_u^+
\tilde{H}_d^0$ production (again for the bulk of the points). The 
corrections to the $\tilde{W}^+ \tilde{H}_u^0$, $\tilde{W}^+
\tilde{W}^3$, $\tilde{W}^+
\tilde{S}$, $\tilde{H}_u^+ \tilde{H}_u^0$, $\tilde{H}_u^+
\tilde{W}^3$, $\tilde{H}_u^+ \tilde{S}$ final
states are somewhat larger but do not exceed what is in general
expected for EW corrections. The relative corrections to the decay widths lie between
about -35\% and +40\% (depending on the specific final state) for the
bulk of the parameter points, and those to the branching ratios between
about -30\% and +30\% apart from a few outliers. The largest
corrections are found for $\tilde{W}^+ \tilde{B}$, $\tilde{H}_u^+
\tilde{B}$ production where the relative corrections to the partial
widths range between -34\% (-34\%) and +77\% (+57\%) for $\tilde{W}^+
\tilde{B}$ ($\tilde{H}_u^+ \tilde{B}$) and to the branching ratios
between -30\% and +40\% barring a few outliers that can go up to
-100\%. These are found, however, for small LO widths and branching
ratios so that the relative correction easily gets enhanced. \s

Let us also briefly comment on the size of the branching ratios. The
largest branching ratios are obtained for the $\tilde{H}_u^+
\tilde{W}^3$ and $\tilde{W}^+ \tilde{W}^3$ final states with 37\% and
30\%, respectively, followed by $\tilde{H}_u^+ \tilde{B}$ (35\%) and
$\tilde{W}^+ \tilde{B}$ (25\%) production. The branching ratios into $\tilde{W}^+
\tilde{H}_d^0$, $\tilde{W}^+ \tilde{H}_u^0$, $\tilde{W}^+ \tilde{S}$,
$\tilde{H}_u^+ \tilde{H}_u^0$, $\tilde{H}_u^+ \tilde{S}$ reach maximum
values between 10 and 20\%. The smallest branching ratio is found for
the $\tilde{H}_u^+  \tilde{H}_d^0$ final state with at most 0.5\%. 
%------------------------------------------------
%%%%%%%%%%%%%%%%%%%%%%%%%%%%%%%%%%%%%%%%%%%%%%%%%%%%%%%%%%%%%%
\subsection{Decays into Sfermions}
%------------------------------------------------
\begin{figure}[t!]
   \centering
   \includegraphics[width=0.48\linewidth, trim=0mm 8mm 0mm 18mm, clip]{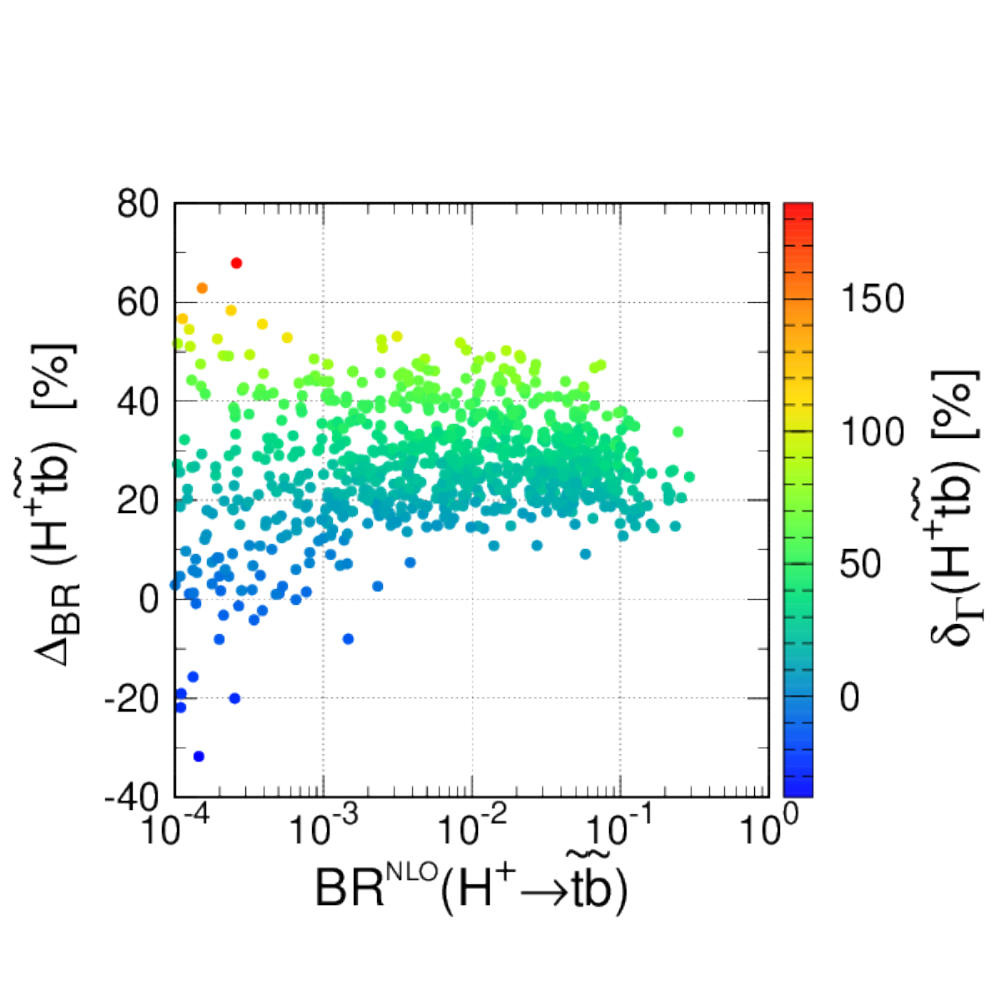}\hspace{0.3cm}
   \includegraphics[width=0.48\linewidth, trim=0mm 8mm 0mm 18mm, clip]{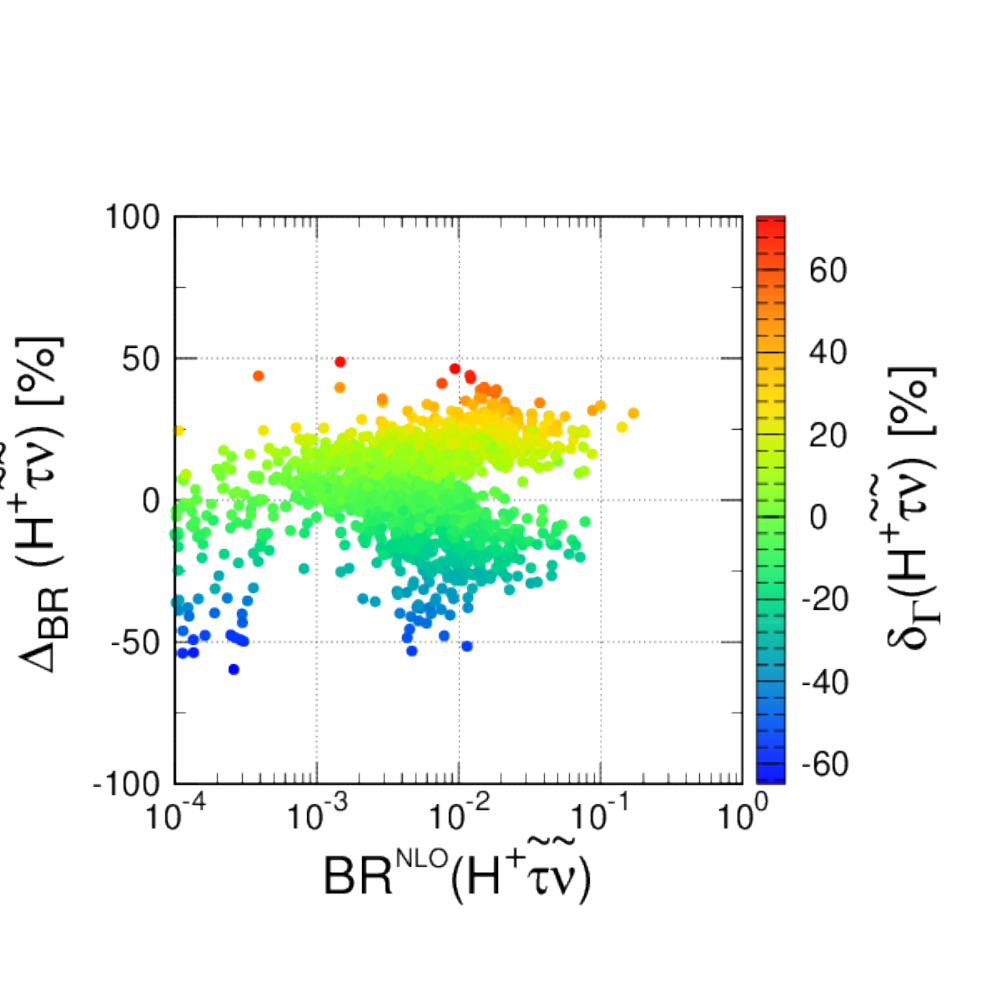}  
   \caption{Relative change in the branching ratio as defined in 
     Eq.~(\ref{eq:relchange}) for the $H^+$
     decays into stop-sbottom (left) and stau-sneutrino (right) as a
     function of their NLO corrected branching ratio for all
     scan parameter points passing the constraints. The color code
     indicates the relative corrections in the partial width as defined in
     Eq.~(\ref{eq:delGamdef}).}
   \label{fig:sfermiondecays}
\end{figure}
%------------------------------------------------

Like for the electroweakino decays also in the decays into sfermions
we apply some cuts to get rid of 
artificially enhanced corrections due to specific corners in the
parameter space. Thus we require for the decays into stop-sbottom pairs, $H^+
\tilde{t}_i \tilde{b}_j$, that the involved masses fulfill
\beq
M_{H^+} - M_{\tilde{t}_i} - M_{\tilde{b}_j} > 40 \mbox{ GeV}
\eeq
so that large corrections originating from threshold effects in 
triangle loop diagrams do not play a role. Furthermore, we apply the
following cuts on the mass differences\footnote{We remind the reader
  that the final state particle masses receive loop corrections as
  they are renormalized $\overline{\mbox{DR}}$ whereas the masses of
the particles in the loop are taken at tree level. Capital letters
denote loop-corrected, small letters denote tree-level masses.}
\beq
|M_{\tilde{t}_i}-m_{\tilde{t}_{i'}}| > 10 \mbox{ GeV} \quad
\mbox{and} \quad 
|M_{\tilde{b}_j}-m_{\tilde{b}_{j'}}| > 10 \mbox{ GeV} \;, \quad
\mbox{with}~
i\ne i', \; j \ne j' \;.
\eeq
This way large corrections arising from degenerate squark mixing contributions
are removed. In case of large loop corrections, we recommend the
user to change the renormalization scheme of the stop/sbottom sector
as in the case of decays into electroweakinos. \s

The relative corrections to the decay widths and branching ratios into
sfermion final states are shown in Fig.~\ref{fig:sfermiondecays} for
the stop-sbottom final state on the left and for the stau-sneutrino
final state on the right. Note that we sum over all possible final
states $\tilde{t}_i \tilde{b}_j^\ast$ ($i,j=1,2$) in the left and
$\tilde{\tau}_{i}^\ast \tilde{\nu}_\tau$ ($i=1,2$) in the right plot. 
The reason why we have less points in the stop-sbottom than in the
stau-sneutrino case is simply because in a lot of the scenarios of 
our scan compatible with the applied constraints the former decays are
kinematically closed. \s

The decays into squarks receive SUSY-EW and SUSY-QCD
corrections. For most
of the points the relative SUSY-QCD corrections to the decay widths
are distributed between 10\% and 90\%,  while the typical size for the
SUSY-EW corrections is between -30\% and +22\%. The
combination of both corrections alters the decay width between 
-37\% and +92\%. Some parameter points can reach
corrections of up to more than 100\% which is due to the
renormalization of the squark wave functions in the SUSY-QCD
corrections and calls for future improvements through the inclusion of
higher-order corrections or resummation. Barring these outliers the
relative corrections in the branching ratios range between -10\% and
+50\%. \s

The relative corrections to the decays into
stau-sneutrino, stemming only from SUSY-EW corrections, lie 
between -64\% and +72\%,
ranging for the bulk of the corrections only between -31\% and +39\%,
however. For $\Delta_{\text{BR}}$ we find values between -25\% and 30\%.

%%%%%%%%%%%%%%%%%%%%%%%%%%%%%%%%%%%%%%%%%%%%%%%%%%%%%%%%%%%%%%
%%%%%%%%%%%%%%%%%%%%%%%%%%%%%%%%%%%%%%%%%%%%%%%%%%%%%%%%%%%%%%
\subsection{The Impact of CP Violation \label{sec:cpviol}}
For the investigation of the impact of CP violation on the higher-order
corrections to the charged Higgs decays we chose the following
parameter point from the set of valid scenarios obtained in our scan, 
\beq
\label{eq:BPCP}
\begin{array}{llllllllllll}
M_{H^\pm}&=&2537\ {\rm GeV}, & \; t_\beta&=&4.84 \;, & \;
                                                       |\lambda|&=&0.590 \;,
& \;  |\kappa|&=&0.339 \;, \\ 
|M_1|&=&764 \ {\rm GeV}, & \; |M_2|&=& 917 \ {\rm GeV}, & \; |M_3|&=&2211 \
{\rm GeV},  \\ 
|A_t|&=& 3.4\ {\rm TeV}, & \; |A_b|&=& 2\ {\rm TeV},& \;  |A_\tau|&=&
                                                                      2\ {\rm TeV}, \\   
|\mu_{\rm eff}|&=&585\ {\rm GeV},& \;  \Re A_\kappa&=&- 16.4\ {\rm GeV},
                                                  & \;  m_{\tilde{Q_{3}}}&=&1.14\ {\rm TeV},  \\
m_{\tilde{t}_R}&=&1.57\ {\rm TeV}, & \;  m_{\tilde{b}_R}&=&1.76\ {\rm
  TeV},& \; m_{\tilde{L}_3}&=&476\ {\rm GeV},& \;
  m_{\tilde{\tau}_R}&=&1.66\ {\rm TeV.}
\end{array} \nonumber
\eeq
The remaining parameters  and phases are fixed as
 \begin{align}
%\notag
& 
m_{\tilde{u}_R,\tilde{c}_R}=m_{\tilde{d}_R,\tilde{s}_R}=m_{\tilde{Q}_{1,2}}=m_{\tilde{L}_{1,2}}=m_{\tilde{e}_R,\tilde{\mu}_R}=3\ {\rm TeV}, \nonumber \\
&\varphi_{M_{1},M_{2},M_{3}}=\varphi_{A_{t},A_{b},A_{\tau}}=\varphi_{\mu}=\varphi_{\kappa}=0 \;,
\end{align}
where for simplicity we choose the notation $\varphi_{\mu_{\rm
    eff}}\equiv \varphi_{\mu}$. 
The lightest CP-even Higgs boson $H_1$ is 
the SM-like Higgs boson and the Higgs mass spectrum is given by 
\beq
M_{H_1} &=&123.97 \mbox{ GeV}, 
\; M_{H_2}= 132.57 \mbox{ GeV}, \; M_{H_3}= 680.45 \mbox{ GeV}, \nonumber \\
M_{A_1} &=& 2536.92 \mbox{ GeV}, 
\; M_{A_2}= 2538.66 \mbox{ GeV} \;.
\eeq
In the following we vary the three phases $\varphi_{\mu}$,
$\varphi_{M_{2}}$, $\varphi_{A_{t}}$ individually away from their
benchmark value zero while keeping all the other phases fixed to zero in
order to quantify the effect induced by CP 
violation through the respective non-zero phase. For the phase of $\lambda$, we set
$\varphi_{\lambda}=2\varphi_{\mu}/3$ so that we do not encounter CP
violation at tree level in the Higgs sector.  Note, that we vary the phases
almost up to their maximum values 
leading to scenarios that are already excluded by the EDM
constraints. For illustrative purposes, we still allow for these
variations, however. We stop all plots at $\pm 0.47\pi$ and not at
$\pm \pi/2$ because for phases $|\varphi_\mu| > 0.47\pi$ the
singlet-like Higgs boson has a negative mass. \s

%%--------------------------------------CPV, tb
\begin{figure}[htbp]\centering
%trim=left bottom right top
\includegraphics[width=0.4\linewidth, trim=0mm 0mm 0mm 0mm,
clip]{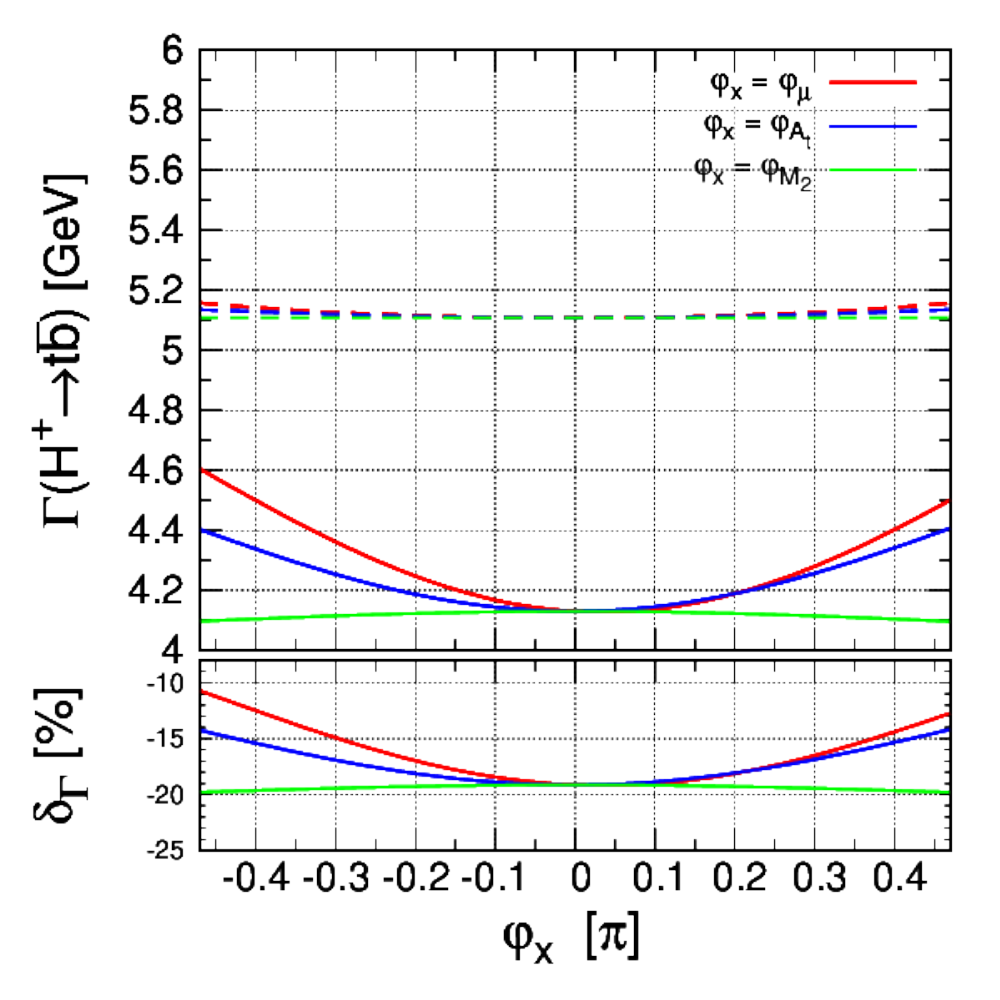}\hspace{10mm}
\includegraphics[width=0.4\linewidth, trim=0mm 0mm 0mm 0mm,
clip]{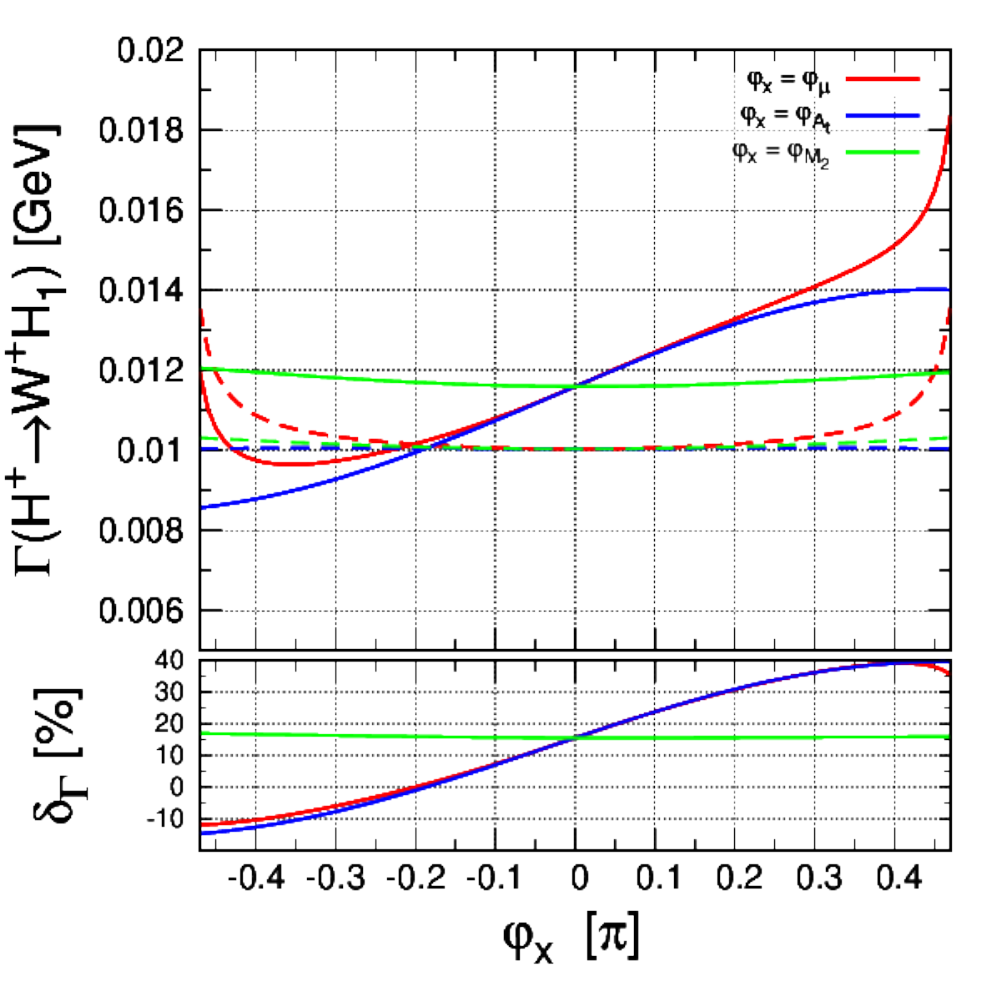}
\caption{Left: $\Gamma (H^+\to t\bar{b})$, right: $\Gamma (H^+\to
  W^+H_1)$  as function of the CP-violating 
  phase for $\mu_{\rm eff}$ (red), $A_{t}$ (blue) and $M_{2}$ (green), respectively.
Full lines correspond to the decay width at NLO, while dashed lines
are those at LO. The lower panels show the relative corrections
$\delta_\Gamma$ as defined in Eq.~(\ref{eq:delGamdef}). 
\label{eq:CPVtb}
}
\end{figure}
%%--------------------------------------
In Fig.~\ref{eq:CPVtb} (left) we show the tree-level and loop-corrected decay width of the
charged Higgs decay into the top-bottom final state as function of a
variation of either $\varphi_\mu$ (red), $\varphi_{A_t}$ (blue) or
$\varphi_{M_2}$ (green) while keeping all other phases to zero and ensuring
a vanishing tree-level CP-violating phase in the Higgs
sector. The lower panels show the
relative corrections of the partial decay width as defined in
Eq.~(\ref{eq:delGamdef}) as a function of the respective non-zero
phase.  Despite the vanishing CP-violating phase at tree level we
still see a small dependence of the tree-level decay width on the
CP-violating phase $\varphi_\mu$ and $\varphi_{A_t}$, respectively. This is due to
the $\Delta_b$ corrections 
included in our definition of the tree-level decay width into quarks,
which depends on the phases of $\mu_{\text{eff}}$ and $A_t$. As for
the loop corrections, the relative correction which is
$\delta_\Gamma = -19\%$ for the chosen parameter
point, barely changes when $\varphi_{M_2}$ is turned on, whereas for
non-zero $\varphi_\mu$ it varies from -19\% to
  $-11$\% for $\varphi_{\mu}=-0.47\pi$ and $-12.5$\% for
  $\varphi_{\mu}=+0.47\pi$. The dependence on $\varphi_{A_t}$ is
given by values ranging from -19\%
at vanishing phase to $-14$\% at
$\varphi_{A_t}=\pm 0.47\pi$. \s

For the decays into a charged $W^+$ plus Higgs boson final state for the
chosen benchmark point we see
the largest impact of the CP-violating phases on the $W^+H_1$ final
state, where $H_1$ is the SM-like Higgs
boson, {\it cf.}~Fig.~\ref{eq:CPVtb} (right). For $\varphi_\mu$ and
$\varphi_{A_t}$ each, the relative corrections vary from
about -10\% to at $-0.47\pi$ to about +40\% at $+0.47\pi$. The
dependence on $\varphi_{M_2}$ is very weak, however. 
The LO dependence on $\varphi_\mu$ and $\varphi_{A_{t}}$ is due to the
inclusion of the two-loop corrections into the final state Higgs boson
masses, where the dependence on $\varphi_\mu$ is
largest. 
For the $W^+H_2$ final state $\varphi_{\mu}$ and $\varphi_{M_2}$ change
$\delta_\Gamma$ from 3\% at zero phase to 4.5\%
(1\% for $\varphi_{\mu}$ and 1.5\% for $\varphi_{M_{2}}$) at $0.47\pi$
($-0.47\pi$). The dependence on 
$\varphi_{A_t}$ is very small. The loop corrections to $H^+ \to W^+A_1$
show a smaller dependence on $\varphi_\mu$, $\varphi_{M_2}$ and again
the dependence on $\varphi_{A_t}$ is almost negligible. The other
decays into charged $W^+$ boson plus Higgs final states are kinematically
closed. \s

%\begin{figure}[htbp]\centering
%\includegraphics[width=0.4\linewidth, trim=0mm 0mm 0mm 0mm,
%clip]{Fig_CPV/scan_phi_WH1.pdf}\hspace{5mm}
%\\
%\includegraphics[width=0.4\linewidth, trim=0mm 0mm 0mm 0mm, clip]{Fig_CPV/scan_phi_WH2.pdf}\hspace{5mm}
%\includegraphics[width=0.4\linewidth, trim=0mm 0mm 0mm 0mm, clip]{Fig_CPV/scan_phi_WA1.pdf}\hspace{5mm}
%\caption{$\Gamma (H^{\pm}\to WH_1)$ as a function of $\varphi_\mu$
%  (red),$\varphi_{A_{t}}$ (blue) and $\varphi_{M_{2}}$ (green) at LO
%  (dashed) and NLO (full). Lower insert: Relative change of the decay
%  width $\delta_\Gamma)$. 
%\label{eq:CPVWHi}
%}
%\end{figure}
%%--------------------------------------

\begin{figure}[t]\centering
\includegraphics[width=0.4\linewidth, trim=0mm 1mm 3mm 1mm,
clip]{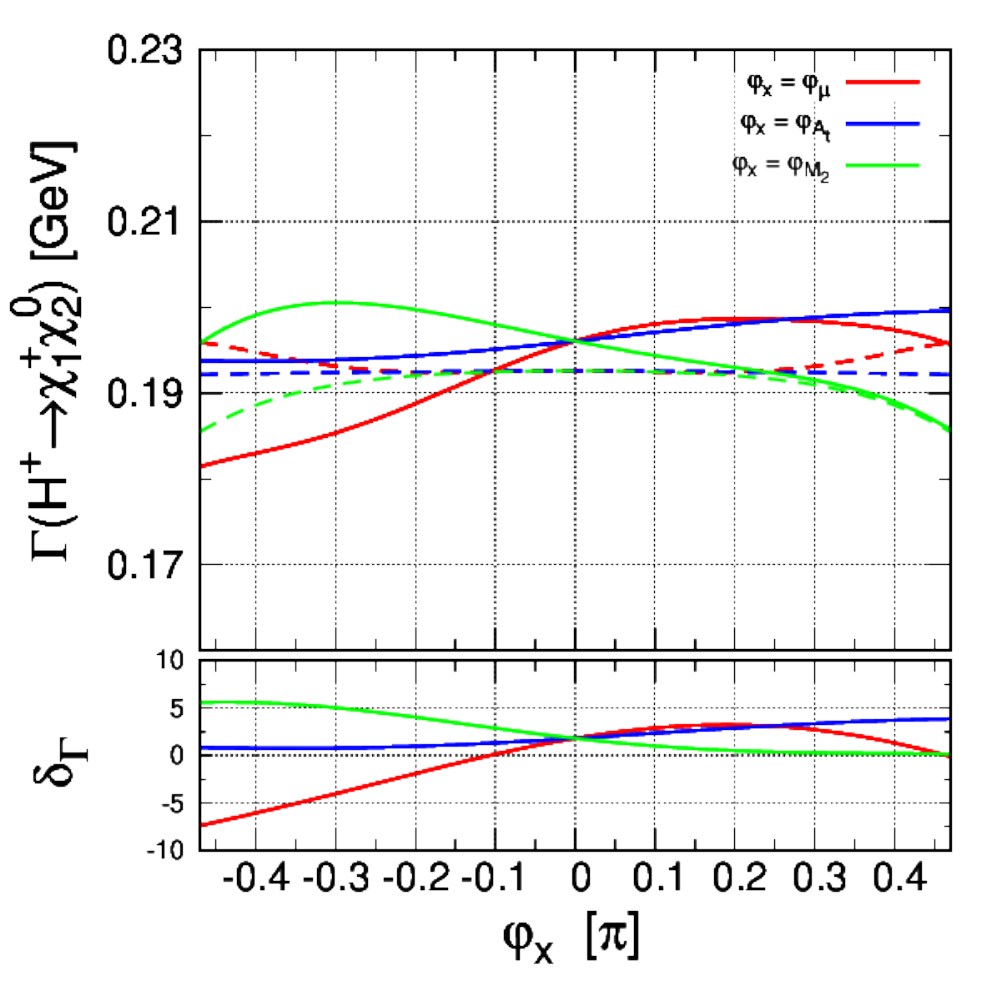}\hspace{10mm} 
\includegraphics[width=0.4\linewidth, trim=0mm 1mm 3mm 1mm, clip]{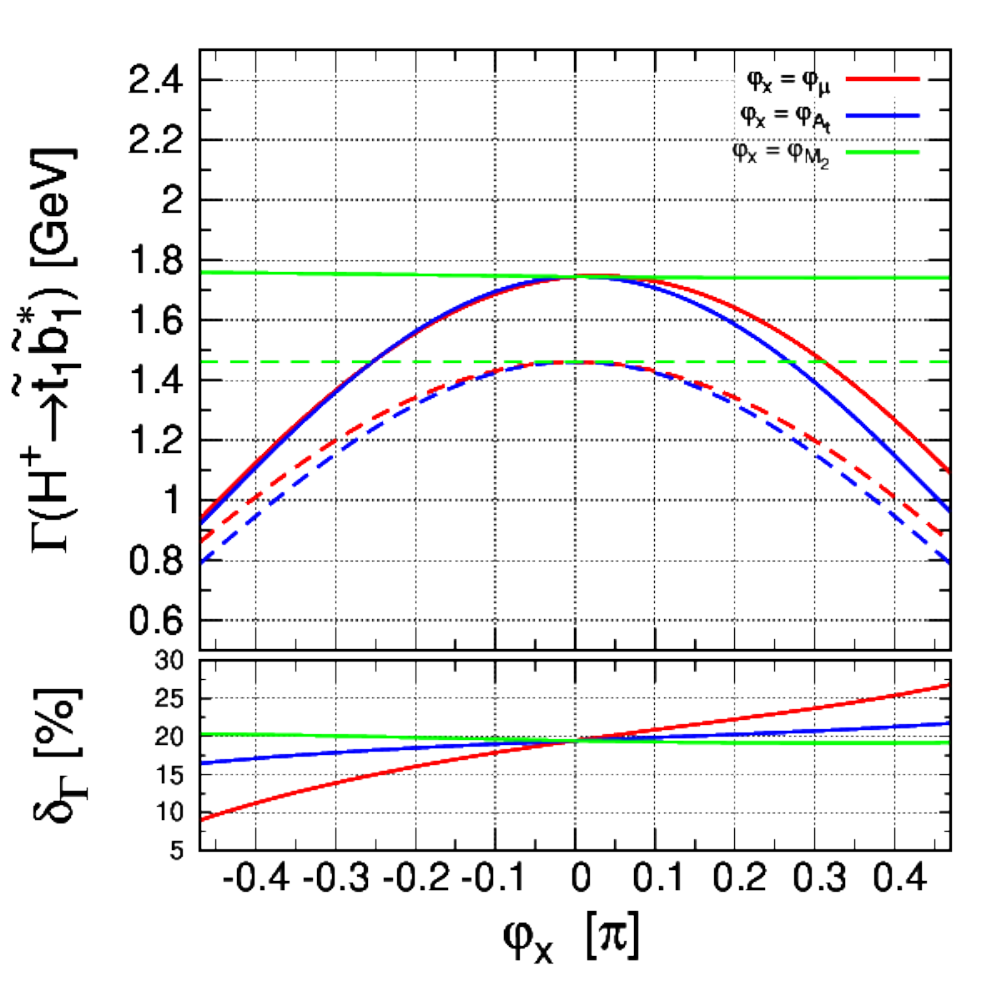}
\caption{Left: $\Gamma (H^+\to \chi^{+}_{1} \chi^{0}_{2})$, right:
  $\Gamma (H^+\to \tilde{t}_1 \tilde{b}_1^*)$ 
  as a function of $\varphi_\mu$
  (red), $\varphi_{A_{t}}$ (blue) and $\varphi_{M_{2}}$ (green) at LO
  (dashed) and NLO (full). Lower insert: Relative correction of the decay
  width $\delta_\Gamma$. 
\label{eq:CPVewikino}
}
\end{figure}

The charged Higgs couplings to the electroweakinos depend on the
phases $\varphi_\mu$ and $\varphi_{M_2}$ so that the 
LO decay widths already show a dependence on these CP-violating
phases. We exemplary show the effect on the decay $H^+ \to \tilde{\chi}_1^+
\tilde{\chi}_2^0$ in Fig.~\ref{eq:CPVewikino} (left). In this case the
$\tilde{\chi}_1^+$ is a higgsino-like chargino  while the
$\tilde{\chi}_2^0$ is a higgsino-like neutralino.  (The CP-violating impact is
found to be less important for the $\tilde{\chi}_1^+
\tilde{\chi}_1^0/\tilde{\chi}_3^0$ final states of this benchmark
point.) The relative correction $\delta_\Gamma$ shows
  a substantial dependence on all three phases. \s 
%%--------------------------------------CPV, electroweakinos
%%--------------------------------------

The tree-level charged Higgs couplings to squarks depend on the
phases $\varphi_\mu$ and $\varphi_{A_t}$ so that the LO
width shows a substantial dependence on these phases that is also
translated to NLO, as can be inferred from Fig.~\ref{eq:CPVewikino}
(right), which shows the decay $H^+ \to \tilde{t}_1 \tilde{b}_1^*$ as
a function of the CP-violating phases at LO and NLO. (The decays into
other stop-sbottom final states are kinematically closed.) The dependence on
$\varphi_{M_2}$ is only radiatively induced and very weak.
%%--------------------------------------CPV, stsb
%\begin{figure}[htbp]\centering
%\includegraphics[width=0.4\linewidth, trim=0mm 1mm 3mm 1mm, clip]{Fig_CPV/scan%_phi_stsb.pdf}\hspace{5mm}\\
%\caption{$\Gamma (H^{\pm}\to tb)$  as function of CPV phase for $\mu_{\rm eff}$,$A_{t}$ and $M_{2}$.
%Solid lines corresponds to the decay width at NLO, while dashed lines are those of LO. 
%\label{eq:CPVstsb}
%}
%\end{figure}
%%--------------------------------------

%%%%%%%%%%%%%%%%%%%%%%%%%%%%%%%%%%%%%%%%%%%%%%%%%%%%%%%%%%%%%%
\subsection{The Impact of the Renormalization Scheme}
%%%
\begin{table}[h]
\begin{center}
 \begin{tabular}{|ll||c|c|c|c|c|c|c|c|c|c|c|c|c|c|}
\hline
~~ & &${M_{\tilde{\chi}_1^0}}$&${M_{\tilde{\chi}^0_2}}$&${M_{\tilde{\chi}_3^0}}$&${M_{\tilde{\chi}_4^0}}$&${M_{\tilde{\chi}_5^0}}$&${M_{\tilde{\chi}_1^+}}$&${M_{\tilde{\chi}_2^+}}$ \\ \hline \hline
 \multirow{2}{*} {OS1}  & tree-level & 544.65 &  592.70 &  699.49 &  771.90 &  960.61 &  574.57 & 960.20  \\  
                       & one-loop & 549.32 &  596.02 &  698.83 &  771.90 &  960.54 &  580.40 &  960.20\\ \hline 
\multirow{2}{*} {$\DRb$}  & tree-level & 543.88 &  592.72 &  699.24 &  771.11 &  931.54 &  573.57 &  930.97\\  
                       & one-loop & 549.40 &  595.97 &  698.83 &  771.81 &  960.01 &  580.48 &  959.57\\ \hline 
& main component & $\tilde H_u^0$ &  $\tilde H_d^0$ &  $\tilde S $ &  $\tilde{B}$ &  $\tilde W_3$  &  $\tilde H^+$ &  $\tilde{W}^+$\\ \hline  
\end{tabular}
\caption{Masses (in GeV) and main components of the
  neutralino and chargino mass eigenstates at  
  tree level and one-loop level in the two renormalization schemes OS1 and $\DRb$.}
\label{tab:massneucha1}
\end{center}
\vspace*{-0.5cm}
\end{table}

\begin{table}[h!]
\begin{center}
 \begin{tabular}{|ll||c c c c|}
 \hline
    ~~&  & $m_{\ti t_1} \,[\mbox{GeV}]$ &$m_{\ti t_2} \,[\mbox{GeV}]$
   & $m_{\ti b_1}\,[\mbox{GeV}]$ & $m_{\ti b_2}\,[\mbox{GeV}]$\\ 
\hline
\multirow{2}{*}  {OS} & tree-level &  1062.22  & 1586.95 & 1146.27  & 1802.81 \\
                      & one-loop & 1062.22 &  1586.95 & 1150.31  & 1801.81 \\ \hline
\multirow{2}{*}  {$\DRb$} & tree-level &  1064.98  & 1626.29 & 1136.82  & 1758.28 \\
                      & one-loop &     1079.94  & 1583.78 & 1158.76 &  1806.97 \\ \hline
& main component & $\tilde t_L$ &  $\tilde t_R$ &  $\tilde b_L $ &  $\tilde b_R$ \\ \hline  
\end{tabular}
\caption{The tree-level and one-loop corrected stop
  and sbottom masses in the $\DRb$ and the OS scheme.} 
\label{tab:StopSbottomMasses}
\end{center}
\vspace*{-0.5cm}
\end{table}

\begin{table}[h!]
\begin{center}
 \begin{tabular}{|ll||c c c |}
 \hline
    ~~&  & $m_{\ti \tau_1} \,[\mbox{GeV}] $ &$m_{\ti \tau_2} \,[\mbox{GeV}]$
   & $m_{\ti \nu_\tau}\,[\mbox{GeV}]$ \\ 
\hline
\multirow{2}{*}  {OS} & tree-level &  496.51  & 1659.16 & 490.50  \\
                      & one-loop & 514.16 &  1659.16 & 490.50  \\ \hline
\multirow{2}{*}  {$\DRb$} & tree-level &  478.48  & 1658.01 & 472.24   \\
                      & one-loop &     496.74  & 1659.17 & 472.46 \\ \hline
& main component & $\tilde \tau_L$ &  $\tilde \tau_R$ &  $\tilde \nu_L $  \\ \hline  
\end{tabular}
\caption{The tree-level and one-loop corrected stau
  and tau sneutrino masses in the $\DRb$ and the OS scheme.} 
\label{tab:StausneuMasses}
\end{center}
\vspace*{-0.5cm}
\end{table}

We now want to discuss the impact of the renormalization scheme. The
renormalization scheme dependence of the decay widths also gives us a
possibility to roughly estimate the remaining theoretical error due to
missing higher-order corrections. The code {\tt NMSSMCALCEW} that we
use for the computation of the decay widths and branching ratios
follows the SLHA conventions where the soft SUSY breaking parameters
are understood to be $\overline{\mbox{DR}}$ input parameters at the
scale $M_{\text{SUSY}}$ which per default is given as in
Eq.~(\ref{eq:renscale}). Consequently, depending on the chosen
renormalization scheme for the computation of the higher-order
corrections the input parameters have to be converted to the applied
scheme where necessary. For example, if we use OS renormalization in
the stop sector then the soft SUSY-breaking parameters ($\ti m_{Q_3}^2$, 
$\ti m_{t_R}^2$, $A_t$) affecting the
stop sector must be converted from the SLHA $\overline{\mbox{DR}}$
parameters to OS parameters. With these converted parameters the
NLO width is calculated then. In Ref.~\cite{Baglio:2019nlc} we outlined in
detail the procedure applied in {\tt NMSSMCALCEW} to convert the input
parameters. In the following we show the change in the LO and NLO
widths for the electroweakino and sfermion decays when applying
different renormalization schemes after consistently converting the
input parameters. The benchmark point used in the plots is the same as
for the investigation of the impact of CP-violation in the previous
subsection~\ref{sec:cpviol}, we only vary $\tan\beta$
  in the following. In order to quantify the renormalization 
scheme dependence we introduce 
\begin{align}
\Delta_{ \Gamma}= 
\left| \frac{\Gamma^{\rm OS}-\Gamma^{\rm \overline{DR}}}{\Gamma^{\rm
  OS}} \right| \;,
\label{eq:rendef}
\end{align}
where $\Gamma^{\rm OS}$ $(\Gamma^{\rm \overline{DR}})$ denotes partial
width evaluated in the OS ($\overline{\rm DR}$) scheme. 
It should be noted that in both renormalization schemes we use
  the loop-corrected masses for the external lines only while we use
  tree-level masses and tree-level couplings for particles inside
  loops.  For our chosen parameter point, we present the
  tree-level and loop-corrected masses for the electroweakinos in
  Table~\ref{tab:massneucha1}, for the stops/sbottoms in
Table~\ref{tab:StopSbottomMasses}, and for the staus/tau-sneutrino in
Table~\ref{tab:StausneuMasses}.

%--------------------------------------
%\subsubsection*{\underline{Decays into electroweakinos} }
%--------------------------------------renormalization scheme dependence, ewikinos
\begin{figure}[htbp]\centering
\includegraphics[width=0.4\linewidth, trim=0mm 1mm 2mm 1mm, clip]{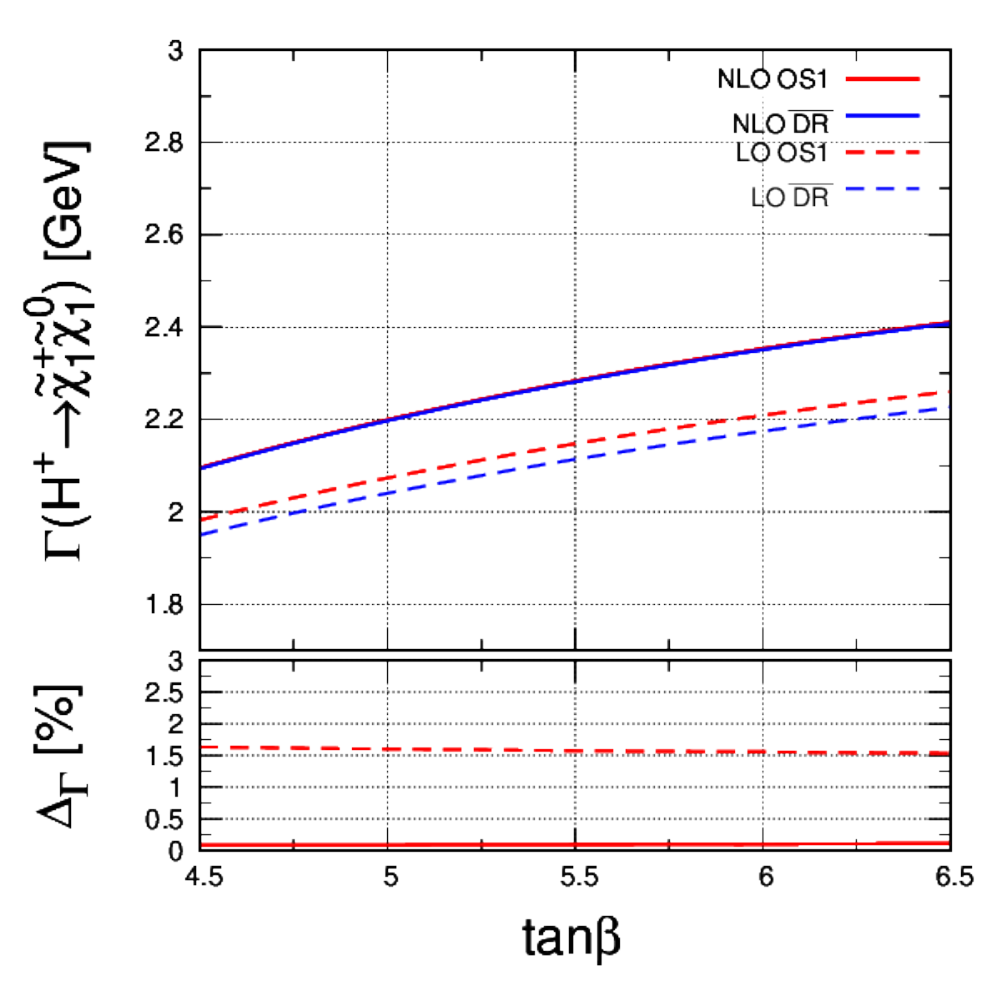}\\
\includegraphics[width=0.4\linewidth, trim=0mm 1mm 2mm 1mm, clip]{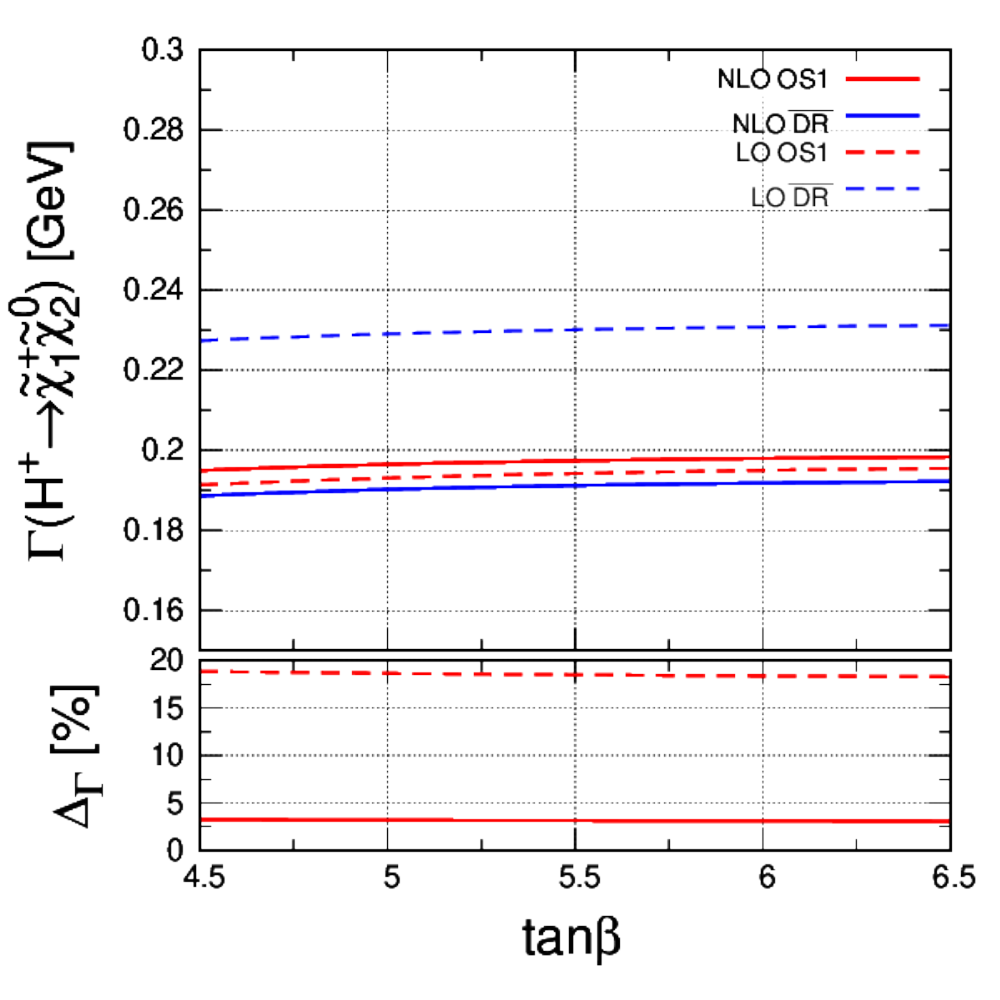}\hspace{5mm}
\includegraphics[width=0.4\linewidth, trim=0mm 1mm 2mm 1mm, clip]{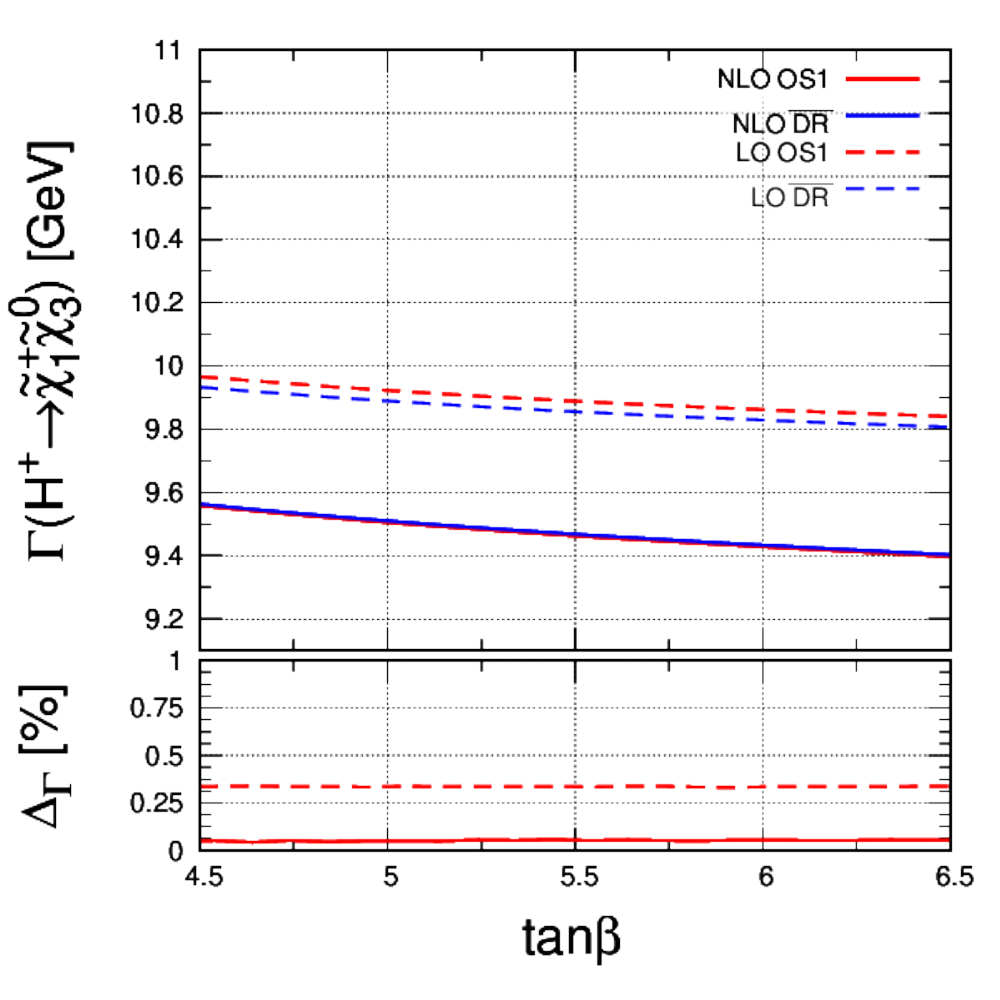}
\caption{The partial decay width for the decays of $H^+$ into the
  electroweakino final states $\tilde{\chi}_1^+\tilde{\chi}_1^0$,
  $\tilde{\chi}_1^+\tilde{\chi}_2^0$, $\tilde{\chi}_1^+\tilde{\chi}_3^0$
  as a function of $\tan\beta$ for OS1 (red) and
  $\overline{\mbox{DR}}$ (blue) renormalization at LO (dashed) and NLO
  (full). Lower insert: The relative difference in the widths due to 
  different renormalization schemes at LO (dashed) and NLO
  (full). 
\label{fig:renodepewikino}
  }
\end{figure}
%--------------------------------------
Figure~\ref{fig:renodepewikino} shows the LO and NLO widths for the
charged Higgs decays into the exemplary electroweakino final states $\tilde{\chi}_1^+
\tilde{\chi}_1^0$, $\tilde{\chi}_1^+\tilde{\chi}_2^0$, and $\tilde{\chi}_1^+
\tilde{\chi}_3^0$, respectively, for OS1
renormalization\footnote{We remind the reader that in the OS1 scheme the
  soft-breaking parameters $M_{1}$ and $M_{2}$ are renormalized by
  the on-shell conditions for the wino-like chargino and the bino-like
  neutralino.}  (red) and for $\overline{\mbox{DR}}$ renormalization (blue)
at LO (dashed) and NLO (full) as a function of
$\tan\beta$. The lower
insert displays $\Delta\Gamma$ where in the definition
Eq.~(\ref{eq:rendef}) OS has to be replaced by OS1. As we can inferred
from the plots for all three decays the dependence on the
renormalization scheme decreases when going from LO to NLO, as
expected. For the $\tilde{\chi}_1^+
\tilde{\chi}_1^0$ and $\tilde{\chi}_1^+\tilde{\chi}_3^0$ final states,
already at LO the dependence is very small with $\Delta_\Gamma \approx
1.5$\% for the former and 0.32\% for the latter,
almost independently of $\tan\beta$. It gets reduced to close to
0\% at NLO. For the $\tilde{\chi}_1^+
\tilde{\chi}_2^0$ final state the LO difference between OS1 and
$\overline{\mbox{DR}}$ renormalization is larger than in the other
decays, as is the relative NLO correction to the decay
width in the OS1 scheme. For this final state the LO dependence on the renormalization
scheme gets reduced from about 18\% to 3\% at NLO.
These results are in accordance with the observed small dependence of the
loop-corrected final state masses on the renormalization scheme, {\it
  cf.}~Table \ref{tab:massneucha1}. \s 

%For the decays into electroweakinos, the renormalization of electroweakinos is relevant, and we have three choices; two OS schemes (OS1 and OS2), and  $\overline{\rm DR}$ scheme. 
%In OS1, soft-breaking parameters $M_{1}$, and $M_{2}$ are renormalized by the on-shell conditions for wino-like chargino and bino-like neutralino. 
%Whereas, in OS2, $M_{1}$, and $M_{2}$ are renormalized by those for wino-like neutralino and bino-like neutralino.
%We have checked that there is a few numerical difference between OS1 and OS2, using the benchmark point given in Eq.~\eqref{eq:BPCP}. 
%In order to describe difference between the OS scheme and $\overline{\rm DR}$ scheme, we compute decay width for electroweakinos in each renormalization scheme, . 
%
%In Fig.~ \ref{fig:renodepewikino}, The decay width and the size of NLO correction for $H^{+}\to \chi^{+}_{1}\chi^{0}_{1}$, $H^{+}\to \chi^{+}_{1}\chi^{0}_{2}$ and $H^{+}\to \chi^{+}_{1}\chi^{0}_{3}$ are shown as a function of $\tan\beta$. 
%
%
%--------------------------------------
%\subsubsection*{\underline{Decays into stop and sbottoms} }
%--------------------------------------renormalization scheme dependence, st sb
\begin{figure}[htbp]\centering
\includegraphics[width=0.4\linewidth, trim=1mm 1mm 1mm 1mm, clip]{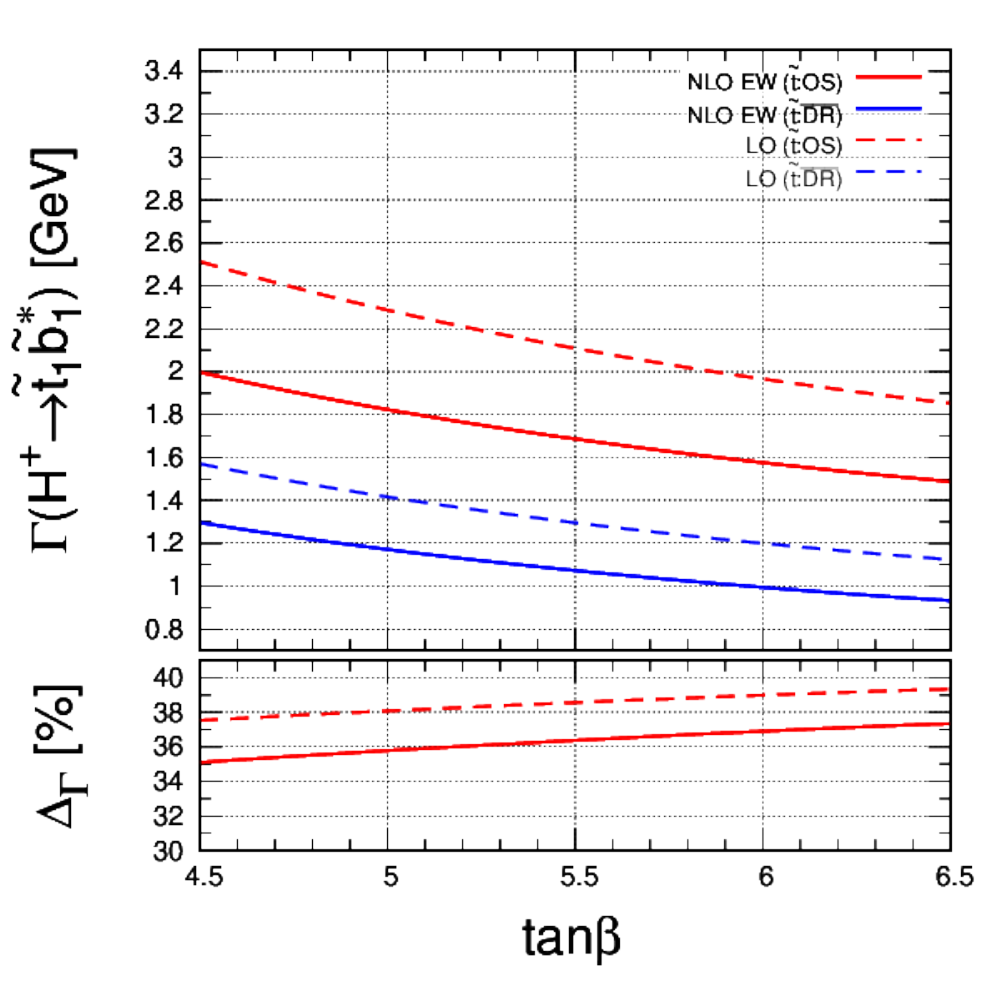}\hspace{5mm}
\includegraphics[width=0.4\linewidth, trim=1mm 1mm 1mm 1mm, clip]{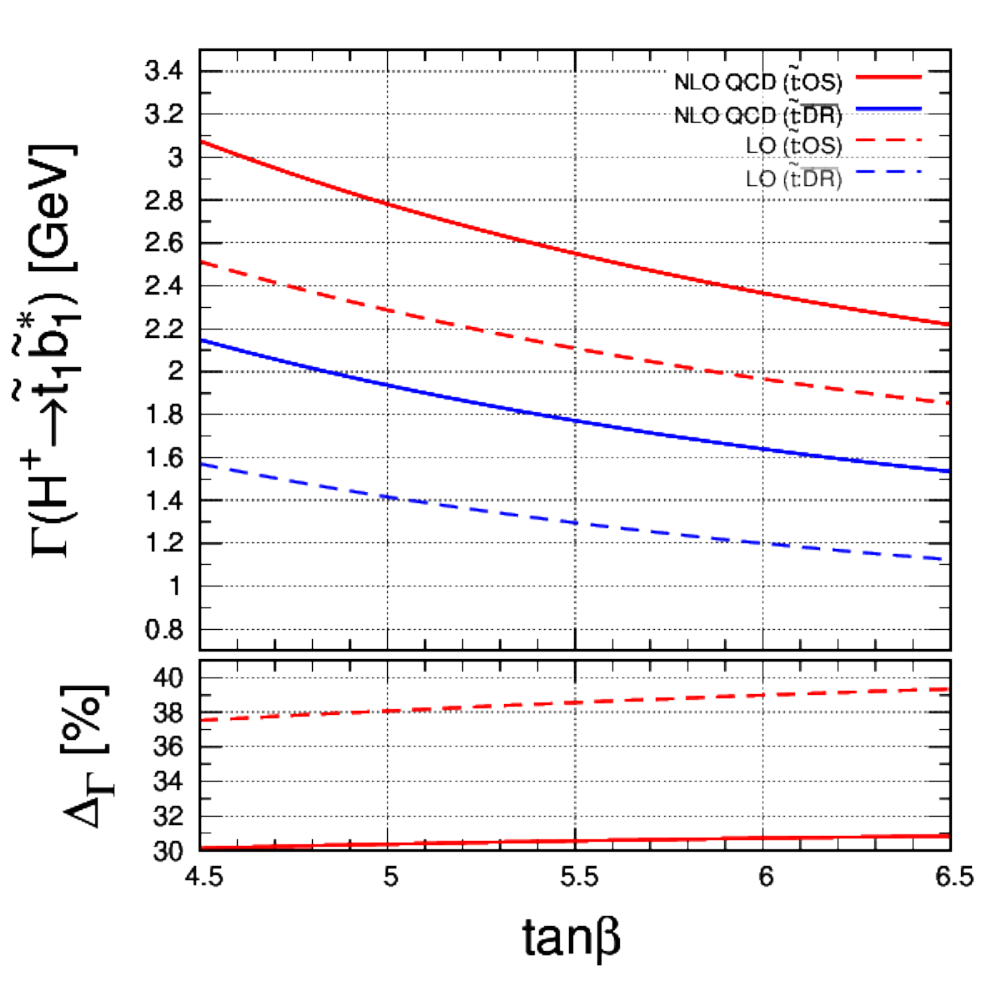}
\caption{The partial decay width for the decay $H^+ \to \tilde{t}_1 \tilde{b}_1^*$
  as a function of $\tan\beta$ for OS (red) and
  $\overline{\mbox{DR}}$ (blue) renormalization in the stop sector at LO (dashed) and NLO
  (full) for the SUSY-EW (left) and the SUSY-QCD corrections (right). Lower insert: The relative difference in the widths due to 
  different renormalization schemes at LO (dashed) and NLO
  (full). }
\label{fig:renodepstsb}
\end{figure}

In Fig.~\ref{fig:renodepstsb} we display the LO and the NLO widths
for the charged Higgs decay into $\tilde{t}_1 \tilde{b}_1^*$ for OS
renormalization (red) and for $\overline{\mbox{DR}}$ renormalization in the stop sector (blue)
at LO (dashed) and NLO (full) for the SUSY-EW (left) and
the SUSY-QCD corrections (right), as a function of $\tan\beta$. The lower
insert displays $\Delta\Gamma$ as defined in
Eq.~(\ref{eq:rendef}). As expected, the plots show that the LO
dependence on the renormalization 
scheme gets reduced when going to NLO, both for the SUSY-EW and
the SUSY-QCD corrections, namely from about 38\%
  (39\%) to 35\% (37\%) 
for the SUSY-EW corrections for $\tan\beta=4.5$ (6.5). For the
SUSY-QCD corrections the reduction is more important, going down to
about 30\%. The overall larger dependence on the renormalization
schemes compared to the electroweakino final states is also reflected
in the larger dependence of the final state squark masses on the
renormalization scheme, {\it cf.}~Table~\ref{tab:StopSbottomMasses}.\s

\begin{figure}[htbp]\centering
\includegraphics[width=0.4\linewidth, trim=1mm 1mm 1mm 1mm, clip]{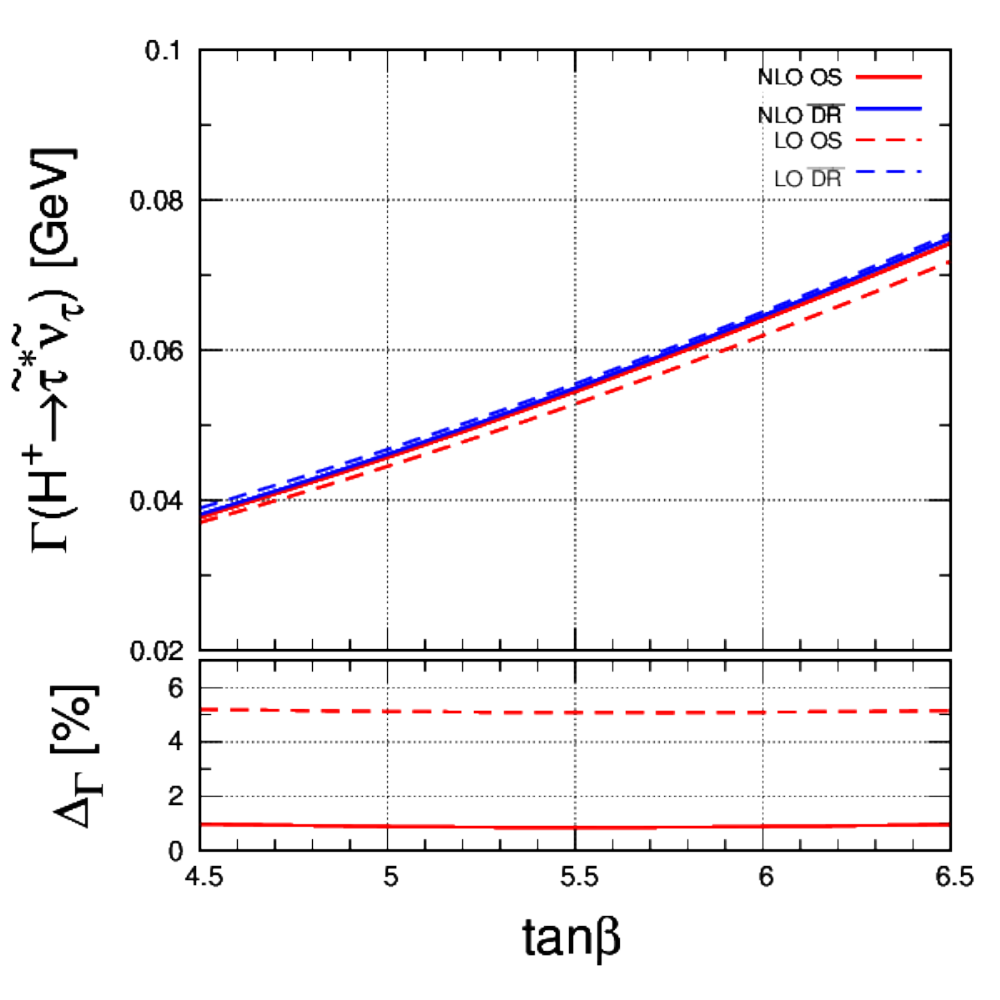}\hspace{5mm}
\caption{The partial decay width for the decays into stau-sneutrino pairs
  as a function of $\tan\beta$ for OS (red) and
  $\overline{\mbox{DR}}$ (blue) renormalization at LO (dashed) and NLO
  (full). Lower insert: The relative difference in the widths due to 
  different renormalization schemes at LO (dashed) and NLO
  (full).}
\label{fig:renodepstausneut}
\end{figure}

Figure \ref{fig:renodepstausneut} shows the $H^+ \to \tilde{\tau}^*
\tilde{\nu}_\tau$ decay (summing over the two possible
$\tilde{\tau}_i$ ($i=1,2$) states) for 
OS (red) and the $\overline{\mbox{DR}}$ (blue) renormalization at LO (dashed)
and SUSY-EW NLO (full) as a function of $\tan\beta$. For the chosen
parameter point the SUSY-EW corrections are rather small, and the
rather small dependence on the renormalization scheme at LO with
$\Delta_\Gamma = 5\%$ gets reduced to 1\% at NLO.
%--------------------------------------
%

%%%%%%%%%%%%%%%%%%%%%%%%%%%%%%%%%%%%%%%%%%%%%%%%%%%%%%%%%%%%%%
\section{Conclusions \label{sec:conclusions}} 
In this paper we complete the computation of the NLO SUSY-EW and
SUSY-QCD corrections to the charged Higgs boson decays in the
CP-conserving and CP-violating NMSSM that was started in a previous paper with the
corrections to the charged $W^+$ plus Higgs boson final states. 
We provide the missing corrections to the on-shell two-body decays
into the SM fermion, the electroweakino and the sfermion final
states. For the decays into electroweakinos and sfermions we provide
the corrections for different renormalization schemes, chosen to
be OS and $\overline{\mbox{DR}}$. This allows us to roughly estimate
the remaining theoretical uncertainty due to missing higher-order
corrections. All corrections have been implemented in the code {\tt
  NMSSMCALCEW} thereby combining them with the already incorporated
state-of-the-art higher-order QCD corrections. \s

In our numerical
analysis we find that the newly computed SUSY-EW and SUSY-QCD
corrections are significant and need to be included for meaningful
predictions of the decay branching ratios. In specific corners of the
parameter space with small LO widths and/or large mixing effects the
corrections can become very large. As for the dependence on the
CP-violating phases of the various parameters
the effects are of typical size for radiatively induced CP
violation. The investigation of the renormalization scheme dependence
shows a good perturbative convergence of the higher-order
corrections. Our results contribute to the improvement of the precision on the
predictions for Higgs boson observables in beyond-the-SM extensions
that is required for the correct interpretation of new physics effects
being looked for at the LHC. 

%%%%%%%%%%%%%%%%%%%%%%%%%%%%%%%%%%%%%%%%%%%%%%%%%%%%%%%%%%%%%%
\section{Acknowledgements}
T.N.D is funded by the Vietnam National Foundation for Science
and Technology Development (NAFOSTED) under grant number 103.01-2020.17.  
M.M. and K.S. acknowledge support by the Deutsche Forschungsgemeinschaft
(DFG, German Research Foundation) under grant  396021762 - TRR
257. 
%%%%%%%%%%%%%%%%%%%%%%%%%%%%%%%%%%%%%%%%%%%%%%%%%%%%%%%%%%
%\newpage

%\bibliography{bibHpm_decay_NMSSM}
%\bibliographystyle{JHEP}

%%%%%%%%%%%%%%%%%%%%%%%%%%%%%%%%%%%%%%%%%%%%%%%%%%%%%%%%%%

\end{document}